\documentclass[11pt,a4]{article}
\pdfoutput=1
\usepackage{jheppub}
\usepackage{graphicx,amssymb,amsmath,amsfonts}
\usepackage{hyperref}
\usepackage[utf8]{inputenc}

\def\be{\begin{equation}}
\def\ee{\end{equation}}

\def\bmat{\begin{pmatrix}}
\def\emat{\end{pmatrix}}

\def\bdet{\begin{vmatrix}}
\def\edet{\end{vmatrix}}

\numberwithin{equation}{section}

\def\bea{\begin{eqnarray}}
\def\eea{\end{eqnarray}}

\newcommand{\dprime}{{\prime\prime}}

\newcommand{\alp}{\ensuremath{\alpha^\prime}}

\newcommand{\pa}{\partial}
\newcommand{\ra}{\rightarrow}
\newcommand{\vac}{\vert 0 \rangle}
\title{Quantizing the rotating string with massive endpoints}

\author{Jacob Sonnenschein and Dorin Weissman}

\affiliation{The Raymond and Beverly Sackler School of Physics and Astronomy,\\
	Tel Aviv University, Ramat Aviv 69978, Israel}

\emailAdd{cobi@post.tau.ac.il}
\emailAdd{dorinw@mail.tau.ac.il}

\abstract{We compute leading order quantum corrections to the Regge trajectory of a rotating string with massive endpoints using semiclassical methods. We expand the bosonic string action around a classical rotating solution to quadratic order in the fluctuations and perform the canonical quantization of the resulting theory. For a rotating string in \(D\) dimensions the intercept receives contributions from \(D-3\) transverse modes and one mode in the plane of rotation, in addition to a contribution due to the Polchinski-Strominger term of the non-critical effective string action when \(D\neq26\). The intercept at leading order is proportional to the expectation value of the worldsheet Hamiltonian of the fluctuations, and this is shown explicitly in several cases. All contributions to the intercept are considered, and we show a simple physical method to renormalize the divergences in them. The intercept converges to known results at the massless limit, and corrections from the masses are explicitly calculated at the long string limit. In the process we also determine the quantum spectrum of the string with massive endpoints, and analyze the asymmetric case of two different endpoint masses.}

\begin{document}

\maketitle

\tableofcontents
\flushbottom
 
\section{Introduction}
More than four decades have passed since A. Chodos and C. Thorn suggested the model of a bosonic string with massive endpoints \cite{Chodos:1973gt}. In spite of the fact that it is a natural model for a QCD meson, with the string describing the chromoelectric flux tube and the massive endpoints the quarks, the string with massive endpoints has not been studied exhaustively and up to date a full quantization of it has not been established.

In the original paper \cite{Chodos:1973gt}, a rotating string solution of the classical equations of motion was written down. This solution implies a modification of the classical Regge trajectory, the relation between the angular momentum and the energy of the system.  We refer to it as the massive modified Regge trajectory of the string. Various other aspects of the model have been studied in the old era, for instance in \cite{Frampton:1975en,Bars:1975dd,Bardeen:1975gx,Kikkawa:1977cga,Ida:1977uy}.

A renaissance of the  model has followed the application of the gauge/string correspondence to QCD and hadron physics. In \cite{Kruczenski:2004me} it was shown that a classical  rotating string in a holographic ten dimensional curved confining  background with its endpoints on flavor branes can be mapped into a rotating sting in four flat dimensions with massive endpoints. Based on this idea  and several other ingredients from gauge/string duality the holography inspired stringy hadron (HISH) model was proposed in \cite{Sonnenschein:2016pim}. This model aims at describing all hadrons: glueballs, mesons, baryons, and exotic  hadrons are all strings, where closed strings correspond to glueballs and open strings with various configurations of particles on their ends to the other hadrons. Since a rotating string in a holographic background is dual to a meson with non-trivial angular momentum, this implies in particular that the holographic meson  trajectories can be  approximated by the modified trajectories of the model of the string with massive endpoints.

There are several definitions of the mass of a quark. In particular there are the QCD (current) quark mass and the constituent quark mass. The mass of the particle at the end of the string, referred to as the string endpoint mass $m_{sep}$, is neither of the above \cite{Kruczenski:2004me}. This is supported by fits of the theoretical Regge trajectory, with the correction for endpoint masses, to experimental data \cite{Sonnenschein:2014jwa}, where the best fits for the $m_{sep}$ show that $m_{sep}$ is generally between the QCD and the constituent quark masses. It was further argued in \cite{Sonnenschein:2014bia}, following the holographic picture, that baryons can also be described as a single string but now with non-symmetric endpoints where on one side there is a quark and on the other a baryonic vertex connected with two short strings to a diquark. The fact that it is a straight string and not a Y-shaped configuration is backed by a theoretical analysis of the stability of the Y-shape string configuration \cite{Sharov:2002ja,tHooft:2004doe} and moreover by the fact that the Regge slope (\(\alp\)) for baryons \cite{Sonnenschein:2014bia} is within 5\% the same as for mesons. Thus the string with two massive endpoints can be used as a model for both mesons and baryons, and also for exotic hadrons \cite{Sonnenschein:2016ibx}.
	
The system of a string with particles on its endpoints can take a variety of different forms, as the endpoint particles, in addition to being massive can also carry electric charge, flavor charge and spin. In this paper we only consider chargeless, spinless endpoints. For this case, with no additional properties to the particles, one can consider the symmetric case where the two masses at the endpoints are the same or the more general asymmetric case where there are two different masses. Strings with electrically charged endpoint particles were analyzed in \cite{BSS} and we currently investigate the role of the spin, which was considered also in \cite{Pisarski:1987jb,Olsson:1992wt}. In \cite{Sever:2017ylk} it was shown that the the asymptotic behavior of the four point amplitude that corresponds to this model is a universal leading order correction of the Veneziano amplitude. Various other aspects of the model were investigated in \cite{Solovev:1998bd,Inopin:1999nf,Allen:2001wu,Baker:2002km,Aharony:2009gg,Aharony:2013ipa,Zahn:2016bam,Zahn:2017gyz}.

For an ordinary bosonic string with no massive endpoints, the passage from the classical Regge trajectory to the leading quantum Regge trajectory is accomplished by adding an intercept,
 \be\label{classicalRt}
 J= \alp M^2 \ra J= \alp M^2 +a 
 \ee 
 In the critical dimension \(D=26\) the intercept can be computed by doing the sum over the eigenfrequencies of the \(D-2\) transverse excitations of the string, with the result
\be a = -\frac{D-2}{2}\sum_{n=1}^\infty n = \frac{D-2}{24} = 1 \label{eq:12} \ee
where the infinite sum is given by the Riemann Zeta function, \(\zeta(-1) = -\frac{1}{12}\).

The main goal of this paper was to determine the corresponding passage from classical to quantum Regge trajectory for the string with massive endpoints. Denoting the quantum corrections to the classical energy and angular momentum by  $E_{cl} \ra E_{cl}+ \delta E$ and  $J_{cl}\ra J_{cl}+ \delta J$, we show in this paper that the following relation holds
\be
 a \equiv \langle \delta\left(J - J_{cl}(E)\right) \rangle = \langle \delta J- \frac{1}{k} \delta E\rangle= -\frac1k\langle \mathcal H_{ws} \rangle
\ee 
where \(J_{cl}(E)\) is the classical relation between \(J\) and \(E\) and $k$ is the angular velocity of the rotating string (as described in section \ref{sec:classical}). This defines for us the intercept in the massive case, as the quantum correction to the relation between \(J\) and \(E\). One can easily verify that for the case of a string without massive endpoints this result coincides with \ref{classicalRt}. Thus the determination of the quantum Regge trajectory translates to computing the intercept for the string with massive endpoints model. As for the last part of the equation, we show explicitly here that the intercept is proportional to the expectation value of the Hamiltonian for the quantum fluctuations around the classical rotating solution, as derived more generally in \cite{Frolov:2002av,PandoZayas:2003yb}.

In $D$ spacetime dimensions the intercept for the rotating string acquires contributions from fluctuations in the $D-3$ directions transverse to the plane of rotation and from one planar mode, which is transverse to the string but in the plane of rotation. One must also consider radial fluctuations of the massive endpoints, which serve to modify the boundary condition for the planar mode. In this paper, we analyze separately the transverse and planar fluctuations and determine their contributions to the intercept.

For a bosonic string in a general number of $D$ dimensions, as for instance in the four dimensional HISH model, that is not the end of the story. In non-critical dimensions 
one is required to incorporate also the Liouville \cite{Polyakov:1981rd} or the Polchinski-Strominger (PS) term of effective string theory \cite{Polchinski:1991ax} in order to render the quantization procedure consistent. This was generally not done in previous papers that considered the quantization of a string with massive endpoints.

The incorporation of the PS term to the string with no massive endpoints was done in \cite{Hellerman:2013kba}. The result of that paper was that the intercept of the bosonic string is $a=1$, not only in the critical dimension but in fact for any spacetime dimension $D$. This is because the contribution of the Liouville mode cancels out the dependence on \(D\), as
\be a(m=0) = \frac{D-2}{24} + \frac{26-D}{24} = 1\ee
This result was obtained by introducing a  particular boundary counterterm to cancel the divergence of the PS term and renormalizing the whole system. For the rotating string with finite endpoint masses the PS term does not diverge, and in fact the endpoint masses can be thought of as a regulator to this divergence.

One of the outcomes of the present paper is  that instead of the procedure used in \cite{Hellerman:2013kba} one can add massive endpoints, perform the calculation of the intercept  and then take the limit of zero endpoint mass. Traditionally, one uses the Riemann Zeta function to perform the renormalization of the intercept. In this paper we further develop a procedure that was proposed in \cite{Lambiase:1995st} by generalizing it from the static to the rotating string. This procedure is based on converting the infinite sum over the eigenfrequencies into a contour integral and subtracting from the corresponding Casimir force of a string of a given length $L$ the Casimir force when $L\to \infty$. In the present paper we apply this method to both the planar and transverse modes of the rotating string, and we use a similar subtraction also for the PS mode.\footnote{In spite of the fact that it does not diverge one has to perform the subtraction used also for the other modes to get the correct finite result.} We show that the renormalized contribution of the PS term to the intercept takes the form
\be
a_{PS} = \frac{26-D}{12\pi} \arcsin\beta
\ee
where $\beta$ is the velocity of the endpoint particles for the case of identical  endpoint masses.

Alternatively we show that one can also renormalize the  various contributions to the intercept by renormalizing the string tension and the endpoint masses. Namely, the divergences are such that can be eliminated by adding counterterms to the action redefining \(T = T_{bare}+\delta T\) and \(m = m_{bare}+\delta m\), with the appropriate choice of the themselves divergent coefficients \(\delta T\) and \(\delta m\).

Altogether, the intercept is given by
\be a = (D-3)a_t + a_p + a_{PS} \ee
where \(a_t\) is the contribution of each transverse mode and \(a_p\) that of the planar mode. All of the different contributions are ultimately given as functions of the endpoint velocity \(\beta\), with the limit \(\beta\to1\) replicating the result of the massless case. The intercept can also be expanded in powers of \(m/TL\), with the final result
\be a = 1 - \frac{26-D}{12\pi}(\frac{2m}{TL})^{1/2} + \frac{199-14D}{240\pi}(\frac{2m}{TL})^{3/2} \ee
accounting for all contributions to the intercept in any spacetime dimension \(D\).

The action of a string in critical dimensions is that of a set of free fields. The action of the endpoint particles is in fact also that of free particles. But once we couple the string and the particles the system is no longer free even in the critical dimensions. Using the orthogonal gauge \eqref{orthogonal}, one can bring the Nambu-Goto action to a quadratic form. However, this will not happen for the action terms of the endpoint particles. To quantize the system, we take the semiclassical approach in which we expand our action to quadratic order in fluctuations around the classical solution (a rotating string), and then canonically quantize the fluctuations. We have found that the truncation to second order in the fluctuations is valid in the regime where the string is long in comparison with all length scales in the problem. Namely, we require that $mL \gg 1$, $TL^2 \gg1$, and also $\frac{TL}{m} \gg 1$, where $T$ is the string tension, $m$ the endpoint masses, and $L$ the length of the string.

The intercept plays an important role in the phenomenology of stringy hadrons. In \cite{Sonnenschein:2014jwa} and \cite{Sonnenschein:2014bia} we have determined the best fit values of the intercepts associated with with the massive modified Regge trajectories of various hadrons. It is clear from the results that the intercept is a function of the endpoint particles' masses, as also the spin of the hadron. One of the goals of this paper is to decipher the dependence of the intercept on the endpoint masses. Furthermore, upon analyzing the experimental data of the trajectories of  all  mesons and baryons, it turns out  that the intercept (defined in the plane of $M^2$ and the orbital angular momentum) is always negative $a<0$. This property is crucial for the description of hadrons in terms of strings. Essentially, the issue is how can one justify using a bosonic string model which usually is known to have a tachyonic ground state. A negative intercept also implies a repulsive Casimir force between the endpoint particles which prevents the string from collapsing to zero size when it does not rotate, hence the tachyonic nature is avoided. Therefore, a natural question for the present work is in under what conditions does the string with massive endpoints admit a negative intercept and whether using this model one can account for the observed negative values of the various intercepts.

This paper does not offer the answer to this question of the phenomenological intercept, as we only compute the corrections due to the mass to the result \(a=1\) corresponding to the string without massive endpoints. On the other hand, one could argue that the intercept we calculate, the asymptotic intercept at high energies is a completely different quantity to the phenomenological intercept of low spins.

In the process of to computing the intercept we also determine the quantum spectrum of the string with massive endpoints. We showed that in for this model the well known linear quantum trajectory for a radially excited state $J + N = \alp E^2  + a$ is generalized to
\be J + \frac{1}{\beta}(N_t+N_p) = J_{cl}(E) + a
\ee
where \(J_{cl}(E)\) is the classical relation between the energy and angular momentum, the massive modified Regge trajectory, and \(N_t\) and \(N_p\) are the  excitation numbers for the transverse and planar modes respectively. The eigenvalues of \(N_t\) and \(N_p\) are not exactly integers as in the massless case and they have to be computed by solving the equations for the allowed eigenfrequencies of the different modes. Like the intercept, they are also a function of \(\beta\), and return to their massless values as \(\beta\to1\).
 
The paper is organized as follows. In section \ref{sec:classical} the classical string with massive endpoints is described. We write the action and the corresponding equations of motion. We describe the rotating string solution. The classical energy and angular momentum are written down, with the relation between them constituting the classical massive modified Regge trajectory. In section \ref{sec:gauge} we show how we add the fluctuations around the classical solution and discuss the choice of gauge to be made before quantizing them.

Section \ref{sec:transverse} in the longest part of this paper. In it we analyze the transverse fluctuations, and show in detail the calculation of their contribution to the intercept. We start by writing down the action and Hamiltonian. We write down the mode expansion for the fluctuation which is followed by the canonical quantization of the modes. The solutions of the equations of motion for the fluctuations are determined. Then we write the expressions for the quantum corrections to the energy and angular momentum, and from them the corresponding correction of the classical Regge trajectory. We show that the intercept is proportional to the sum of the eigenfrequencies of the modes, or the expectation value of the worldsheet Hamiltonian of the fluctuations. Finally, we discuss the renormalization of this infinite sum and find the finite answer for the intercept. We show how we renormalize first for the massless case, then the static (non-rotating) massive string, and finally the rotating string with massive endpoints.

The subject of section \ref{sec:planar} is the planar mode. This section mirrors the section preceding it as we write down the action and Hamiltonian for the fluctuations, the mode expansion, the equation of motion and boundary conditions (corrected by the radial mode living on the boundary), and finally determine the contribution to the intercept. Section \ref{sec:noncritical} discusses the quantization of the string in non-critical dimensions. We write down the Polchinski-Strominger term in the action and the associated intercept. We renormalize the term and determine the finite contribution to the total intercept.

Following the determination of the intercept, we discuss the quantum massive modified Regge trajectory in section \ref{sec:quantumRegge}. That is, we discuss the spectrum of states and the radial trajectories. In section \ref{sec:asym} we summarize and generalize the results of the previous sections from the symmetric to the asymmetric case where the masses on the two ends of the string are not equal. In section \ref{sec:nextorder} we examine the range of validity of the quadratic approximation, and argue that it is a long string approximation. Section \ref{sec:summary} is a summary of the results of the paper, including a list of open questions.
%
%

\section{The classical string with massive endpoints} \label{sec:classical}
\subsection{Action and equations of motion}
We describe the string with massive endpoints by combining the Nambu-Goto string action,
\be S_{st} = -T\int d\tau d\sigma \sqrt{-h} = -T\int d\tau d\sigma \sqrt{\dot X^2 X^\prime{}^2 - (\dot X\cdot X^\prime)^2} \ee
with the point particle action
\be S_{pp} = -m\int d\tau \sqrt{-\dot X^2} \ee
In the string action \(h_{\alpha\beta} = \eta_{\mu\nu}\pa_\alpha X^\mu \pa_\beta X^\nu\) is the induced metric on the worldsheet and \(h = \det h_{\alpha\beta}\) is its determinant, The indices \(\alpha\) and \(\beta\) being either \(\tau\) or \(\sigma\). To keep track of units, we take the worldsheet coordinates \(\tau\) and \(\sigma\) to have dimension of length, with \(-\infty<\tau<\infty\) and \(-\ell\leq\sigma\leq\ell\). Two copies of the point particle action are inserted at the boundaries, the two string endpoints \(\sigma=\pm\ell\), so
\be S = S_{st} + S_{pp}\vert_{\sigma=-\ell} + S_{pp}\vert_{\sigma=\ell} \ee
 For this section and most of the following, we will assume a symmetric string with two equal endpoint masses, generalizing only in section \ref{sec:asym}, after we have computed all of our results for the symmetric case.

Varying \(X^\mu\) we find the bulk equations of motion
\be \pa_\alpha(\sqrt{-h}h^{\alpha\beta}\pa_\beta X^\mu) = 0 \ee
and also the endpoint particles equations of motion/boundary conditions of the string
\be T\sqrt{-h}\pa^\sigma X^\mu \pm m \pa_\tau\left(\frac{\dot X^\mu}{\sqrt{-\dot X^2}}\right) = 0\ee
where the plus sign should be taken at \(\sigma = \ell\) and minus at \(\sigma=-\ell\).

\subsection{The classical rotating solution}
We now define the classical rotating solution for the string. We pick a solution rotating in the \(12\) plane, and find its classical energy and angular momentum. It should be noted here that when the spacetime dimension is \(D > 4\) what we describe is not the most general rotating solution. Rather, for \(D>4\) the general rotating solution is described by two angular momenta in two different planes (this follows from the fact that the rotation group \(SO(D-1)\) for \(D>4\) contains \(SO(4)\sim SU(2)\times SU(2)\)). In this paper we pick the case of rotation in a single plane, given our eventual interest in \(D=4\). The general case was analyzed in \cite{Hellerman:2013kba} for the string without masses.

To describe the string rotating in the \(12\) plane, we use the configuration given by
\be\label{classicalconf} X^0 = \tau, \qquad X^1 = R(\sigma)\cos(k\tau), \qquad X^2 = R(\sigma)\sin(k\tau) \ee
which is a solution to the bulk equations of motion for any choice of \(R(\sigma)\). For this solution the induced metric is
\be h_{\alpha\beta} = \bmat -1+k^2R^2 & 0 \\ 0 & R^\prime{}^2 \emat \label{eq:hab}\ee
To be a solution, we must also pick the parameters that will satisfy the boundary conditions at \(\sigma = \pm \ell\):
\be T\frac{\sqrt{(1-k^2R^2)R^\prime{}^2}}{R^\prime} \mp m \frac{k^2R}{\sqrt{1-k^2R^2}} = 0\ee

The world sheet parameters \(\ell\) and \(k\) can be related to more physically meaningful (that is parametrization independent) target space parameters. First, the length of the string in target space is defined by
\be L = \int_{-\ell}^\ell d\sigma |\frac{d \vec X}{d\sigma}| = \int_{-\ell}^\ell d\sigma\sqrt{R^{\prime2}} = R(\ell)-R(-\ell) = 2R(\ell)\ee
We assumed \(R(\sigma)\) is monotonous, and that we picked an antisymmetric solution, as we would for an open string with two identical masses on its endpoints. The endpoint velocity \(\beta\) is defined using
\be \gamma^{-1} = \sqrt{1-\beta^2} \equiv \sqrt{-\dot X^2} = \sqrt{1-k^2 R^2} \ee
evaluated at \(\sigma = \pm\ell\). We can see that in terms of these target space parameters, the boundary condition can be written as
\be \frac{T}{\gamma} = \frac{2\gamma m \beta^2}{L} \qquad \Rightarrow \qquad \frac{TL}{2m} = \gamma^2\beta^2 \label{eq:bdc}\ee
regardless of the choice of \(R(\sigma)\). This has the very simple interpretation as the requirement that the centrifugal force acting on the massive endpoint particle be balanced by the string tension, with appropriate relativistic factors of \(\gamma\) on both sides of the equation.

There are two choices of \(R(\sigma)\) we can take to simplify the expressions. These are \(R(\sigma) = \frac{1}{k}\sin(k\sigma)\) and \(R(\sigma) = \sigma\). The two choices will both be utilized later on. The boundary conditions, string length and endpoint velocity in each case are
\begin{align} R(\sigma) &= \sigma: \qquad &\frac{T}{mk} &= \frac{k\ell}{1-k^2\ell^2} \qquad &L &= 2\ell \qquad &\beta &= k\ell \label{eq:conditions_lin}\\
R(\sigma) &= \frac{1}{k}\sin(k\sigma): \qquad &\frac{T}{mk} &= \frac{\sin(k\ell)}{\cos^2(k\ell)} \qquad &L &= \frac{2}{k}\sin(k\ell) \qquad &\beta &= \sin(k\ell) \label{eq:conditions_sin}\end{align}
In the latter case we define the useful parameter \(\delta \equiv k\ell = \arcsin\beta\), which ranges from \(0\) to \(\frac\pi2\).

\subsection{Energy and angular momentum} \label{sec:classical_EJ}
The action for the string with massive endpoints naturally has the full Poincar\'e symmetry \(X^\mu\to \Lambda^{\mu}_\nu X^\nu + a^\mu\). In the coordinate system where
\be (x^0, x^1, x^2, \ldots) = (t, \rho\cos\theta, \rho\sin\theta,\ldots)\,, \ee
we define the energy \(E\) and angular momentum \(J\) as the Noether charges associated with the translation symmetries in \(t\) and rotations in the \(12\)-plane (which are translations in \(\theta\)) respectively. The contribution of the string to each is
\be E_{st}  = -T\int_{-\ell}^\ell d\sigma \sqrt{-h}h^{\tau\alpha}\pa_\alpha t \qquad
 J_{st}  = -T \int_{-\ell}^\ell d\sigma \sqrt{-h} \rho^2 h^{\tau\alpha}\pa_\alpha \theta \ee
The contributions of the point particles,
\be E_{pp} = m\frac{\dot t}{\sqrt{-\dot X^2}} \qquad J_{pp} = m\frac{\rho^2\dot\theta}{\sqrt{-\dot X^2}} \ee

For the rotating string solution, the classical energy and angular momentum are expressible as functions of \(T\), \(m\), \(L\) and \(\beta\):
\begin{align} E &= \frac{2m}{\sqrt{1-\beta^2}} + TL\frac{\arcsin\beta}{\beta} \label{eq:classical_E}\\
J &= \frac{m L \beta}{\sqrt{1-\beta^2}}+\frac14TL^2\frac{\arcsin\beta-\beta\sqrt{1-\beta^2}}{\beta^2} \label{eq:classical_J}\end{align}
Using the boundary condition \ref{eq:bdc} one can eliminate the string length from the equations and write both \(E\) and \(J\) as a function of \(T\), \(m\), and the single continuous parameter \(0\leq\beta<1\). The resulting parametric relation between \(J\) and \(E\) defines the classical Regge trajectory of the string with massive endpoints, or what we call the \emph{massive modified Regge trajectory}. The limit \(\beta\to1\) is the massless limit where one obtains the linear Regge trajectory \(E^2 = 2\pi T J\). For \(\beta\) close to one we can write expansions in \(\frac{1}{\gamma}=\sqrt{1-\beta^2}\) for \(E\) and \(J\) and find the approximate relation
\be J = \frac{1}{2\pi T}E^2\left(1 -\frac{8\sqrt{\pi}}{3}\left(\frac m E\right)^{3/2}+\frac{2\pi^{3/2}}{5}\left(\frac m E\right)^{5/2}+\ldots\right) \label{eq:JclE} \ee
Our goal in this paper is to find the quantum correction to this classical trajectory. To do that, we will introduce fluctuations around the rotating solution.

\section{Fluctuations and gauge choice} \label{sec:gauge}
We start by defining
\be X^\mu = X^\mu_{cl}+\delta X^\mu = \left(t\,,\rho\,,\theta\,,z^i\right) = \left(\tau + \lambda\delta t,R(\sigma) + \lambda\delta \rho,k \tau+\lambda\delta\theta,\lambda\delta z^i\right)\,.\ee
That is, we introduce fluctuations around the rotating solution defined in the previous section. For convenience, we work with polar coordinates in the plane of rotation. The fluctuations are all multiplied by a formal expansion parameter \(\lambda\). This parameter can be ultimately absorbed into the definition of the fluctuations, and should not appear in expressions for physical quantities. Later, in section \ref{sec:nextorder}, we will see that the small parameter in the expansion is actually \(\frac{1}{mL}\), meaning that the next to leading order corrections are suppressed by this factor.

The Nambu-Goto action is diffeomorphism invariant, that is it is symmetric under
\be \tau \to \tilde\tau(\tau,\sigma) \qquad \sigma\to\tilde\sigma(\tau,\sigma)\qquad \tilde X^\mu(\tilde\tau,\tilde\sigma) = X^\mu(\tau,\sigma) \ee
and we can use this gauge freedom to impose two conditions on \(X^\mu\) in the bulk. There are two useful and common gauge choices one might use in this case. First is the \textbf{orthogonal gauge}, defined by
\begin{align} &\frac12(h_{\tau\tau}+h_{\sigma\sigma})  = \frac12(\dot X^2+X^{\prime2}) = 0 \nonumber  \\
&h_{\tau\sigma}  = h_{\sigma\tau} = \dot X \cdot X^\prime = 0 \label{orthogonal}\end{align}
The induced metric can then be written in the form \(h_{\alpha\beta} = e^\phi \eta_{\alpha\beta}\). This gauge has the advantage of linearizing the equations of motion of the string, leaving the NG action quadratic. However, it is not particularly helpful in the case of the rotating string with massive endpoints. This is because the boundary condition due to the mass term remains non-linear, and the gauge constraint itself has a non-trivial form when the string is rotating.

Instead, when quantizing the fluctuations we will pick the \textbf{static gauge}, where we use diff invariance to set the fluctuations in the time direction to zero by fixing
\be \tau = X^0 \qquad \Rightarrow \qquad \delta t = 0\, \ee
Similarly by setting \(\sigma\) we use it to specify a choice of \(R(\sigma)\) for the rotating solution. We will solve explicitly the system for two different choices of \(R(\sigma)\) and show that they are equivalent. We will also see later on that \(\delta\rho\), the perturbation added to \(R(\sigma)\) is not dynamic in the bulk because of the remaining reparametrization invariance in \(\sigma\), but it does introduce some boundary terms which affect the dynamics of the other fluctuation in the plane of rotation, \(\delta\theta\). This is expected since on the boundary the choice \(X^0 = \tau\) uses up the reparametrization freedom on the worldline (\(\tau\to\tilde\tau(\tau)\)), and we have to take account also of \(\delta\rho\).

Later, we will also look at the contribution to the intercept from the Polchinski-Strominger term of non-critical effective string theory. To write the PS term we will use the orthogonal gauge, where we have a simple form for the PS action. On the other hand, in leading order the PS intercept depends only on the classical solution, so we can can satisfy the gauge choice by picking the rotating solution with \(R(\sigma) = \frac{1}{k}\sin(k\sigma)\), which obeys the orthogonal gauge constraints at the classical level (see eq. \ref{eq:hab}). The result for the PS intercept and the intercept from the different fluctuations should be separately gauge independent, so no problems will arise if we use for each part of the calculation the most appropriate gauge choice for that part.

To summarize, for the quantization of the fluctuations we use the gauge freedom to set \(\delta t = 0\) and then set \(R(\sigma)=\sigma\) or \(R(\sigma)=\frac1k\sin(k\sigma)\). In both cases we can solve the system, and the two choices will be shown to be equivalent. For the PS term, where the other fluctuations do not enter the picture, it suffices to set \(R(\sigma)=\frac1k\sin(k\sigma)\) to compute the PS intercept.


\section{Transverse fluctuations} \label{sec:transverse}
In this section we will compute the contribution to the intercept from the transverse modes \(\delta z^i\). These are all the mode orthogonal to the plane of rotation. For a general number of dimensions there are \(D-3\) of these modes.

We will review in detail all the necessary steps. First is the expansion of the original Nambu-Goto plus point particle action around the classical solution, then the solution of the resulting equations of motion. This will give us the spectrum of the transverse modes. Then, we show that the intercept is given by the sum of eigenfrequencies, and show how to compute and renormalize the infinite sum to obtain a finite result.

\subsection{Action and Hamiltonian}
We expand our action around the rotating solution of section \ref{sec:classical} with the gauge choice explained in section \ref{sec:gauge}, taking terms up to quadratic order in the fluctuations, that is to order \(\lambda^2\).

The resulting action for each of the transverse perturbations (since all transverse modes are identical we omit the index \(i\) from \(\delta z^i\) from now on) is
\begin{align} S_{st,\delta z} &= -T\lambda^2\int d\tau d\sigma \left[\frac12\left(\sqrt{R^{\prime2}}g\right)^{-1}\delta z^{\prime2}-\frac12\left(\sqrt{R^{\prime2}}g\right)\delta \dot z^2\right] \\
S_{pp,\delta z} &= m\lambda^2\int d\tau \frac{1}{2}\gamma\delta\dot z^2 \end{align}
We have defined \(g(\sigma) = (1-k^2R^2)^{-1/2}\) (position dependent time dilation along the rotating string) and \(\gamma = g(\ell)\) is its value on the boundary. We have a copy of \(S_{pp}\) on each of the boundaries, \(\sigma=\pm\ell\).

To get properly normalized kinetic terms, we define
\be f_t \equiv (\sqrt{R^{\prime2}}g)^{1/2}\delta z \ee
One can write the general expression for the action for \(f_t\), but it is simpler to proceed after specifying the function \(R(\sigma)\). For \(R(\sigma) = \frac{1}{k}\sin(k\sigma) \), then \(f_t = \delta z\) and the action reduces to the simple form
\begin{align} S_{st,\delta z} &= -T\lambda^2\int d\tau d\sigma \left(\frac12f_t^{\prime2}-\frac12\dot f_t^2\right) \\
S_{pp,\delta z} &= m\lambda^2\int d\tau \frac{1}{2}\gamma\dot f_t^2 \end{align}
For \(R(\sigma) = \sigma\), which is just a different parametrization of the same classical solution, we get a different picture for the transverse modes, which now have added position dependent mass terms:
\begin{align*} S_{st,t} &= -T\lambda^2\int d\tau d\sigma \left(-\frac12\dot f_t^2+\frac12 g^{-2}f_t^{\prime2}+\frac18k^2(1+g^2)f_t^2\right) \\ 
 S_{pp,t} &=  -\lambda^2\int d\tau \left(-\frac14Tk^2\ell f_t^2 - \frac12m\dot f_t^2\right) \end{align*}
The advantage of the latter picture is the straightforward relation between the worldsheet parameters \(k,\ell\) to the target space length (\(L=2\ell\)) and velocity (\(\beta = k\ell\)), as seen in section \ref{sec:classical}. Since we can find exact solutions for \(f_t\) using both choices for \(R(\sigma)\), we can show explicitly that the two formulations give identical results.

\subsubsection{The worldsheet Hamiltonian}
Also of interest is the Hamiltonian derived from the above action. We have defined the modes in such a way that the conjugate momentum to the coordinate \(f_t\) is just its time derivative, up to some constants. For \(R(\sigma)=\frac{1}{k}\sin(k\sigma)\), 
\be \pi_t = \frac{\pa \mathcal L}{\pa \dot f_t} = \lambda^2 (T + \gamma m\delta(\sigma\pm \ell))\dot f_t \ee
The worldsheet Hamiltonian is then
\be H = \frac12T\lambda^2 \left(\int_{-\ell}^{\ell} d\sigma (\dot f_t^2 + f_t^{\prime2}) + \frac{\gamma m}{T} \dot f_t^2|_{\pm\ell}\right)\ee
Similarly, for \(R(\sigma) = \sigma\) we have
\be \pi_t = \frac{\pa \mathcal L}{\pa \dot f_t} = \lambda^2 (T + m\delta(\sigma\pm \ell))\dot f_t \ee
\be H = \frac12T\lambda^2 \left[\int_{-\ell}^{\ell} d\sigma \left(\dot f_t^2+\frac12 g^{-2}f_t^{\prime2}+\frac14k^2(1+g^2)f_t^2\right) + \left(\frac{m}{T} \dot f_t^2 -\frac12k^2\ell f_t^2\right)|_{\pm\ell}\right]\ee

\subsection{Mode expansion and canonical quantization}
Before writing and solving the equations of motion we outline the procedure used to solve the system. We write a mode expansion for the solution as follows.
\be f_t = f_0 + i \sqrt{\mathcal N} \sum_{n\neq0}\frac{\alpha_n}{\omega_n}e^{-i\omega_n\tau}f_n(\sigma) \ee
To keep \(f_t\) real, we require \(\alpha_{-n}=\alpha_n^*\), and additionally we can use \(\omega_{-n} = -\omega_n\) and \(f_{-n}(\sigma) = f_n(\sigma)\). The zero mode \(f_0\) is not relevant to our calculation so we omit it from now on.

For the functions \(f_n\), we will need to solve a Sturm-Liouville problem. Most generally, we will have a differential equation of the form
\be \frac{d}{dx}\left(p(x)\frac{df_n}{dx}\right) - q(x)f_n(x) = -\lambda_n w(x) f_n(x) \label{eq:SturmLiouville}\ee
for \(a\leq x \leq b\) and with boundary conditions relating \(f\) and \(f^\prime\) at the two boundary points \(x = a, b\). One property which we will need to use is the orthogonality relation for eigenmodes with \(m \neq n\).
\be (\lambda_m-\lambda_n)\int_a^b dx w f_m f_n = \left(f_m p \frac{d f_n}{dx}-f_n p \frac{d f_m}{dx}\right)\vert_{a}^b \label{eq:SturmLiouvilleOrtho} \ee
This can be derived directly from eq. \ref{eq:SturmLiouville}. For Neumann or periodic boundary conditions the RHS is zero. For other types of boundary conditions, we need to include the boundary terms of the RHS in the orthogonality relation.

Now we turn to solving the boundary conditions. The most general procedure is as follows. Since the equation of motion for a given eigenmode \(f_n(\sigma)\) is a linear second order differential equation, its general solution is of the form
\be f_n(\sigma) = c_1 f_n^{(1)}(\sigma) + c_2 f_n^{(2)}(\sigma) \ee
for two linearly independent functions \(f_n^{(1)}\) and \(f_n^{(2)}\). The boundary condition is of the form \(\mathcal O_{a,b} f|_{x=a,b} = 0\), for some linear differential operator. Using this notation, the two boundary conditions can be combined into the equation
\be \mathcal M \bmat c_1 \\ c_2 \emat \equiv \bmat \mathcal O_a f_n^{(1)}|_{a} & \mathcal O_af_n^{(2)}|_{a} \\ \mathcal O_b f_n^{(1)}|_{b} & \mathcal O_b f_n^{(2)}|_{b} \emat \bmat c_1 \\ c_2 \emat = 0 \label{eq:eigenfrequencies} \ee
The eigenfrequencies are obtained by requiring that 
\be \det \mathcal M = 0\ee
so that there are non-trivial solutions for \((c_1,c_2)\). We will use this requirement to write the equation determining the allowed eigenfrequencies in all the following cases.

\subsubsection{Quantizing the modes} \label{sec:canonical_quantization}
To quantize the modes, we want to impose the commutation relation
\be [f_t(\sigma),\pi_t(\sigma^\prime)] = i\delta(\sigma-\sigma^\prime) \ee
for canonical quantization of the fluctuations. One can show that this holds if we require that
\be [\alpha_m,\alpha_n] = \omega_m\ell \delta_{n+m} \ee
We include a factor of \(\ell\) in the commutator since \(\omega_n\) has in our units dimensions of mass.
Inserting the mode expansions for \(f_t\) and \(\dot f_t\),
\be [f_t(\sigma),\dot f_t(\sigma^\prime)] = i \mathcal N \ell \sum_{n} f_n(\sigma) f_n(\sigma^\prime) \ee
Then, for the conjugate momentum (we pick the case \(R(\sigma) = \frac{1}{k}\sin(k\sigma)\) to illustrate this):
\be [f_t(\sigma),\pi_t(\sigma^\prime)] = i N T\lambda^2 \sum_n \left(1+\frac{\gamma m}{T}\delta(\sigma^\prime\pm\ell)\right)f_n(\sigma) f_n(\sigma^\prime) \ee
If we multiply the RHS by some eigenfunction \(f_k(\sigma^\prime)\) and integrate over \(\sigma^\prime\), we find that
\begin{align} \int d\sigma^\prime & f_m(\sigma^\prime)[f_t(\sigma),\pi_t(\sigma^\prime)] = \nonumber\\ &i N T\ell\lambda^2 \sum_n f_n(\sigma) \int_{-\ell}^\ell d\sigma^\prime \left(1+\frac{\gamma m}{T}\delta(\sigma^\prime\pm\ell)\right)f_m(\sigma^\prime) f_n(\sigma^\prime) = 2 i N T\ell^2 \lambda^2 f_m(\sigma) \end{align}
where in the last step we use the orthogonality relation of the eigenmodes (as derived from eq. \ref{eq:SturmLiouvilleOrtho}) to carry out the integral. The above is sufficient to prove that the commutator is a delta function up to constants, namely that
\be [f_t(\sigma),\pi_t(\sigma^\prime)] = 2i \mathcal N T\ell^2\lambda^2 \delta(\sigma-\sigma^\prime) \ee
So, with the normalization constant \(\mathcal N = \frac{1}{2T\ell^2\lambda^2}\), we have the properly normalized commutator between \(f_t\) and its conjugate momentum. This will be true in the other cases we examine as long as we normalize the modes the same way.

\subsection{Equations of motion and their solutions}
We can derive the equations of motion and boundary condition for the fluctuations in two ways. One is to take the equations of motion derived from the original Nambu-Goto plus point particle action and expand them to linear order in the fluctuations. The second way, which yields the same results, is to derive them by varying the quadratic action for the fluctuations.

\subsubsection{First formulation}
For \(R(\sigma) = \frac{1}{k}\sin(k\sigma)\), the bulk equation of motion of a mode with frequency \(\omega_n\) is simple,
\be f_n^\dprime + \omega_n^2 f_n = 0 \ee
The boundary condition at \(\sigma=\pm\ell\) is
\be T f_n^\prime \mp \gamma m \omega_n^2 f_n = 0 \ee
The general solution to the bulk equation of motion is
\be f_n(\sigma) = C_1 \sin(\omega_n \sigma) + C_2 \cos(\omega_n\sigma) \ee
The equation for the eigenfrequencies is derived using eq. \ref{eq:eigenfrequencies}
\be 2\delta\cot(\delta) x\cos(2x) + (\delta^2-\cot^2(\delta) x^2)\sin(2x) = 0 \label{eq:w_n_sin}\ee
where we have defined the dimensionless parameters
\be x \equiv \omega_n\ell\qquad \delta \equiv k \ell = \arccos(\gamma^{-1})\ee
We can also write the equation as
\be \left(x\sin x\cot\delta+ \delta \cos x\right)\left(x\cos x\cos\delta - \delta \sin x\right) = 0 \ee
For the symmetric string, with identical masses at the two endpoints, the eigenmodes are either odd - the term in the left brackets is zero and the solution is just \(\sin(\omega_n\sigma)\) - or vice versa for an even solution, \(\cos(\omega_n\sigma)\).

\subsubsection{Second formulation}
With the choice \(R(\sigma) = \sigma\), the bulk equation of motion can be written as
\be (1-x^2) f^\dprime_n(x) -2x f^\prime(x) + \left(\frac{\omega_n^2}{k^2} - \frac14 - \frac{1}{4(1-x^2)}\right)f_n(x) = 0 \ee
where \(x = k\sigma\), and derivatives are done with respect to \(x\). The boundary conditions at \(x=\pm k\ell\) are
\be \frac{T k}{\gamma^2} f^\prime_n \mp (m\omega_n^2+\frac12 Tk^2\ell)f_n = 0 \label{eq:boundary_t_lin}\ee

The equation of motion is an instance of the general Legendre equation, whose solutions are given in terms of the Legendre \(P\) and \(Q\) functions. The general solution for the \(n\)-th mode is then
\be f_n(\sigma) = c_1 P_{\nu_n}^{1/2}(k\sigma) + c_2 Q_{\nu_n}^{1/2}(k\sigma)\qquad\nu_n \equiv \frac{\omega_n}{k}-\frac12\ee
The Legendre functions of this order are given explicitly by
\begin{align} P_\nu^{1/2}(x) = \sqrt{\frac2\pi}(1-x^2)^{-1/4}\cos\left((\nu+\frac12)\arccos x\right) \\
Q_\nu^{1/2}(x) = -\sqrt{\frac\pi2}(1-x^2)^{-1/4}\sin\left((\nu+\frac12)\arccos x\right) \label{eq:LegendrePQ}\end{align}
and these functions indeed satisfy the equation of motion.
The eigenfrequency equation derived from this solution using eq. \ref{eq:eigenfrequencies} can be ultimately reduced to the simple form
\be 2 x \beta^2\sqrt{1-\beta^2} \cos\left(\frac{2 x \arcsin\beta}{\beta}\right) + (
\beta^4-(1-\beta^2)x^2)\sin\left(\frac{2 x \arcsin\beta}{\beta}\right) = 0 \label{eq:w_t_lin}\ee
Where
\be x=\omega_n\ell \qquad \beta = k\ell \ee
And, like in the previous case the equation can be factorized to two separate equations for odd and even modes:
\begin{align} 
&\left[\sqrt{1-\beta^2}x\sin\left(\frac{x \arcsin\beta}{\beta}\right)-\beta^2\cos\left(\frac{ x \arcsin\beta}{\beta}\right)\right]  \times \nonumber \\ 
&\qquad\left[\sqrt{1-\beta^2}x\cos\left(\frac{ x \arcsin\beta}{\beta}\right)+\beta^2\sin\left(\frac{x \arcsin\beta}{\beta}\right)\right]  = 0
\end{align}
Up to the factor of \(\frac{\arcsin\beta}{\beta}\) multiplying \(x=\omega_n\ell\), this is the exact same equation as the one we got in the previous subsection (remember that \(\beta=\sin\delta\)). In fact, we have the same exact set of eigenfrequencies \(\omega_n\) in both cases. The factor of \(\frac{\arcsin\beta}{\beta} = \frac\delta\beta\) enters because of the different ways the parameter \(\ell\) and consequently the variable \(x\) are defined in each case.

\begin{figure} \centering
\includegraphics[width=0.48\textwidth]{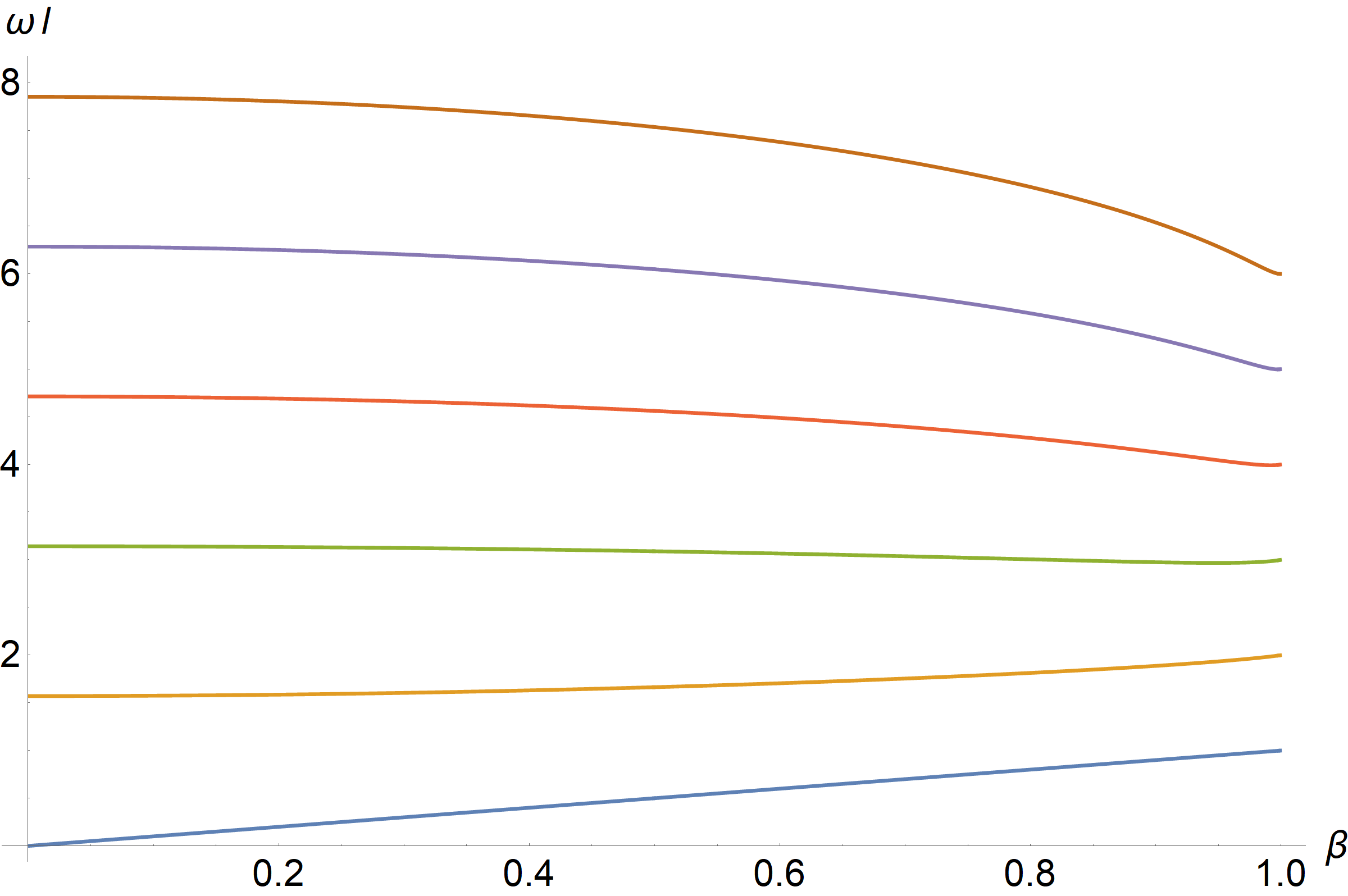}
\includegraphics[width=0.48\textwidth]{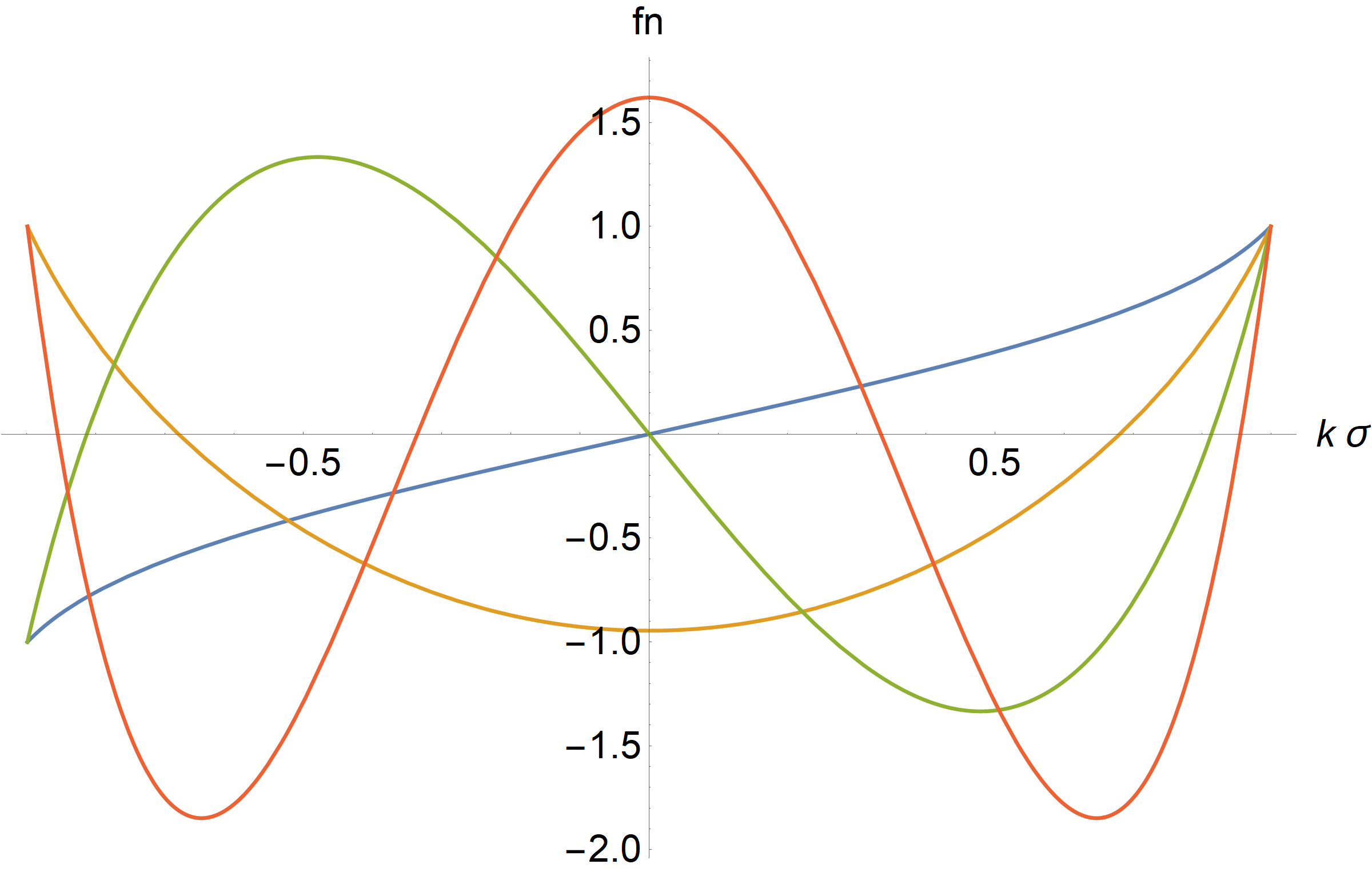}
\caption{\label{fig:spectrum_t} The first few eigenfrequencies \((\omega_n\ell)\) for the transverse mode as a function of \(\beta\) (left), and the first few eigenfunctions plotted for \(\beta=0.9\) (right). The modes are plotted in the gauge with \(R(\sigma)=\sigma\).}
\end{figure}

We plot the first few values of \(\omega_n\) as a function of \(\beta\) in figure \ref{fig:spectrum_t}. We also plot the first few eigenmodes \(f_n\) for a specific value of \(\beta\). The modes are seen to be either even or odd.

\subsection{Energy, angular momentum, and the corrected Regge trajectory}
Expanding the general expressions for \(E\) and \(J\) (eqs. \ref{eq:classical_E} and \ref{eq:classical_J}) to quadratic order in \(\lambda\), and isolating the contribution from the transverse modes, we get that for \(R(\sigma) = \frac{1}{k}\sin(k\sigma)\),
\begin{align} E_{st,f_t} & = T\lambda^2\int_{-\ell}^\ell d\sigma \frac12\frac{1}{\cos^2(k\sigma)}(f_t^{\prime2}+\dot f_t^2 ) \label{eq:Estt}\\
E_{pp,f_t} &= \frac12m\lambda^2 \gamma^3\dot f_t^2 \label{eq:Eppt} \\ 
J_{st,f_t} &= T\lambda^2\int_{-\ell}^\ell d\sigma \frac12\frac{\tan^2(k\sigma)}{k}(f_t^{\prime2}+\dot f_t^2 ) \label{eq:Jstt}\\
J_{pp,f_t} &= \frac12m\lambda^2 \gamma^3\frac{\sin^2(k\ell)}{k}\dot f_t^2 
\label{eq:Jppt}\end{align}
For \(R(\sigma) = \sigma\), the expressions are
\begin{align} E_{st,f_t} & = T\lambda^2\int_{-\ell}^\ell d\sigma \left[\frac12g^2\dot f_t^2 +\frac12 f_t^{\prime2} -\frac12 g^2 k^2\sigma f_t f_t^\prime + \frac18 k^4\sigma^2 g^4 f_t^2 \right] \label{eq:Estt2}\\
E_{pp,f_t} &= \frac12m\lambda^2\gamma^2\dot f_t^2 \label{eq:Eppt2}\\ 
J_{st,f_t} &= Tk\lambda^2\int_{-\ell}^\ell d\sigma \sigma^2\left[\frac12g^2\dot f_t^2 +\frac12 f_t^{\prime2} -\frac12 g^2 k^2\sigma f_t f_t^\prime + \frac18 k^4\sigma^2 g^4 f_t^2 \right] \label{eq:Jstt2} \\
J_{pp,f_t} &=  \frac12m\lambda^2 \gamma^2 k \ell^2 \dot f_t^2  \label{eq:Jppt2}
\end{align}

What we are interested in calculating is the correction to the Regge trajectory of the string. Rather than the energy and angular momentum themselves, we look at the correction to the classical relation between them. The classical energy and angular momentum can be written (see section \ref{sec:classical}) as functions of the string tension, endpoint masses, and one continuous parameter, which now we take to be \(\gamma = (1-\beta^2)^{-1/2}\):
\be E = E(m,T,\gamma) \qquad J = J(m,T,\gamma) \ee
This parametric relation defines the classical trajectory for given \(m\) and \(T\),
\be J = J_{cl}(E) \ee
Even though we cannot write the closed form of \(J_{cl}(E)\) in the general case, we can use to the parametric relation to compute the first correction to the Regge trajectory, which we define as
\be \delta\left(J-J_{cl}(E)\right) = \delta J - \frac{\pa J}{\pa E} \delta E \ee
Now, using the classical expressions for \(J\) and \(E\) and the classical boundary condition we can write 
\be \frac{\pa J}{\pa E} = \frac{\pa J_{cl}/\pa \gamma}{\pa E_{cl}/\pa \gamma} = \frac{m}{T}\gamma^2\beta = \frac{1}{k} \ee
Then, we can \emph{define} the intercept to be the expectation value of the above combination,
\be a = \langle\delta J - \frac1k \delta E\rangle \ee
This is a generalization to the rotating case of the static case of \cite{PandoZayas:2003yb}, where one looks at \(\delta J - L\delta E\). Here 
\be \frac1k = \frac12 \frac{L}{\beta} \ee
takes the role of a corrected length of the string. Now we can write the expressions for the intercept in terms of the fluctuations. We will see explicitly that they take the form of the worldsheet Hamiltonian.

\subsubsection{First formulation}
When \(R(\sigma) = \frac{1}{k}\sin(k\sigma)\), using the expressions for the contribution of \(f_t\) to the energy and angular momentum of eqs. \ref{eq:Estt}-\ref{eq:Jppt}, we can see that the correction to the trajectory coming from the transverse modes takes the form
\be \delta J-\frac{1}{k}\delta E = -\frac12\frac{T\lambda^2}{k}\left(\int_{-\ell}^{\ell} d\sigma(\dot f_t^2+ f_t^{\prime2}) + \frac{\gamma m}{T} \dot f_t^2|_{\pm\ell}\right)\ee
which is exactly the form of the worldsheet Hamiltonian. Therefore the intercept is seen to be
\be a = -\frac{1}{k} \langle H \rangle \ee

Now we insert the mode expansion for the fluctuations
\be f_t = \frac{1}{\sqrt{2 T\ell^2\lambda^2}}i\sum_{n\neq0}\frac{\alpha_n}{\omega_n} e^{-i\omega_n\tau} f_n(\sigma) \ee
into the Hamiltonian. The eigenmodes obey the orthogonality relation
\be \frac{1}{\ell}\int_{-\ell}^\ell d\sigma f_m(\sigma)f_n(\sigma) + \frac{\gamma m}{T\ell}\left[f_m^+f_n^+ + f_m^-f_n^-\right] = (\delta_{m+n}+\delta_{m-n}) \ee
Where we use the notation \(f^\pm_n \equiv f_n(\pm\ell)\). This is the equation derived from the general equation \ref{eq:SturmLiouvilleOrtho} for the present equation of motion and boundary condition. We have two deltas on the RHS because \(f_{-n} \equiv f_n\).

Plugging the expansion into the expression for the intercept:
\begin{align} & \int_{-\ell}^{\ell} d\sigma(\dot f_t^2+ f_t^{\prime2}) + \frac{\gamma m}{T} \dot f_t^2|_{\pm\ell} = \nonumber \\
& \frac{1}{2T\ell^2\lambda^2}\sum_{n,m\neq0}\alpha_m\alpha_n e^{-i(\omega_n+\omega_m)\tau}\left(\frac{\gamma m}{T} (f^+_m f^+_n+f^-_m f^-_n) +\int_{-\ell}^\ell d\sigma(f_m f_n - \frac{f^\prime_m}{\omega_m}\frac{f^\prime_n}{\omega_n}) \right) = \nonumber \\
& \frac{1}{2T\ell^2\lambda^2}\sum_{n,m\neq0}\alpha_m\alpha_n e^{-i(\omega_n+\omega_m)\tau}\left(\frac{\gamma m}{T} (f^+_m f^+_n+f^-_m f^-_n) - \frac{f^+_m}{\omega_m}\frac{f^{\prime+}_n}{\omega_n} + \frac{f^-_m}{\omega_m}\frac{f^{\prime-}_n}{\omega_n} + \int_{-\ell}^\ell d\sigma(f_m f_n + \frac{f_m}{\omega_m}\frac{f^\dprime_n}{\omega_n})\right) 
\end{align}
From the second to third line we have integrated the last term by parts. Next we use the equation of motion and boundary condition to rewrite the terms including derivatives of the eigenfunctions in terms of the eigenfunctions themselves. This results in (for the terms in the brackets)
\begin{align} & \frac{\gamma m}{T} (f^+_m f^+_n+f^-_m f^-_n - \frac{\omega_n}{\omega_m}f^+_m f^{+}_n - \frac{\omega_n}{\omega_m}f^-_m f^-_n) + \int_{-\ell}^\ell d\sigma(f_m f_n - \frac{\omega_n}{\omega_m} f_m f_n) = \nonumber\\ & (1-\frac{\omega_n}{\omega_m})\ell(\delta_{n+m}+\delta_{n-m}) = 2\ell\delta_{n+m} 
\end{align}
And therefore
\be \int_{-\ell}^{\ell} d\sigma(\dot f_t^2+ f_t^{\prime2}) + \frac{\gamma m}{T} \dot f_t^2|_{\pm\ell} = \frac{1}{T\ell\lambda^2} \sum_{n\neq0}\alpha_{-n}\alpha_n \ee
and
\be H = -\frac{1}{2}\frac{T\lambda^2}{2}\frac{1}{T\ell\lambda^2}\sum_{n\neq0} \alpha_{-n}\alpha_n = \frac{1}{2\ell}\sum_{n\neq0} \alpha_{-n}\alpha_n \ee
The intercept, as the expectation value of \(H\) is obtained as the normal ordering constant in the Hamiltonian, which is
\be a_t = -\frac{1}{k}\langle H \rangle = -\frac12\sum_{n>0} \frac{\omega_n \ell}{\delta} \ee
At the massless limit \(\delta = \frac\pi2\) and \(\omega_n\ell = \frac{\pi}{2}n\), so the contribution of a single transverse mode to the intercept is
\be a_t(m=0) = -\frac{1}{2}\sum_{n>0} n = \frac{1}{24} \ee
where the finite result can be obtained using Riemann Zeta function regularization of the sum. For corrections at finite mass we will need to use some other method of computing the sum.

\subsubsection{Second formulation}
When \(R(\sigma) = \sigma\), then again we can use the expressions for \(E\) and \(J\) to recover the worldsheet Hamiltonian
\begin{align} \delta J - \frac{1}{k}\delta E =
-\frac{T\lambda^2}{2k}\int_{-\ell}^\ell d\sigma\left(\dot f_t^2 + g^{-2} f_t^{\prime2}+\frac14(1+g^2)k^2 f_t^2\right) - \left(\frac{m\lambda^2}{2k}\dot f_t^2 - \frac14T\lambda^2k\ell f_t^2\right)\vert_{\pm\ell} \end{align}
We have the same mode expansion, this time with the orthogonality relation
\be \frac{1}{\ell}\int_{-\ell}^{\ell} d\sigma f_m f_n + \frac{m}{T\ell}(f_m^+ f_n^+ + f_m^- f_n^-) = \delta_{m+n}+\delta_{m-n}\ee

The calculation of the Hamiltonian is similar. We insert the mode expansion, first into the bulk part of the Hamiltonian
\begin{align} & \int_{-\ell}^\ell d\sigma\left(\dot f_t^2 + g^{-2} f_t^{\prime2}+\frac14(1+g^2)k^2 f_t^2\right) = \nonumber \\
& \frac{1}{2T\ell^2\lambda^2}\sum_{n,m\neq0}\alpha_m\alpha_n e^{-i(\omega_n+\omega_m)\tau}\int_{-\ell}^\ell d\sigma\left(f_m f_n - g^{-2}\frac{f^\prime_m}{\omega_m}\frac{f^\prime_n}{\omega_n} -\frac14 (1+g^2)k^2 f_m f_n \right)  \end{align}
We integrate the second term by parts, and use the equation of motion and the boundary conditions to get that the integral is
\be (1-\frac{\omega_m}{\omega_n})\int d\sigma f_m f_n - \frac{m}{T}\frac{\omega_m}{\omega_n}(1+\frac12 \frac{Tk^2\ell}{m\omega_n^2})(f_m^+ f_n^+ + f_m^- f_n^-) \ee
Then, plugging in the expansion into the boundary terms in \(H\) gives
\be -(\frac{m\lambda^2}{2k}+ \frac14 T\lambda^2 k\ell \frac{1}{\omega_m\omega_n})(f_m^+ f_n^+ + f_m^- f_n^-) \ee
Summing both contributions, we get
\be -\frac{T\lambda^2}{2k}(1-\frac{\omega_n}{\omega_m})\left(\int_{-\ell}^\ell d\sigma f_m f_n + \frac{m}{T}(f_m^+ f_n^+ + f_m^- f_n^-)\right) = -\frac{T\ell\lambda^2}{2k}(1-\frac{\omega_n}{\omega_m})(\delta_{n-m}+\delta_{n+m}) \ee
So that, ultimately
\be \delta J - \frac{1}{k}\delta E = -\frac{1}{2k\ell}\sum_{n\neq0}\alpha_{-n}{\alpha_n} \ee
with the expectation value
\be a_t = -\frac{1}{2}\sum_{n>0} \frac{\omega_n\ell}{\beta} \ee
At the \(\beta\to1\) limit, \(\omega_n\ell = n\), and again
\be a_t(m=0) = -\frac12\sum_{n>0}n = \frac{1}{24} \ee

The eigenfrequencies in the present case with \(R(\sigma) = \sigma\) differed from the previous case of \(R(\sigma)=\frac1k\sin(k\sigma)\) by a factor of \(\frac{\arcsin\beta}{\beta} = \frac{\delta}{\beta}\). This same factor multiplies the expression for the intercept here compared with the intercept in the previous case. One can see then that the two expressions are equivalent. In both cases we can write
\be a = -\frac12\sum_{n=1}^\infty \frac{\omega_n}{k} \ee

The overall factor of \(\frac{1}{\beta}\) in the expression for the intercept causes the intercept to diverge for \(\beta\to0\), but this is not much of a problem since we consider the expansion to be valid only for long strings, which necessarily means relativistic endpoint velocities. This will be shown more explicitly in section \ref{sec:nextorder}.


\subsection{Renormalization of the sum over the eigenfrequencies} \label{sec:transverse_sum}
Our goal now is to calculate the renormalized sum of the eigenfrequencies and to get the contribution of the transverse modes to the intercept for any finite value of the mass. We use the approach of \cite{Lambiase:1995st,Lambiase:1998tf} to replace the infinite sum with a calculable contour integral, and renormalize the result by subtracting the result for an infinitely long string.

\begin{figure}[ht!] \centering
\includegraphics[width=0.40\textwidth]{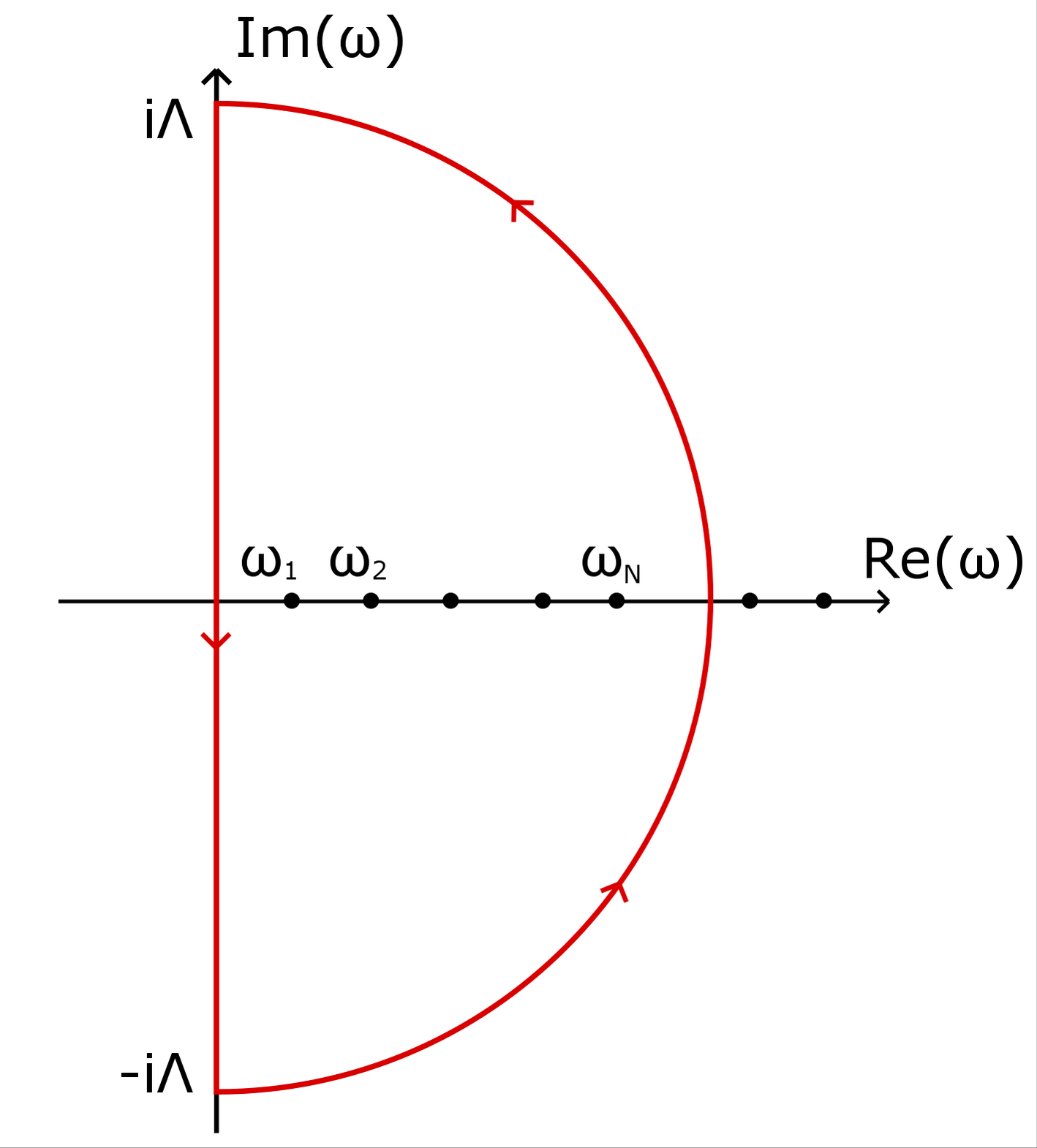}
\caption{\label{fig:contour} The eigenfrequencies \(\omega_n\) are the zeroes of an analytic function \(f(\omega)\) and are all on the real axis. To perform the sum \(\sum_{n=1}^\infty \omega_n\) we use formula \ref{eq:Cauchy} with an appropriate choice of \(f(\omega)\), and integrate on the contour drawn. We insert a cutoff \(\Lambda\) which will be later be sent to infinity in order to calculate the infinite sum over all \(\omega_n\).} 
\end{figure}

The conversion of the sum into a contour integral is done using the following formula, which can be easily derived using the Cauchy integral formula:
\be \frac{1}{2\pi i}\oint dz z \frac{d}{dz}\log f(z) = \frac{1}{2\pi i}\oint dz z \frac{f^\prime(z)}{f(z)} = \sum_j n_j z_j - \sum_k \tilde n_k \tilde z_k \label{eq:Cauchy}\ee
That is, for an analytic function \(f(z)\) with zeroes of order \(n_j\) at \(z = z_j\) and poles of order \(\tilde n_k\) at \(z = \tilde z_k\), the contour integral defined above is equal to a weighted sum over its poles and zeroes inside the contour. It follows that if we define a function \(f(\omega)\) which has no poles and only simple zeroes at \(\omega=\omega_n\), then the above formula can be utilized to compute the infinite sum over \(\omega_n\). The zeroes we want to sum over will be on the real positive semi-axis. Therefore, we will take the contour including the segment \((i\Lambda,-i\Lambda)\) and the semicircle of radius \(\Lambda\), where the radius \(\Lambda\) will be taken to infinity for the contour to encircle the entire half-plane \(Re(\omega)>0\) (see figure \ref{fig:contour}). We will then need to offer a prescription for subtracting the divergences that appear as we take \(\Lambda\to\infty\).

The zero point energy of the string is the result of a Casimir effect, and the sum over the eigenfrequencies is the \emph{Casimir energy} of the string. We define it by
\be E_C = \frac12\sum_{n=1}^\infty \omega_n \ee
This definition is independent of the gauge, and we have seen that for both choices of \(R(\sigma)\) we find the same set of \(\omega_n\). The intercept is related to the Casimir energy by
\be E_C = -\frac{2\beta a}{L} \ee
We will use the notation
\be E_C^{(reg)} = \frac12\sum_{n=1}^{N(\Lambda)}\omega_n \ee
to denote the regularized Casimir energy with the cutoff at finite \(\Lambda\). The renormalization we will perform on the Casimir energy can be written in the form (in the notation of \cite{Lambiase:1995st})
\be E_C^{(ren)} = \lim_{\Lambda\to\infty}\left(E_C^{(reg)}(m,T,L)-E_C^{(reg)}(m,T,L\to\infty)\right) \ee
This notation is not precise, however, as the limit \(L\to\infty\) has to be taken in a particular way which we will show below. On the other hand, it demonstrates the physical interpretation of the procedure which we will follow here, of renormalizing by the subtracting the divergent contribution from the infinitely long string.

In the following we will use the formulation with \(R(\sigma) = \sigma\), only because it is slightly simpler to work with given the straightforward relation between the worldsheet and target space parameters, specifically the length of the string which is then \(L = 2\ell\).

\subsubsection{String with massless endpoints}
We start with the massless limit (which is equivalent to the \(\beta\to1\) limit of the string with masses). The equation for the eigenfrequencies can be then written simply as
\be f(\omega) = \sin(\pi \omega \ell) = 0 \ee
The eigenvalues are of course \(\omega_n\ell = n\). We want to show how the contour integral method can replicate the answer of the Zeta function regularization of the sum \(\sum_{n=1}^\infty n = \zeta(-1) = -\frac{1}{12}\). According to our contour integral formula the Casimir energy of the string is
\be E_C(m=0) = \frac12\sum_{n=1}^{\infty}\omega_n = \frac{1}{4\pi i}\oint \omega \frac{f^\prime(\omega)}{f(\omega)} d\omega = \frac{1}{4 i}\oint \omega\ell \cot(\pi \omega \ell) d\omega \ee
We start by taking the contour integral for finite \(\Lambda\). The integral is comprised of two pieces, one along the imaginary axis, another along the semicircle of radius \(\Lambda\):
\be \frac{1}{4 i}\oint \omega\ell \cot(\pi \omega \ell) d\omega = -\frac14\int_{-\Lambda}^\Lambda y\ell \coth(\pi y\ell) dy + \frac14\Lambda^2\ell \int_{-\frac{\pi}{2}}^{\frac{\pi}{2}}e^{2i\theta}\cot(\pi \Lambda \ell e^{i\theta})d\theta \ee
Up to terms that vanish as \(\Lambda\to\infty\), the result is
\be E_C^{(reg)} = \left(-\frac{\Lambda^2\ell}{4}-\frac{1}{24\ell}\right)+\frac12\Lambda^2\ell = \frac{\Lambda^2\ell}{4}-\frac{1}{24\ell} = \frac{\Lambda^2 L}{8}-\frac{1}{12L} \label{eq:478}\ee
The first two terms in the first equation are those from the imaginary axis integral, and the third term is the result of the second integral. The divergent term is quadratic is proportional to \(L\), and consequently it can be absorbed into a redefinition of the string tension. We will look at a different prescription next. When renormalizing the Casimir effect, one method is to look at the Casimir force rather than the energy,
\be F_C^{(reg)} = -\frac{d}{dL}E_C^{(reg)} = -\frac{\Lambda^2}{8}+\frac{1}{12L^2} \ee
and the renormalization is done by subtracting the constant force that is left when taking the length \(L\) to infinity. Hence the finite result is
\be F_C^{(ren)} = \lim_{\Lambda\to\infty}\left(F_C^{(reg)}(L)-F_C^{(reg)}(L\to\infty)\right) = \frac{1}{12 L^2} \ee
The renormalized energy is given as the integral of the force, and hence defined up to an integration constant. If we take that the Casimir energy goes to zero at infinite length, then we recover the known result
\be E_C^{(ren)} = -\frac{1}{12L} \qquad \Rightarrow \qquad a = \frac{1}{24}\ee
An alternative way to obtain the same finite result is the following. First write the contour integral as
\be E_C^{(reg)} = \frac{1}{4\pi i}\oint \omega \frac{d}{d\omega}\log f(\omega) = -\frac{1}{4\pi}\int_{-\Lambda}^\Lambda y\frac{d}{dy}\log f(iy)dy + \Lambda I_{sc} \ee
where \(I_{sc}\) is the value of the integral along the semicircle, up to an overall factor of \(\Lambda\) which we have taken out, in this case
\be I_{sc}(x) = \frac14 x \int_{-\frac{\pi}{2}}^{\frac{\pi}{2}}e^{2i\theta}\cot(\pi x e^{i\theta})d\theta \ee
 The integral along the imaginary axis we can integrate by parts and get
\be E_C^{(reg)} = \frac{1}{4\pi}\int_{-\Lambda}^\Lambda \log f(iy)dy - \frac{\Lambda}{4\pi}\left(\log f(i\Lambda)+\log f(-i\Lambda)\right) + \Lambda I_{sc} \label{eq:CasimirParts}\ee
This last formula is independent of the choice of \(f(\omega)\), and we will make use of it again later on. Now take the function for the massless case (we use the fact that \(\Lambda\) is large when writing the boundary term),
\be E_C^{(reg)} = \frac{1}{4\pi}\int_{-\Lambda}^\Lambda \log (i \sinh(\pi \ell y)) dy - \frac{\Lambda}{4\pi}(2\pi \Lambda \ell-\log 4) + \Lambda I_{sc}(\Lambda \ell) \label{eq:479}\ee
We can subtract the contribution from an infinitely long string when the integral is written in this form simply by formally subtracting the asymptotic form of the expression for large \(\ell\), which is
\be \frac{1}{4\pi}\int_{0}^\Lambda \log (\frac{i}{2}\exp(\pi \ell y)) dy + \frac{1}{4\pi}\int_{-\Lambda}^0 \log (-\frac{i}{2}\exp(-\pi \ell y)) dy - \frac{\Lambda}{4\pi}(2\pi \Lambda \ell-\log 4) + \Lambda I_{sc}(\Lambda \ell) \label{eq:480}\ee
We make use of the fact that along the semicircle the large \(\ell\) limit is the same as the large \(\Lambda\) limit.\footnote{The only exception to this would have been if \(I_{sc}(\Lambda \ell)\) contained a term that goes like \(\frac{1}{\Lambda\ell}\), but we have computed the integral for large \(\Lambda \ell\) for eq. \ref{eq:478} and seen that there are no such terms.} This form is constructed to capture all the divergent terms in \(E_C^{(reg)}\) - and only the divergent terms. In this case, it is not too difficult to calculate the integrals and see that the result (for large \(\Lambda\)) is simply \(\frac14\Lambda^2\ell\), which is the divergent part of the full integral (eq. \ref{eq:478}). Therefore subtracting \ref{eq:480} from \ref{eq:479} leads to a finite answer when we take \(\Lambda\to\infty\),
\be E_C^{(ren)} = \frac{1}{2\pi}\int_{0}^\infty \log \left(1-e^{-2\pi \ell y}\right) dy = -\frac{1}{12L} \label{eq:481}\ee
which is the result we were after.

To recapitulate, we first compute explicitly the integral and show that the divergent pieces can be eliminated by subtracting the contribution from an infinitely long string to the resulting Casimir force. Then, we show how one can subtract the divergent terms directly, by writing a second integral which contains only the divergent parts and subtracting it from the original contour integral. After that step, we can write the finite answer as a simple calculable integral as in eq. \ref{eq:481}. 
We will use this same approach in the following.

\subsubsection{Non-rotating string with massive endpoints}
We start by reviewing the calculation for a non-rotating string. In that case the eigenfrequencies are given by the solutions to the equation \cite{Lambiase:1995st}
\be f(\omega) = 2 \frac{m \omega}{T} \cos(\omega L) + (1-\frac{m^2\omega^2}{T^2})\sin(\omega L) = 0 \label{eq:w_n_stat_eq}\ee
We want to compute the sum of the eigenfrequencies and renormalize it in the same way as before. We make use of the form of eq. \ref{eq:CasimirParts} to write the necessary integral, and calculate the divergent terms from the asymptotic form as we take the length \(L\) to be large. It is (generalizing eq. \ref{eq:480})
\begin{align} \frac{1}{4\pi}\int_{0}^\Lambda \log \left(\frac{i}{2}\exp(L y)(1+\frac{my}{T})^2\right) dy + \frac{1}{4\pi}\int_{-\Lambda}^0 \log \left(-\frac{i}{2}\exp(-L y)(1-\frac{my}{T})^2\right) dy - \nonumber\\ - \frac{\Lambda}{2\pi}(\Lambda L - \log \left(\frac{2T^2}{(T+m\Lambda)^2}\right)) + \Lambda I_{sc}(\Lambda L,\frac{TL}{m}) \label{eq:483}\end{align}
where the semicircle integral is
\be I_{sc}(x,q) = \frac{x}{2\pi} \int_{-\frac\pi2}^{\frac\pi2} e^{2 i \theta } \frac{\left(q^2+2
   q-e^{2 i \theta } x^2\right) \cos\left(e^{i \theta } x\right)-2 e^{i \theta} (q+1) x \sin \left(e^{i \theta } x\right)}
	{\left(q^2-e^{2 i \theta }x^2\right) \sin \left(e^{i \theta }x\right)+2 e^{i \theta } q x \cos
   \left(e^{i \theta } x\right)}d\theta \ee
for \(x=\Lambda L\) as before and defining \(q = TL/m\). The integrals on the imaginary axis can be solved analytically, while the semicircle integral we compute numerically to capture its large \(x\) behavior up to and including terms of order \(1/x\), and in this case they are all terms which are observed to not depend on \(q\). The result is that the two contributions from the two regions are, up to terms that vanish for \(\Lambda\to\infty\),
\be \left[-\frac{\Lambda^2 L}{4\pi} - \frac{\Lambda}{\pi}+\frac{T}{\pi m}\log\left(\frac{m\Lambda}{T}\right)\right] + \left[\frac{\Lambda^2 L}{2\pi} + \frac{\Lambda}{\pi}\right] = \frac{\Lambda^2 L}{4\pi} +\frac{T}{\pi m}\log\left(\frac{m\Lambda}{T}\right) \label{eq:484}\ee
where the terms in the right brackets are \(\Lambda I_{sc}\) and the terms in the left brackets come from the integral on the imaginary axis and its boundary terms. These are all terms that we can subtract by looking at the force as before and doing the subtraction then for infinite string length. Alternatively, the quadratic divergence can be absorbed into a redefinition of the string tension, while the logarithmic divergence can be dealt with by redefining the mass.

So, if we subtract the divergent parts in their integral form of eq. \ref{eq:483} from the full contour integral, we are left with only
\be E_C^{(ren)} = \frac{1}{2\pi}\int_{0}^\infty \log \left(1-e^{-2 L y}\frac{(T-m y)^2}{(T+my)^2}\right) dy = \frac{1}{2\pi L}\int_0^\infty \log\left(1-e^{-2 x}\frac{(q-x)^2}{(q+x)^2}\right) dx\ee
where \(q = \frac{TL}{m}\). This is the result of \cite{Lambiase:1995st}. The integral can be easily computed numerically and plotted a function of \(q\). In this case, we recover the by now familiar result of \(a = \frac{1}{24}\) in two opposing limits. When \(q \to \infty\) (infinitely long string or zero mass) and when \(q = 0\) (infinite mass), then
\be E_C^{(ren)}(q=0) = E_C^{(ren)}(q\to\infty) = \frac{1}{2\pi L}\int_0^\infty dx \log(1-e^{-2x}) = -\frac{\pi}{24 L}\ee
In the static case with no masses we can define the intercept as the coefficient of \(1/L\) in the Casimir energy, or more precisely as
\be a = -\frac{1}{\pi}L\delta E \label{eq:a_stat}\ee
This is consistent with the long string (\(TL\gg2m\)) limit where we can use \(E = TL\) to write
\be \delta (\alp E^2) = 2\alp E \delta E = \frac{1}{\pi T} T L \delta E = -a \ee
so that \(a\) is indeed the intercept in that case. At the \(q\to\infty\) limit the intercept is \(a_t = \frac{1}{24}\) as before. For finite \(q\) we keep using eq. \ref{eq:a_stat} to define the intercept and get that \(a_t = \frac{1}{24}\) also in the infinite mass limit of \(q\to0\), which is equivalent to the string with Dirichlet boundary conditions. We draw the result for all \(q\) in figure \ref{fig:a_t}.

\subsubsection{Rotating string}
We now turn to the case of the transverse mode in a rotating string. For the rotating string, we now we most naturally write our equations in terms of the endpoint velocity \(\beta\) rather than the length \(L\). The velocity and length are of course related through the classical boundary condition.

With the parametrization \(R(\sigma) = \sigma\), the eigenfrequencies are given by zeros of the function (per eq. \ref{eq:w_t_lin})
\be f(z) = 2 z \beta^2\sqrt{1-\beta^2} \cos\left(\frac{2 z \arcsin\beta}{\beta}\right) + (
\beta^4-(1-\beta^2)z^2)\sin\left(\frac{2 z \arcsin\beta}{\beta}\right) \ee
where \(z = \omega \ell\) is complex. A transverse mode's contribution to the intercept is proportional to the sum of zeroes of \(f(z)\). The Casimir energy can be written as the same contour integral discussed above,
\be E_C = \frac12\sum_{n>0}\omega_n = \frac1\ell\frac{1}{4\pi i}\oint dz z \frac{d}{dz}\log f(z) \ee
As before, we separate the contour integral into two parts and integrate by parts the integral along the imaginary axis. We use the asymptotic forms of \(f(iy)\) as in eq. \ref{eq:483} to find the divergent parts. To do that we need to compute the integral
\begin{align} \frac{1}{4\pi\ell}\int_{0}^\Lambda \log \left(\frac{i}{2}\exp(\frac{2y \arcsin\beta}{\beta})(\beta^2+y\sqrt{1-\beta^2})^2\right) dy + \nonumber \\ + \frac{1}{4\pi\ell}\int_{-\Lambda}^0 \log \left(-\frac{i}{2}\exp(\frac{-2y \arcsin\beta}{\beta})(\beta^2-y\sqrt{1-\beta^2})^2\right) dy - \nonumber\\ - \frac{\Lambda}{4\pi}\left(\frac{4\Lambda \ell\arcsin\beta}\beta-\log4+4\log(\beta^2+\sqrt{1-\beta^2}\Lambda\ell)\right) + \Lambda I_{sc}(\Lambda \ell,\beta) \label{eq:497}\end{align}
Here the semicircle integral is
\begin{align} &I_{sc}(x,\beta) = \frac{x}{2\pi\beta}\int_{-\frac\pi2}^{\frac\pi2} d\theta \nonumber \\ &\left(\frac{\cos \left(e^{i \theta } \hat x\right) \left(\sqrt{1-\beta ^2}\beta ^3+\arcsin\beta \left(\beta ^4+\left(\beta^2-1\right) e^{2 i \theta } x^2\right)\right)+ \sin \left(e^{i \theta}\hat x\right)\left(\beta ^2-2 \sqrt{1-\beta ^2} \beta \arcsin\beta-1\right) \beta  e^{i\theta } x}
{2 \cos \left(e^{i \theta }\hat x\right) \sqrt{1-\beta
   ^2} \beta ^2 e^{i \theta } x+\sin \left(e^{i \theta } \hat x\right)\left(\beta ^4+\left(\beta ^2-1\right) e^{2 i \theta} x^2\right)} \right)\end{align}
where \(\hat x =x\arcsin\beta/\beta\).
As with the static case, we look at the behavior of the integral for large \(\Lambda\). The divergent parts from the two parts of the contour integral are
\begin{align} &\left[-\frac{\Lambda^2 L \arcsin\beta}{\pi\beta} -\frac{2\Lambda}{\pi} +\frac{T}{2\gamma m}\log\frac{2\gamma m \Lambda}{T}\right] + \left[\frac{2\Lambda^2 L \arcsin\beta}{\pi\beta} + \frac{2\Lambda}{\pi}\right] = \nonumber \\ \qquad & = \frac{\Lambda^2 L \arcsin\beta}{\pi\beta} + \frac{T}{2\gamma m}\log\frac{2\gamma m \Lambda}{T} \label{eq:4100} \end{align}
The terms in the right brackets are \(\Lambda I_{sc}\), which were found using a numerical computation, while in the left there are the rest of terms of eq. \ref{eq:497}. The form is like before, but now the terms are \(\beta\) dependent. 

We cannot repeat the procedure of differentiating with respect to \(L\) and subtracting the force, since \(\beta\) is implicitly \(L\)-dependent through the boundary condition. On the other hand, we note that if we look at \(\tilde L = L\frac{\arcsin\beta}{\beta}\) as the effective length of the string, and \(\tilde m = \gamma m\) the mass of the particle, then we have that the energy of the string is \(T\tilde L+2\tilde m\) (compared with \(TL+2m\) for a non-rotating static string). And moreover, in terms of \(\tilde L\) and \(\tilde m\), the last equation is precisely of the form of the equation from the static case \ref{eq:484}, as it is now
\be \frac{\Lambda^2 \tilde L}{\pi} + \frac{T}{2\tilde m}\log\frac{2\tilde m \Lambda}{T} \ee
Therefore, the rotating string looks like a static string with effective length and endpoint masses due to the rotation, and we will use this to perform the subtraction in the same way we did before. We define the Casimir force by differentiating the energy with respect to \(\tilde L\) (and treating \(\tilde m\) as a constant), and subtract the result from \(L\to\infty\).

In other words, the form of the divergences is significant because it is the same form as the classical energy. Because of that, the divergences can be again eliminated by renormalizing the string tension and the endpoint masses and absorbing the divergences into the coefficients. The \(\Lambda^2\) divergence can be eliminated by adding a counterterm to the action redefining \(T = T_{bare}+\delta T\), while the logarithmic divergence, which is proportional to \(1/(\gamma m)\), can be renormalized by an appropriate mass counterterm, \(m = m_{bare}+\delta m\).


If we subtract from the full contour integral over \(f(z)\) the divergent parts in their integral form, represented as the integral over the asymptotic form of \(f\) for large \(\ell\), then we eliminate the divergences, and can safely take \(\Lambda\) to infinity and write:
\be E_C^{ren} = \frac{1}{\pi}\int_0^\infty \log\left(\frac{2 y \beta^2\sqrt{1-\beta^2} \cosh\left(\frac{2 y \arcsin\beta}{\beta}\right) + (
\beta^4+(1-\beta^2)y^2)\sinh\left(\frac{2 y \arcsin\beta}{\beta}\right)}{\frac12\left((1-\beta^2)y^2+2\beta^2\sqrt{1-\beta^2}+\beta^4\right)\exp\left(\frac{2 y \arcsin\beta}{\beta}\right)}\right) dy \ee
Again the boundary terms and the semicircle integral of eq. \ref{eq:497} have dropped. After simplifying the integrand and multiplying by constants to get to the intercept, the contribution of a single transverse mode to the intercept for the rotating string is given by the expression
\be a_t = -\frac{1}{2\pi \beta} \int_0^\infty \log\left[1-\exp\left(-\frac{4\arcsin\beta}{\beta}y\right)\left(\frac{y-\gamma\beta^2}{y+\gamma\beta^2}\right)^2\right] dy\label{eq:a_t_finite}\ee
For \(\beta = 1\), we recover the result from the massless case:
\be a_t(\beta\to1) = -\frac{1}{2\pi}\int_0^\infty \log\left(1-e^{-2\pi y}\right) = \frac{1}{24}\ee
When \(\beta\) is close to 1 we can write the result as an expansion in \(\frac1\gamma\), or alternatively in \(\frac{2m}{TL}=\frac{1}{\gamma^2-1}\), and find that
\be a_t = \frac{1}{24}-\frac{11}{360\pi}\frac{1}{\gamma^3} + \ldots = \frac{1}{24}-\frac{11}{360\pi}(\frac{2m}{TL})^{3/2} + \ldots \label{eq:a_t_approx}\ee
where the next terms are of order \(\gamma^{-5}\) or \((2m/TL)^{5/2}\).

\begin{figure} \centering
\includegraphics[width=0.48\textwidth]{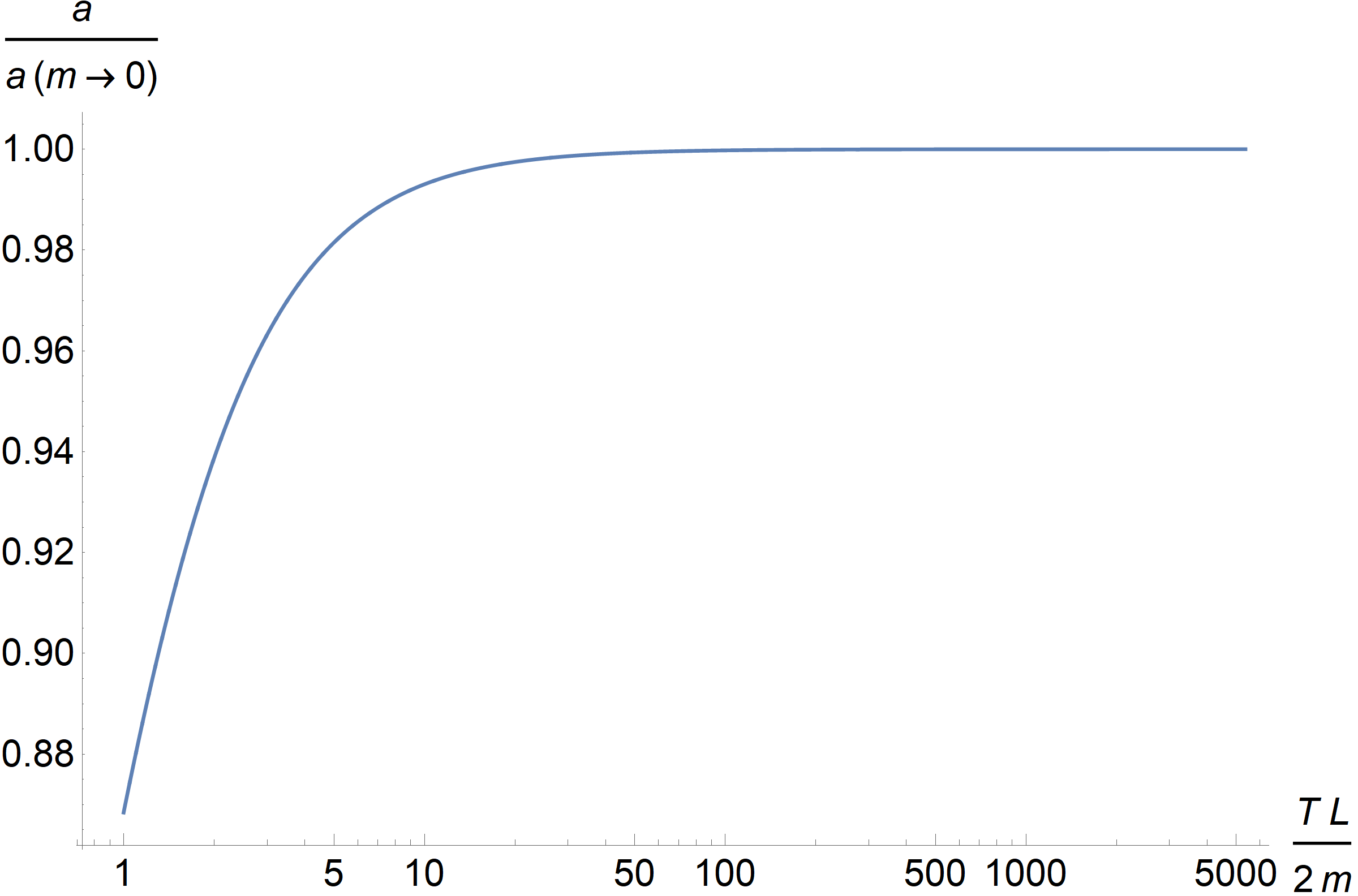}
\includegraphics[width=0.48\textwidth]{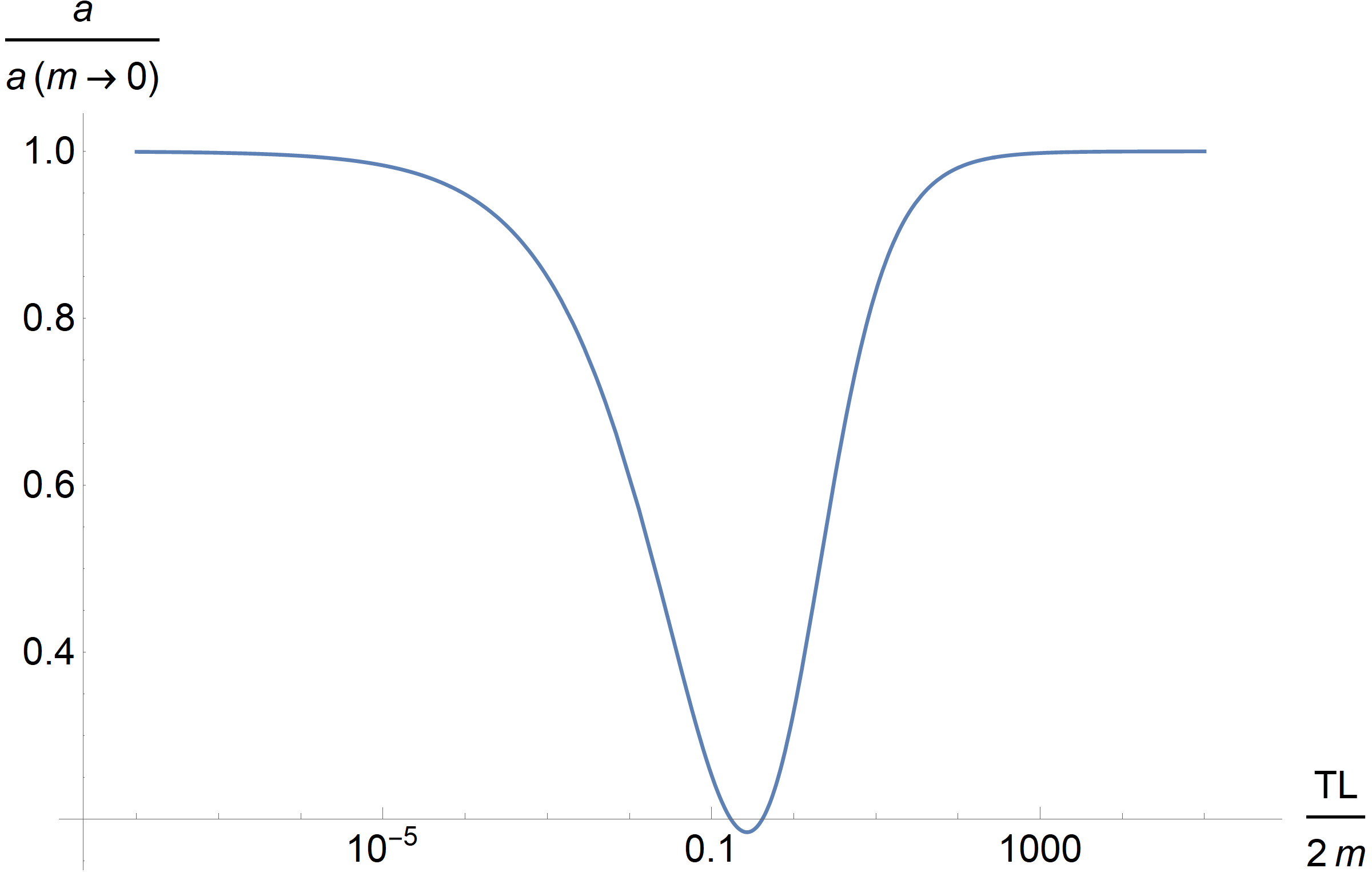}
\caption{\label{fig:a_t} \textbf{Left:} The intercept of the rotating string as a function of \(\frac{TL}{2m}=\gamma^2\beta^2\) (eq. \ref{eq:a_t_finite}). The intercept is normalized by the value at \(m\to0\), which is \(\frac{1}{24}\). \textbf{Right:} The intercept for the static case as defined in eq. \ref{eq:a_stat}, similarly normalized.}
\end{figure}

For any value of \(\beta\) we can evaluate the integral numerically. The result is plotted in figure \ref{fig:a_t} as a function of \(\frac{TL}{2m}\). We only plot the range \(\frac{TL}{2m}>1\), since the expansion is expected to break down when the string is short.

\subsubsection{Zeta function regularization}
\begin{figure} \centering
\includegraphics[width=0.48\textwidth]{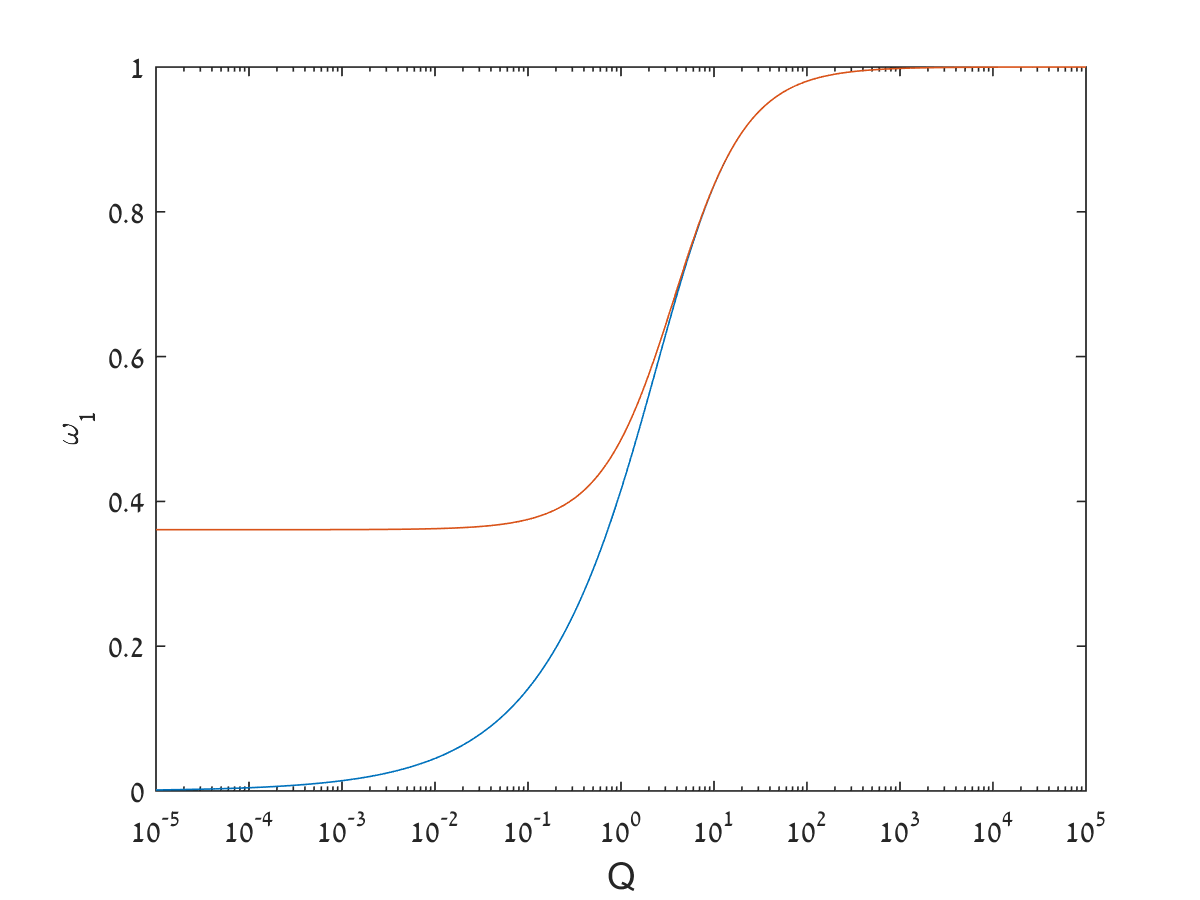}
\includegraphics[width=0.48\textwidth]{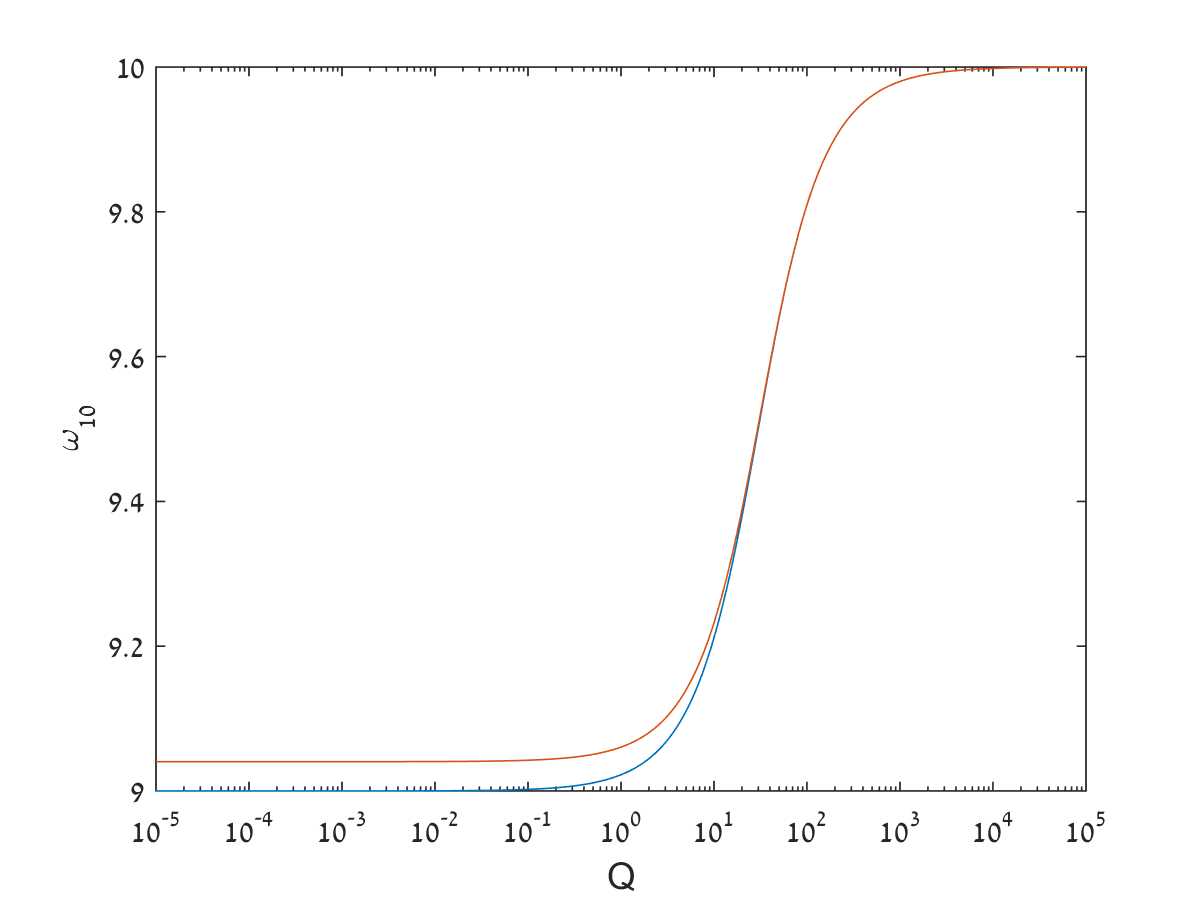} \\
\caption{\label{fig:w_approx} The exact solution of eq. \ref{eq:y_n} for \(y_n(Q)\) (blue) versus the approximate solution of eq. \ref{eq:w_approx} (red), plotted for \(n=1\) and 100. For \(Q\gg1\) the approximation is valid for any \(n\). As \(n\) increases the approximation is improving for smaller \(Q\) as well.}
\end{figure}

To finish this section, we show that the leading term correction to the intercept could have been calculated in a different approach using the Zeta function. The key is finding an approximate solution to the eigenfrequency equation. To do that, we rewrite eq. \ref{eq:w_n_sin} by switching variables from \(2\omega\ell = 2x = \pi y\) and defining \(Q = 2\delta\tan\delta\), as
\be \tan(\pi y) = \frac{2\pi Q y}{\pi^2y^2-Q^2} \label{eq:y_n}\ee
Now note that
\be Q = 2\delta\tan\delta = 2\gamma\beta\arcsin\beta \ee
so at high energies, when \(\gamma \gg 1\), then also \(Q\gg1\). When \(Q\) is infinite we have \(y_n = n\) as for the massless string, and when \(Q\) is large we find an approximate solution of the form
\be y_n = n+\delta_n = n+\frac2\pi\arctan{\frac{Q+2}{n\pi}}-1 \label{eq:w_approx}\ee
The easiest way to see that this is indeed a good solution is to plot it versus the exact numerical one, and we do so in figure \ref{fig:w_approx}. To see also analytically that this is a good approximation of the solution, we will plug it into the equation for \(y\)
\be \tan(\pi \delta_n) = \frac{2\pi Q (n+\delta_n)}{\pi^2(n+\delta_n)^2-Q^2} \ee
and show that it holds when \(Q \gg 1\). Since \(n\) can take any integer value, there are three limits that one should check separately: \(\pi n \gg Q \gg 1\), \(Q\gg\pi n\), and \(Q \sim \pi n\).

To quantify how well the equation holds we can look at the expression \(\eta=(LHS-RHS)/(LHS+RHS)\), where \(LHS\) and \(RHS\) are the left and right hand sides of the last equation. When \(Q \gg \pi n\), 
then the deviation is \(\eta \sim n^2/Q^3\ll1\), while for \(\pi n \gg Q\) it is of order \(1/Q \ll 1\). In the intermediate region \(\pi n\approx Q\) then \(\eta\) is not a reliable metric since one of the sides will go to infinity at some point. In that case we will just look at the point \(Q= \pi(n-1/2)\). There we know the exact solution of eq. \ref{eq:y_n} is \(\delta_n = -1/2\), while our approximate solution gives \(\delta_n = -1/2+\frac{4+\pi}{2\pi Q}\), so the deviation is again small when \(Q\) is large.

Now, given the approximate solution for \(\omega_n\), we can use it to obtain the leading order mass correction to the rotating string intercept through simple Zeta function regularization.

Plugging in the approximate \(\omega_n\) into our expression for the intercept, we have
\be a_t = -\frac{1}{2\delta}\sum_{n=1}^\infty \omega_n\ell = -\frac{\pi}{4\delta}\sum_{n=1}^\infty y_n  \approx -\frac{\pi}{4\delta}\sum_{n=1}^\infty \left(n-1+\frac2\pi\arctan\frac{Q+2}{n\pi}\right) \ee
Writing all the variables in terms of \(\beta\),
\be a_t = -\frac{\pi}{4\arcsin\beta}\sum_{n=1}^\infty\left(n-1+\frac2\pi\arctan\frac{2\gamma\beta\arcsin\beta+2}{n\pi}\right) \ee
Now we take the \(\beta\to1\) limit, by replacing \(\beta = \sqrt{1-\epsilon^2}\) and \(\gamma = \frac{1}{\epsilon}\) and expanding in \(\epsilon\). Note that in this step we actually take \(n\epsilon \ll 1\) for any \(n\), which is clearly not the case for finite \(\epsilon\). However, upon regularization the contribution to the finite part of the sum from large \(n\) is not significant (we have already seen that in a way when performing the subtraction of the integral in the previous section). If we allow ourselves to use this expansion in \(\epsilon\) as stated then
\be a_t \approx \sum_{n=1}^\infty (-\frac12 n + \frac{n-n^3}{3\pi}\epsilon^3 + \ldots)\ee
which we can now regularize using the Zeta function:
\be a_t = -\frac{1}{2}\zeta(-1) + \frac{1}{3\pi}(\zeta(-1)-\zeta(-3))\epsilon^3 = \frac{1}{24}-\frac{11}{360\pi}\epsilon^3 \ee
and find the same result that we found from the contour integral, with the coefficient of the \(\epsilon^3\) correction now seen to be given by the Zeta function.

We can also repeat the same exercise for the static case, noting that the eigenfrequency equation has the same form with \(Q\to q = TL/m\) (see eq. \ref{eq:w_n_stat_eq}). The intercept is
\be a \equiv = -\frac{1}{\pi}L\delta E = -\frac12\sum_{n=1}^\infty \omega_n L \approx -\frac12\sum_{m=1}^\infty(n-1+\frac2\pi \arctan\frac{q+2}{n\pi})\ee
here we just take the \(q\to\infty\) limit directly and find
\be a_{stat} = -\frac12\sum_{n=1}^\infty n(1-\frac2q+\frac{4}{q^2}) = \frac{1}{24}-\frac{1}{12q}+\frac{1}{6q^2}\ee
which is the same result we found by expanding the contour integral to the same order.

The Zeta function regularization gives us then the answer for the leading order corrections for small endpoint masses in both the rotating and non-rotating cases. With the contour integral approach we have seen explicitly the form of the divergences and how they can be absorbed into the physical parameters.

\subsection{Comparison of static and rotating string intercept}
It is instructive to compare the result we have obtained here for the transverse intercept of the rotating string,
\be a_{rot} = -\frac{1}{2\pi \beta} \int_0^\infty \log\left[1-\exp\left(-\frac{4\arcsin\beta}{\beta}y\right)\left(\frac{y-\gamma\beta^2}{y+\gamma\beta^2}\right)^2\right] dy \label{eq:a_rot_q}\ee
to the static string result (previously obtained in \cite{Lambiase:1995st}),
\be a_{stat} = -\frac{1}{2\pi^2}\int_0^\infty \log\left[1-e^{-2 x}\left(\frac{q-x}{q+x}\right)^2\right] dx \label{eq:a_stat_q}\ee
where \(q = TL/m\).

The equation of motion and boundary condition for the transverse fluctuations are actually the same whether one expands around a static string or a long rotating string,\footnote{One has to work in a similar gauge in both cases to end up with the same equations of motion. In this case we use the gauge with \(R(\sigma)=\frac1k\sin(k\sigma)\) for the rotating string, where the gauge that used for the static string is the orthogonal gauge.}, with the difference being that the parameter \(q\) is generalized to
\be q \to \tilde q = \frac{T \tilde L}{\tilde m} = \frac{\arcsin\beta}{\gamma\beta}\frac{T L}m\ee
Another difference, which is important, is in the definition of the intercept. We have
\be a_{stat}(T,L,m) \equiv -\frac{L}{\pi} \delta E = -\frac12\sum \omega_n(q) \ee
\be a_{rot}(T,L,m) \equiv \delta J-\frac1k \delta E = -\frac{\pi}{4\delta} \sum \omega_{n}(\tilde q) \ee
where \(\omega_n(x)\) is the same function in both cases.

We can see that the relation implied by this equation holds between the two integrals above just by changing integration variables in eq. \ref{eq:a_rot_q}. Defining
\be x = \frac{2\arcsin\beta}{\beta}y \ee
leads to
\be a_{rot}(T,L,m) = -\frac{1}{4\pi\arcsin\beta}\int_0^\infty \log\left[1-\exp\left(-2x\right)\left(\frac{x-2\gamma\beta\arcsin\beta}{x+2\gamma\beta\arcsin\beta}\right)^2\right] dx\ee
Using the boundary conditions of the rotating string we can show that
\be 2\gamma\beta\arcsin\beta = \frac{TL \arcsin\beta}{\gamma m \beta} = \frac{T\tilde L}{\tilde m}\ee
with \(\tilde m\) and \(\tilde L\) as defined in the previous section, so in total we can write the relation
\be a_{rot}(T,L,m) = \frac{\pi}{2\arcsin\beta}a_{stat}(T,\tilde L,\tilde m)\ee
The extra factor of \(\arcsin\beta\) is quite crucial when checking the high energy/small endpoint mass limit of \(TL/m \gg 1\).

For the static string we find this correction by expanding the integrand in \ref{eq:a_stat_q} in powers of \(1/q\) and solving the resulting integral for each coefficient in the expansion. The result is
\be a_{stat} = \frac{1}{24} - \frac{1}{12q} + \frac{1}{6q^2} \label{eq:a_stat_q_a} \ee
The first correction is simply linear in \(m/TL\) and each 

For the rotating string this is not the case. The natural expansion parameter at high energies is \(\epsilon = \frac{1}\gamma = \sqrt{1-\beta^2}\), which actually goes like \(\sqrt{m/TL}\). Moreover, we find that the leading order term is actually \(\epsilon^3\). We can see how the \(\epsilon\) and \(\epsilon^2\) terms vanish by evaluating (using \(T\tilde L/\tilde m=2\gamma\beta\arcsin\beta\) as above)
\be a_{rot}(T,L,m) = \frac{\pi}{2\arccos\epsilon}a_{stat}(q=\frac{\epsilon}{2\sqrt{1-\epsilon^2}\arccos{\epsilon}}) \ee
to order \(\epsilon^2\). First, the prefactor is
\be \frac{\pi}{2\arccos\epsilon} \approx 1-\frac{2}{\pi}\epsilon + \frac{4}{\pi^2}\epsilon^2 \ee
while \(a_{stat}\) is evaluated using eq. \ref{eq:a_stat_q_a},
\be a_{stat}(T,\tilde L,\tilde m) \approx \frac{1}{24}\left(1-\frac{\epsilon}{\sqrt{1-\epsilon^2}\arccos\epsilon} + (\frac{\epsilon}{\sqrt{1-\epsilon^2}\arccos\epsilon})^2\right) \ee
When rewritten as a series in \(\epsilon\) this becomes
\be a_{stat}(T,\tilde L,\tilde m) \approx 1+\frac{2}{\pi}\epsilon \ee
(with no \(\epsilon^2\) term) and by multiplying the two series
\be a_{rot}(T,L,m) = \frac{1}{24} + \mathcal O(\epsilon^3) \ee
and the first non-vanishing correction is indeed of the order \((m/TL)^{3/2}\). Note that this mirrors what happens already at the classical level, where the Regge trajectory in the presence of masses is corrected by a term of the order \((m/E)^{3/2}\). In fact we can rewrite eq. \ref{eq:JclE} as
\be \frac{\alp E^2}{J} \approx 1 + \frac{8\sqrt{\pi}}{3}\left(\frac m E\right)^{3/2} \approx 1 + \frac{8}{3\pi}\epsilon^3 \ee
with \(\epsilon=1/\gamma\) as above.

To summarize, one could have used the result from the static string to compute the contribution to the intercept of the rotating string. While the leading order correction is linear in \(m/TL\) for the static string, in the rotating case the parameter \(m/TL\) is transmuted into \(\gamma m/T \ell \sim \sqrt{m/TL}\). When including the factor of \(\arcsin\beta\) in the intercept, which ultimately accounts for the contraction of the string when moving between the rotating and the lab frame, we find that the first non-vanishing term in the correction to the intercept is of the order \((m/TL)^{3/2}\).

In the next section we will analyze the planar mode which has no analogue in the static string, but we will find that it also has the same behavior, giving \(\frac{1}{24}\) with a correction of the order \((m/TL)^{3/2}\).

\clearpage

\section{Planar mode} \label{sec:planar}
In this section, we will follow the same methods from the previous section to calculate the contribution of the planar mode to the intercept. What we call the planar mode is the single mode of fluctuations in the direction orthogonal to the string and in the plane of rotation.

In addition to the planar mode, one should take into account the longitudinal mode. In the bulk action the longitudinal mode can be gauged away, but not so on the boundary. As we shall see, the longitudinal mode changes the boundary condition for the planar mode.

Apart from the changed equation of motion and boundary condition, the structure of the calculation of the contribution of the planar mode to the intercept is the same as that in section \ref{sec:transverse}. We write the mode expansion, obtain the spectrum of eigenfrequencies, and show that the intercept is equal to the sum over all eigenfrequencies times some constants. The sum is then computed and renormalized using the same contour integral approach.

\subsection{Action and Hamiltonian}
We begin by expanding the Nambu-Goto plus point particle action to quadratic order in the relevant fluctuations. These are \(\delta\theta\) and the longitudinal \(\delta\rho\). In the case of \(\delta\rho\), one finds that its action in the bulk is a total derivative, so it only contributes to the boundary action, as expected.

As before we write the action for the two different choices of parametrization of the classical rotating solution. However, we shall see that in the case of the planar mode, the choice \(R(\sigma) = \frac{1}{k}\sin(k\sigma)\) is not as helpful as it was for the transverse modes, and we will focus on \(R(\sigma) = \sigma\) in later subsections.

\subsubsection{First formulation}
For \(R(\sigma)=\frac{1}{k}\sin(k\sigma)\), we define
\be f_p = \frac{1}{k}\tan(k\sigma)\delta\theta \qquad f_r = \delta\rho \ee
In terms of which, the action is
\be S_{st,p} = T\lambda^2\int d\tau d\sigma \left(\frac12\dot f_p^2-\frac12 f_p^{\prime2}-\frac{k^2}{\cos^2(k\sigma)}f_p^2\right) \ee
\be S_{pp,p} = \gamma m\lambda^2\int d\tau \left(\frac12\dot f_p^2+\frac12\gamma^2k^2 f_p^2+\frac12\dot f_r^2+\frac12k^2(2\gamma^2-1)f_r^2-2\gamma k f_r \dot f_p\right) \ee
where \(\gamma = 1/\cos(k\ell)\). The boundary action includes terms obtained from integrating by parts terms from the string action. Those terms are rewritten by using the classical boundary condition \(T = mk\frac{\tan(k\ell)}{\cos(k\ell)}\) to simplify the action.

The Hamiltonian derived from the above action is
\begin{align} H &= T\lambda^2 \int_{-\ell}^{\ell} d\sigma \left(\frac12\dot f_p^2+\frac12 f_p^{\prime2} +  \frac{2}{\cos^2(k\sigma)}f_p^2\right) + \nonumber \\
&+\gamma m\lambda^2\left(\frac12\dot f_p^2+\frac12\dot f_r^2-\frac12\gamma^2k^2 f_p^2-\frac12k^2(2\gamma^2-1)f_r^2\right)|_{\pm\ell} \end{align}

\subsubsection{Second formulation}
For \(R(\sigma) = \sigma\), we define the modes as
\be f_p = (1-k^2\sigma^2)^{-3/4}\sigma \delta \theta\,,\qquad f_r = (1-k^2\sigma^2)^{-1/4}\delta\rho \ee
In terms of \(f_p\) the bulk action for the planar fluctuations is nearly identical to the transverse fluctuation action
\be S_{st,p} = T\lambda^2\int d\tau d\sigma \left(\frac12\dot f_p^2-\frac12 g^{-2}f_p^{\prime2}-\frac18k^2(1+9g^2)f_p^2\right) \ee
The only difference being the factor of nine in the last term. The boundary action for the planar mode
 \be S_{pp,p} =  \lambda^2\int d\tau \left(\frac12 m\dot f_p^2+\frac14\frac{T}{k\ell}(3-\gamma^{-2}) f_p^2\right) \ee
is supplemented by the action for radial mode and the interaction between the two
\be S_{pp,r} = m\lambda^2 \int d\tau \left(\frac12 \dot f_r^2 + \frac12 k^2(2\gamma^2-1) f_r^2-2 \gamma k f_p \dot f_r\right) \label{eq:action_r}\ee
The Hamiltonian in this case is
\begin{align} H &= T\lambda^2 \int_{-\ell}^{\ell} d\sigma \left(\frac12\dot f_p^2+\frac12 g^{-2}f_p^{\prime2}+\frac18k^2(1+9g^2)f_p^2\right) + \nonumber \\
&+m\lambda^2\left(\frac12\dot f_p^2+\frac12\dot f_r^2-\frac14k^2(3\gamma^2-1) f_p^2-\frac12 k^2(2\gamma^2-1) f_r^2\right)|_{\pm\ell} \label{eq:H_p}\end{align}

\subsection{Eigenfrequencies and eigenfunctions}
The mode expansion of the planar mode is of the same form as before
\be f_p = i\sqrt{\mathcal{N}}\sum_{n\neq0}\frac{\alpha_n}{\omega_n} e^{-i\omega_n\tau} f_n(\sigma) \label{eq:f_p_expansion}\ee
The equation of motion for the \(n\)-th mode in the case \(R(\sigma)=\frac{1}{k}\sin(k\sigma)\) is
\be f^{\dprime}_n(\sigma) + \left(\omega_n^2 - \frac{2k^2}{\cos^2(k\sigma)}\right)f_n(\sigma) = 0 \ee
The general solution is given in terms of the Gauss hypergeometric function \({}_2 F_1\). Alternatively, as was done in \cite{Zahn:2013yma}, one can define \(x = \sin(k\sigma)\) and \(g(x) = (1-x^2)^{-1/4}f(x)\) and reach the Legendre equation as the equation of motion for \(g(x)\). This reparametrization actually takes us to the second case, of \(R(\sigma) = \sigma\), where we get the same Legendre equation.

We will continue using the choice \(R(\sigma) = \sigma\) from here on. The equation of motion for the planar mode is then
\be (1-x^2) f^\dprime_n(x) -2x f^\prime(x) + \left(\frac{\omega_n^2}{k^2} - \frac14 - \frac{9}{4(1-x^2)}\right)f_n(x) = 0 \ee
for \(x = k\sigma\). The \(n\)-th planar eigenmode is given by the Legendre \(P\) and \(Q\) functions
\be f_n(\sigma) = c_1 P_{\nu_n}^{3/2}(k\sigma) + c_2 Q_{\nu_n}^{3/2}(k\sigma)\qquad\nu_n \equiv \frac{\omega_n}{k}-\frac12\ee
We can write the Legendre functions of order \(3/2\) explicitly using the known recursion relation,
\be \sqrt{1-x^2}P_l^{m+1}(x)=(l-m)x P_l^m(x)-(l+m)P_{l-1}^m(x) \ee
which applies to \(Q\) as well as \(P\) and where the functions of order \(m=1/2\) are exactly those of eq. \ref{eq:LegendrePQ} (the transverse eigenmodes). The explicit forms of the solutions of the planar mode equation of motion are then
\begin{align}
P_{\nu_n}^{3/2}(x) &= -\sqrt{\frac{2}{\pi}}(1- x^2)^{-3/4}\left[\frac{\omega_n}{k}\cos\left((\frac{\omega_n}{k}-1)\arccos x\right)+ x(1-\frac{\omega_n}{k})\cos\left(\frac{\omega_n}{k}\arccos x\right)\right] \\
Q_{\nu_n}^{3/2}(x) &= \sqrt{\frac\pi2}(1- x^2)^{-3/4}\left[\frac{\omega_n}{k}\sin\left((\frac{\omega_n}{k}-1)\arccos x\right)+ x(1-\frac{\omega_n}{k})\sin\left(\frac{\omega_n}{k}\arccos x\right)\right] \end{align}
and one can verify by substitution that these indeed solve the equation of motion.

\subsubsection{Boundary conditions}
\begin{figure} \centering
\includegraphics[width=0.48\textwidth]{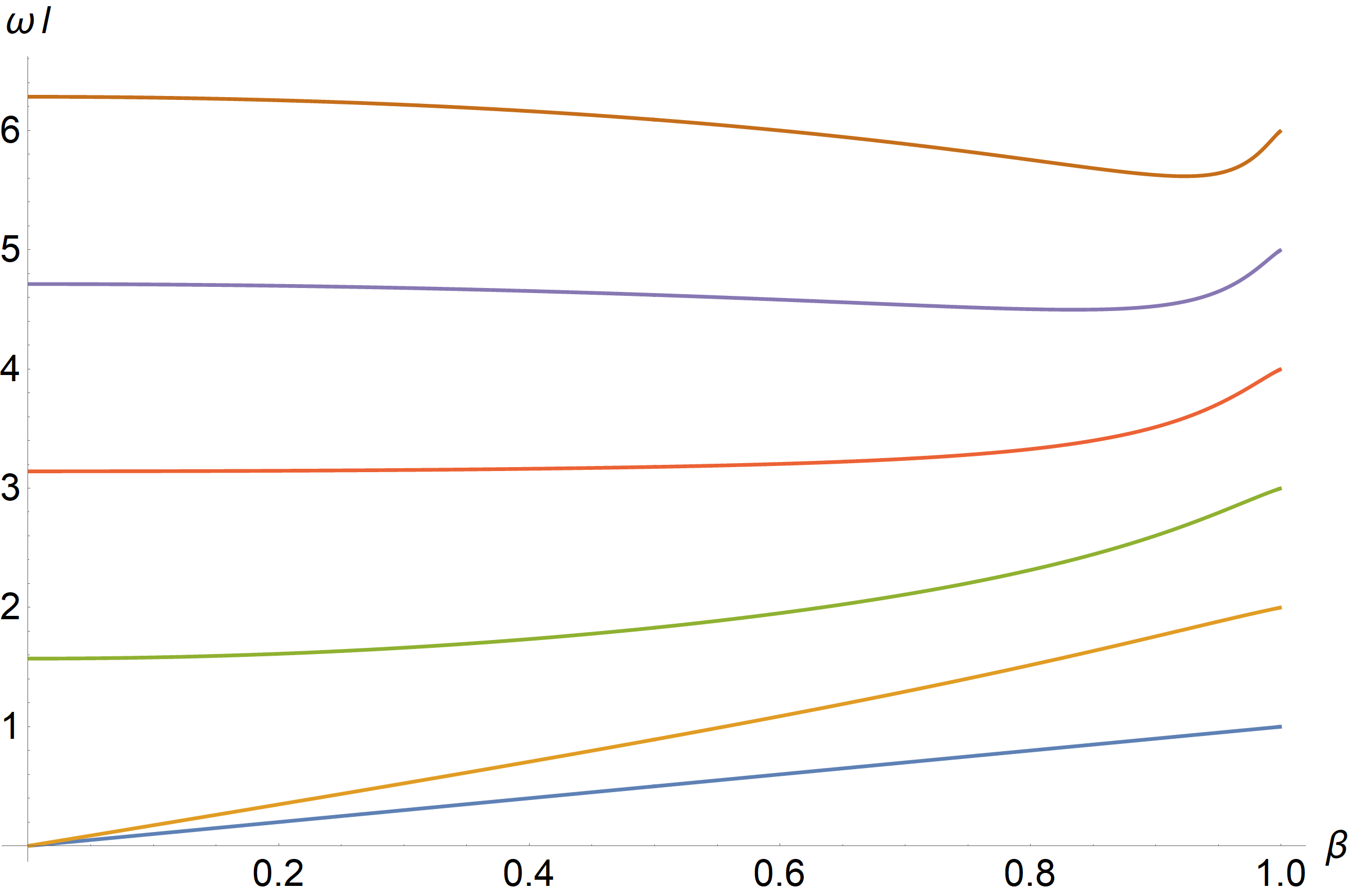}
\includegraphics[width=0.48\textwidth]{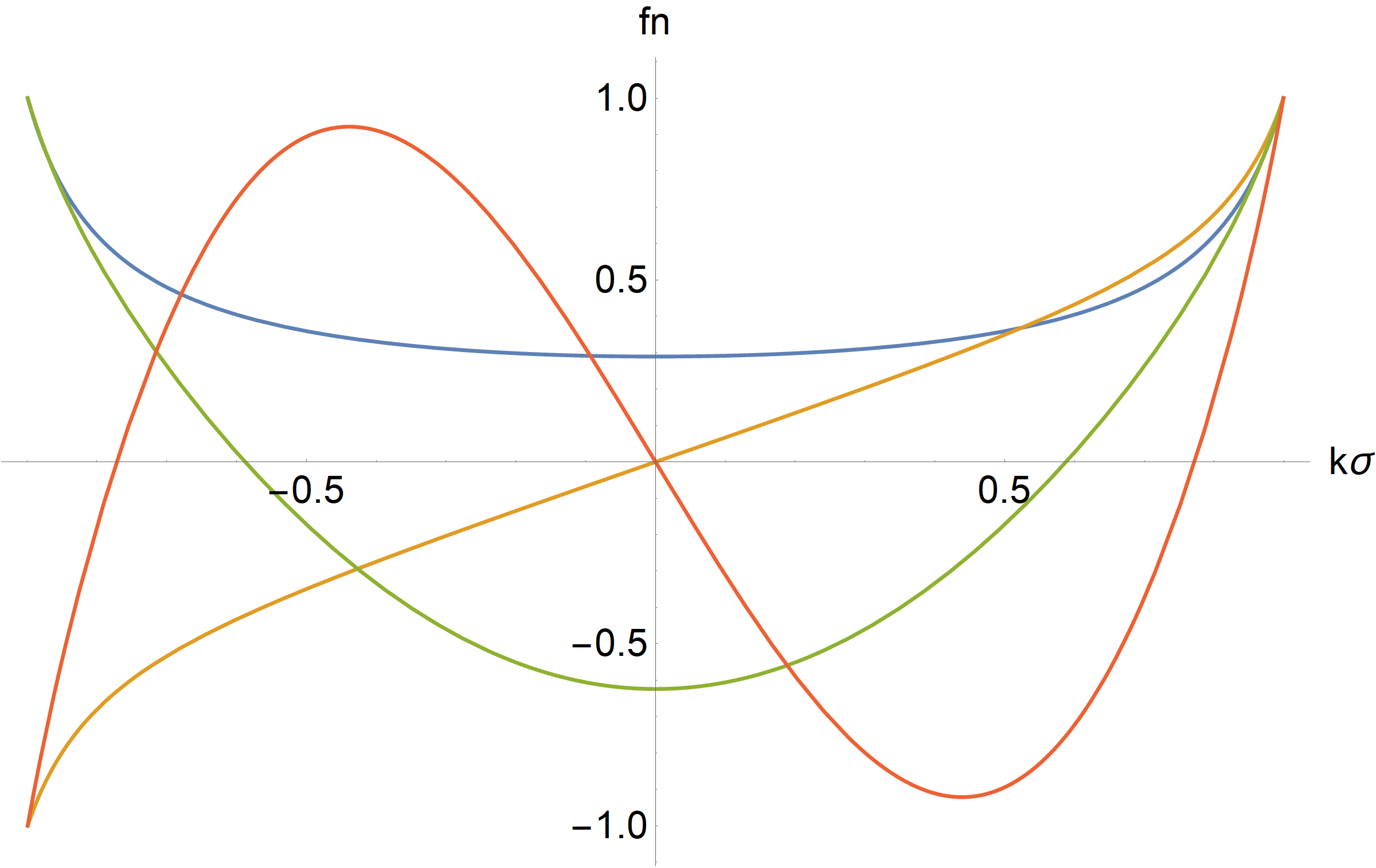}
\caption{\label{fig:spectrum_p} The first few eigenfrequencies \((\omega_n\ell)\) for the planar mode as a function of \(\beta\) (left), and the first few eigenfunctions plotted for \(\beta=0.9\) (right).}
\end{figure}

From varying the action of eq. \ref{eq:action_r} we derive the equation of motion for the radial mode \(f_r(\tau)\), which is defined on the boundary alone,
\be \ddot f_r + (1-2\gamma^2)k^2 f_r = 2\gamma k\dot f_p\,. \ee
This equation tells us that we can write an expansion for \(f_r\) using the same eigenfrequencies and oscillators as appear in \(f_p\), so the radial mode has no independent dynamics. We can write the following expansion for each of the radial modes \(f^{\pm}_r\) at \(\sigma=\pm\ell\)
\be f^\pm_r = i\sqrt{\mathcal N}\sum_{n\neq0} \frac{\alpha_n}{\omega_n} e^{-i\omega_n\tau}f_r^{\pm(n)} \ee
where \(\alpha_n\) and \(\omega_n\) are the same as in eq. \ref{eq:f_p_expansion}and \(f_r^{\pm(n)}\) are just constants. Then, we simply solve the equation of motion of the radial mode by taking
\be f_r^{\pm(n)} = \frac{2i\gamma k\omega_n}{\omega_n^2-(1-2\gamma^2)k^2} f_p^{(n)}(\sigma=\pm\ell) \ee
The boundary condition for the mode \(f_p^{(n)}\) is
\be \frac{T}{\gamma^2}f_p^{(n)\prime} \mp \left[\left(m\omega_n^2+\frac12\frac{T}{\ell}(3-\gamma^{-2})\right)f_p^{(n)}+2i\gamma m k \omega_n f_r^{(n)}\right] = 0 \ee
Inserting the solution for \(f_r\), we find the full boundary condition
\be \frac{T}{\gamma^2}f_p^{(n)\prime} \mp \left(m\omega_n^2+\frac12\frac{T}{\ell}(3-\gamma^{-2})-\frac{4 \gamma^2 m k^2\omega_n^2}{\omega_n^2-(1-2\gamma^2)k^2}\right)f_p^{(n)} = 0 \ee
Using the same procedure as we used for the transverse modes, we obtain the equation for the eigenfrequencies given the equation of motion and the boundary condition. We find that the eigenfrequencies are the zeros of the function
\begin{align}
f(\omega) = &\left[x^4(1-\beta^2)^2-x^2(1-\beta^2)(2\beta^2+6\beta^4)+\beta^4(1+\beta^2)^2\right]\sin\left(\frac{2x\arcsin\beta}{\beta}\right)- \nonumber \\
&\qquad -4\beta^2\sqrt{1-\beta^2}\left[x^3(1-\beta^2)-x\beta^2(1+\beta^2)\right]\cos\left(\frac{2x\arcsin\beta}{\beta}\right) = 0 \label{eq:w_p_lin}\end{align}
where \(x=\omega\ell\). The equation is of a similar form to that of the transverse modes (eq. \ref{eq:w_t_lin}). One can easily see that for \(\beta\to1\), the eigenfrequencies are again \(\omega_n\ell = n\), and there is no difference then between the transverse and planar modes. We plot the first few eigenfrequencies and eigenmodes in figure \ref{fig:spectrum_p}.

\subsection{The intercept}
As in the transverse case, we can write the expressions for the contribution of the planar and radial modes to the energy and angular momentum defined in section \ref{sec:classical}, expanding around the classical solution to quadratic order. And as before, we can use them to see that the combination \(\delta E - k\delta J\) is equal to the Hamiltonian for the fluctuations, in this case the one of eq. \ref{eq:H_p}. The expressions of \(E\) and \(J\) themselves contain terms not found in the Hamiltonian, for instance terms linear in the fluctuations,
\be \delta E_{1} = T\lambda \int_{-\ell}^\ell d\sigma \left(g^{3/2} k\sigma \dot f_p + \frac12 g^{5/2} k^2\sigma f_r + g^{1/2} f_r^\prime\right) + \gamma^{3/2} m\lambda k \ell (\gamma k f_r+\dot f_p)\vert_{\pm\ell}\ee
\be \delta J_{1} = T\lambda \int_{-\ell}^\ell d\sigma g^{1/2}\sigma\left(g \dot f_p + \frac12 (g^{2}+3) k f_r + k\sigma f_r^\prime\right) + \gamma^{1/2} m\lambda \ell ((\gamma^2+1) k f_r+\gamma \dot f_p)\vert_{\pm\ell}\ee
When taking the combination \(\delta J_1 - \frac{1}{k}\delta E_1\) one finds that all the terms in the bulk which did not vanish become a total derivative which, when integrated, cancels out the remaining boundary terms. Thus, the linear parts of \(\delta E\) and \(\delta J\) are irrelevant to the intercept. The quadratic part of \(\delta J - \frac{1}{k}\delta E\) similarly ends up taking the form of the Hamiltonian of eq. \ref{eq:H_p}.

The next part of the calculation is to insert the mode expansion into the Hamiltonian, and use the boundary condition and relation between the planar and radial modes to show that in this case also, the Hamiltonian is diagonal.
Defining the parameters
\be M_p(\sigma)^2 = \frac14k^2(1+9g^2)\,\qquad m_p^2 = \frac12k^2(3\gamma^2-1)\,,\qquad m_r^2 = k^2(2\gamma^2-1)\,,\qquad c = 2\gamma k \ee
one can write more compactly the Hamiltonian,
\begin{align*} H =
T\lambda^2\int_{-\ell}^\ell d\sigma\left(\frac12\dot f_p^2 + \frac12g^{-2} f_p^{\prime2}+\frac12M_p(\sigma)^2 f_p^2\right) + m\lambda^2\left(\frac12\dot f_p^2 - \frac12m_p^2f_p^2+\frac12\dot f_r^2-\frac12m_r^2\right)\vert_{\pm\ell} \end{align*}
the solution for the radial mode equation of motion,
\be f_r^{(n)} = \frac{i\omega_n c}{\omega_n^2-m_r^2}f_p^{(n)} \ee
and the full boundary condition for \(f_p\)
\be \frac{T}{m}\gamma^{-2}f_p^{(n)\prime} = \pm \left((\omega_n^2-m_p^2)f_p^{(n)}+i\omega_n c f_r^{(n)}\right) = \pm \left(\omega_n^2-m_p^2- \frac{c^2\omega_n^2}{\omega_n^2-m_r^2}\right)f_p^{(n)}\,. \label{eq:boundary_p}\ee

\begin{figure} \centering
\includegraphics[width=0.48\textwidth]{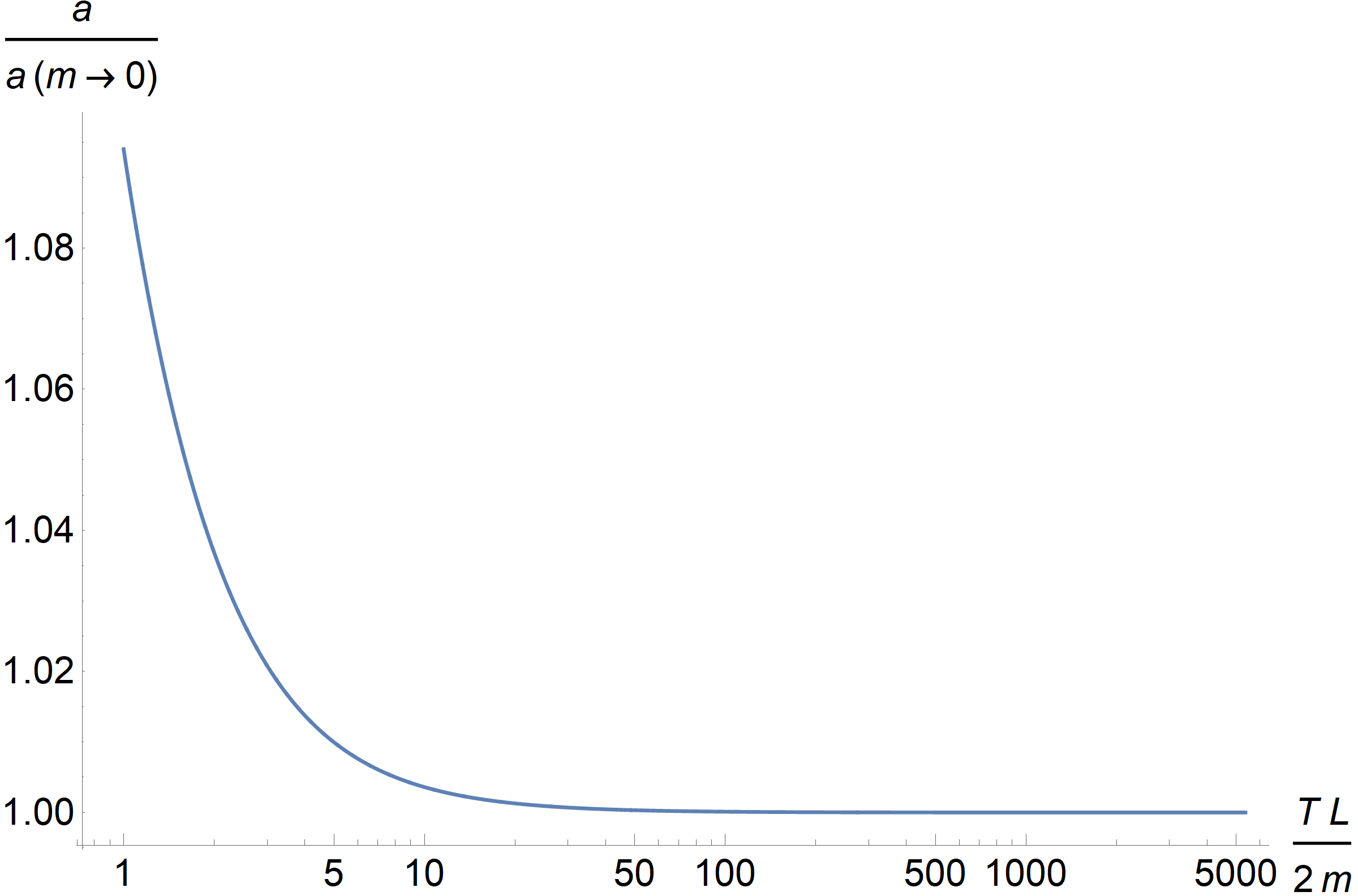}
\caption{\label{fig:intercept_p} The contribution to the intercept of the rotating string as a function of \(\frac{TL}{2m}=\gamma^2\beta^2\) from the planar mode. The intercept is normalized by the value at \(m\to0\), which is \(\frac{1}{24}\).}
\end{figure}

Inserting the expansions for \(f_p\) and the related \(f_r\) into \(H\):
\begin{align} H = \frac12 T\lambda^2 \mathcal N \sum e^{-i(\omega_m+\omega_n)\tau}\alpha_m \alpha_n (1-\frac{\omega_n}{\omega_m}) \times \nonumber \\ \times \left[\int_{-\ell}^\ell d\sigma f_m f_n+ \frac{m}{T}\left(1+\frac{c^2m_r^2}{(\omega_m^2-m_r^2)(\omega_n^2-m_r^2)}\right)(f_m^+ f_n^+ + f_m^- f_n^-)\right]
\end{align}
The integral plus the boundary terms are exactly of the form of the orthogonality relation between \(f_m\) and \(f_n\) that can be derived using the boundary condition \ref{eq:boundary_p} and the general orthogonality equation of the modes \ref{eq:SturmLiouvilleOrtho}. Therefore,
\be H = \frac12 T\lambda^2 \mathcal N \sum e^{-i(\omega_m+\omega_n)\tau}\alpha_m \alpha_n (1-\frac{\omega_n}{\omega_m})\ell(\delta_{n+m}+\delta_{n-m}) = \frac12 \frac{1}{\ell}\sum_n \alpha_{-n}\alpha_n
\ee
And the planar mode's contribution to the intercept has the same form as that of the transverse modes,
\be a = -\frac{1}{2\beta}\sum_{n>0} \omega_n\ell \ee
We can now calculate the intercept using the contour integral representation of the infinite sum. We use the function \(f(\omega)\) defined in eq. \ref{eq:w_p_lin}, and perform on it the same contour integral defined in section \ref{sec:transverse_sum}. We find that we again get the same form of the divergences, a quadratic divergence that goes like \(\Lambda^2 L \frac{\arcsin\beta}{\beta}\), and logarithmic divergences proportional to \(\frac{T}{\gamma m}\). These are all terms that can be eliminated by either redefining the physical \(T\) and \(m\), or renormalizing the Casimir force as in section \ref{sec:transverse_sum}. Using the same regularization method used before of subtracting the divergent terms in their integral form, the final (and finite) result is
\be a_p = -\frac{1}{2\pi \beta} \int_0^\infty dy\log\left[1-\exp\left(-\frac{4\arcsin\beta}{\beta}y\right)\left(\frac{y^2-2y\gamma\beta^2+\gamma^2\beta^2(1+\beta^2)}{y^2+2y\gamma\beta^2+\gamma^2\beta^2(1+\beta^2)}\right)^2\right] \label{eq:a_p_finite}\ee
The \(\beta\to1\) limit is again \(\frac{1}{24}\). Around it we can expand the result as
\be a_p = \frac{1}{24} + \frac{11}{720\pi}\frac{1}{\gamma^3} + \ldots = \frac{1}{24} + \frac{11}{720\pi}\left(\frac{2m}{TL}\right)^{3/2} + \ldots \label{eq:a_p_approx} \ee
The intercept as a function of \(\beta\) is plotted in figure \ref{fig:intercept_p}.

\section{Quantizing the non-critical string} \label{sec:noncritical}
The quantization of Polyakov's bosonic string in non-critical dimensions is achieved by adding a Liouville term. In Polyakov's original paper \cite{Polyakov:1981rd}, when picking the conformal gauge for the world sheet metric \(\gamma_{\alpha\beta} = e^\phi \eta_{\alpha\beta}\). Classically, the action is independent of \(\phi\), but the Liouville field action for \(\phi\) emerges in the careful analysis of the measure in the path integral, where it is needed to preserve the Weyl invariance quantum mechanically. The Liouville term in the action is proportional to \(26-D\).

Polchinski and Strominger suggested \cite{Polchinski:1991ax} to cure the non-critical Nambu-Goto action by similarly adding a term to the action that changes the corresponding Virasoro central charge from $c=D$ to $c=26$. This can be achieved \cite{Hellerman:2014cba} by identifying the Liouville field with the composite $\phi= -\frac12 \log\left(\pa_+X\cdot\pa_-X\right)$, which is the same as identifying the worldsheet metric in the Polyakov action with the induced metric \(h_{\alpha\beta} = \pa_\alpha X^\mu \pa_\beta X_\mu\) of the NG action. This new term, the Polchinski-Strominger (PS) term, is in fact the first term added to the NG action in the effective string theory, where the effective action is written in a long string expansion \cite{Aharony:2010cx,Aharony:2011ga,Aharony:2013ipa}.

We assume that the PS term  is what is needed for the consistency of the quantum picture also for the case of the string with massive endpoints. In what follows we analyze the PS term for our case.

\subsection{The Polchinski-Strominger term and the intercept}

The PS term in the action in the orthogonal gauge is
\be S_{PS} = \frac{B}{2\pi}\int d^2\sigma \pa_+ \phi \pa_- \phi = \frac{B}{2\pi} \int d^2\sigma \frac{(\pa_+^2 X\cdot \pa_-X)(\pa_-^2 X\cdot \pa_+ X)}{(\pa_+ X\cdot \pa_- X)^2} \ee
The derivatives are
\be \pa_\pm X = \frac12 (\pa_\tau \pm \pa_\sigma) X \ee
The coefficient \(B\) is fixed by requiring that the  energy momentum tensor  has the appropriate  conformal symmetry  OPE  in any dimension \(D\), and it is \(B = (26-D)/12\). In terms of derivatives in \(\sigma\) and \(\tau\),
\be S_{PS} = \frac{B}{2\pi}\int d^2\sigma \frac{(\dot X\cdot(\ddot X + \dot X^\prime))(\dot X\cdot(\ddot X - \dot X^\prime))}{(-\dot X^2)^2} \label{eq:PS_term2}\ee

We compute the PS term's contribution to the intercept by inserting the classical rotating solution into the PS Hamiltonian \(H_{PS} = -\int d\sigma \mathcal L_{PS}\) to find its expectation value at leading order. We use the rotating solution with \(R(\sigma) = \frac1k\sin(k\sigma)\) to be consistent with the gauge choice in which the action above is written. The result is
\be\label{PSm} E_{PS} = \langle H_{PS} \rangle = -\int d\sigma \mathcal L_{PS} = \frac{B}{2\pi} \int_{-\ell}^\ell d\sigma k^2\tan^2(k\sigma) = \frac{B}{\pi}k(\tan\delta-\delta)\ee
where \(\delta = k\ell\) as defined in section \ref{sec:classical}. The term \(k\tan\delta\) diverges in the massless case where \(\delta = \pi/2\). For small finite masses, it is finite but large.

\subsection{Renormalization of the Polchinski-Strominger term}
In \cite{Hellerman:2013kba} it was found that, for the \(m=0\) case, the divergence in the PS term can be renormalized by first modifying
\be S_{PS}\to S_{PS}^{(reg)} = \frac{B}{2\pi} \int d^2\sigma \frac{(\pa_+^2 X\cdot \pa_-X)(\pa_-^2 X\cdot \pa_+ X)}{(\pa_+ X\cdot \pa_- X)^2+\alp \epsilon^4 (\pa_+^2 X\cdot \pa_-^2 X)} \ee
Then, the divergence as \(\epsilon\) is taken to zero can be canceled by a boundary counterterm of the form 
\be S_{ct} \propto \frac{1}{\epsilon} \int d\tau (\ddot X\cdot \ddot X)^{1/4} \ee
The counterterm is one of the operators that one would expect to find on the boundary in the effective string theory \cite{Aharony:2013ipa,Hellerman:2016hnf}, and resembles a mass term.

The result of this regularization is that the PS contribution to the intercept in the massless case is
\be a_{PS}(m=0) = \frac{26-D}{24} \ee
which corresponds to taking \(\delta = \pi/2\) in eq. \ref{PSm} and dismissing the infinite part that goes like \(\tan\delta\). The full intercept in that case, including the contribution from the fluctuations turns out to be \(a=1\), independently of the spacetime dimension \(D\).

For finite masses, the endpoint masses themselves act as the regulator, giving a finite answer for finite masses. This can be seen clearly from eq. \ref{eq:PS_term2}. The denominator goes like \(\dot X^2\), which is zero in the massless case (when the endpoints move at the speed of light), but finite with masses. On the other hand, the fact that the result does not diverge for finite masses is not enough. Since we would like a result that can be smoothly continued to the massless result of \cite{Hellerman:2013kba}, the term that diverges at the massless limit should be subtracted. We now ask what is the correct way of doing this subtraction.

We note that with masses, using the classical boundary condition, we can write
\be \frac{T}{mk} = \frac{\sin\delta}{\cos^2\delta} \qquad\Rightarrow\qquad k\tan\delta = \frac{T\cos\delta}{m}=\frac{T}{\gamma m} \ee
and define the effective length of the string and mass of the endpoint in the rotating frame as we did in section \ref{sec:transverse_sum},
\be \tilde m = \gamma m \qquad \tilde L = L\frac{\arcsin\beta}{\beta} = \frac{2}{k}\sin\delta \times \frac{\delta}{\sin\delta} = \frac{2\delta}{k}\ee
to write the PS energy as
\be E_{PS} = \frac{B}{\pi}\left(\frac{T}{\tilde m}-\frac{2\delta^2}{\tilde L}\right) \ee
Then, it seems that we can renormalize \(E_{PS}\) using the same kind of subtraction that we used to regularize the contributions from the fluctuations, subtracting from the force the term that remains when the string length is taken to infinity, and integrating back to leave only the ``finite'' term that goes like \(1/L\) (we assume no finite integration constant is left as \(L\to\infty\)).


The finite (at all \(m\)) PS intercept is then simply
\be a_{PS} = -\frac{1}{k}E_{PS}^{(ren)} = \frac{26-D}{12\pi}\delta = \frac{26-D}{12\pi}\arcsin\beta \ee
For \(\beta\to1\) then \(a_{PS} = \frac{26-D}{24}\) as was found in \cite{Hellerman:2013kba}.
As a function of \(2m/TL\), it is
\be a_{PS} = \frac{26-D}{12\pi}\arccos\left(\sqrt{\frac{2m}{2m+TL}}\right) = \frac{26-D}{24}\left[1 - \frac{2}{\pi}\left(\frac{2m}{TL}\right)^{1/2} + \frac{2}{3\pi}\left(\frac{2m}{TL}\right)^{3/2} + \ldots\right] \label{eq:a_PS_approx} \ee

The subtraction from the PS term can also be achieved through a renormalization of the mass. The term that diverges at the massless limit is proportional to \(\frac{1}{\gamma m}\), as was the divergent term we subtracted from the fluctuations' contributions to the intercept (see eq. \ref{eq:4100}), and all these divergences are subtracted by adding appropriate mass counterterms. To see explicitly how we do it in this case, we add to the action a boundary counterterm
\be S_{ct} = -\delta m \int d\tau \sqrt{-\dot X^2} \ee
which, evaluated for the classical rotating solution is
\be S_{ct} = -\delta m \int d\tau \cos\delta \ee
Then, to eliminate the divergent part in the PS term, we can take 
\be 2\delta m \cos\delta = -k \tan\delta \qquad \Rightarrow \qquad \delta m = \frac{B}{2\pi}k\frac{\sin\delta}{\cos^2\delta} = -\frac{B}{4\pi}\frac T m \ee
The last equality is due to the classical boundary condition, and the factor of two is there because we take equal contributions from the two boundaries. The counterterm coefficient \(\delta m\) is proportional then to \(T/m\), and it diverges as \(m\) itself is taken to zero in such a way that it cancels the divergence of the PS term. 

\begin{figure}[ht!] \centering
\includegraphics[width=0.48\textwidth]{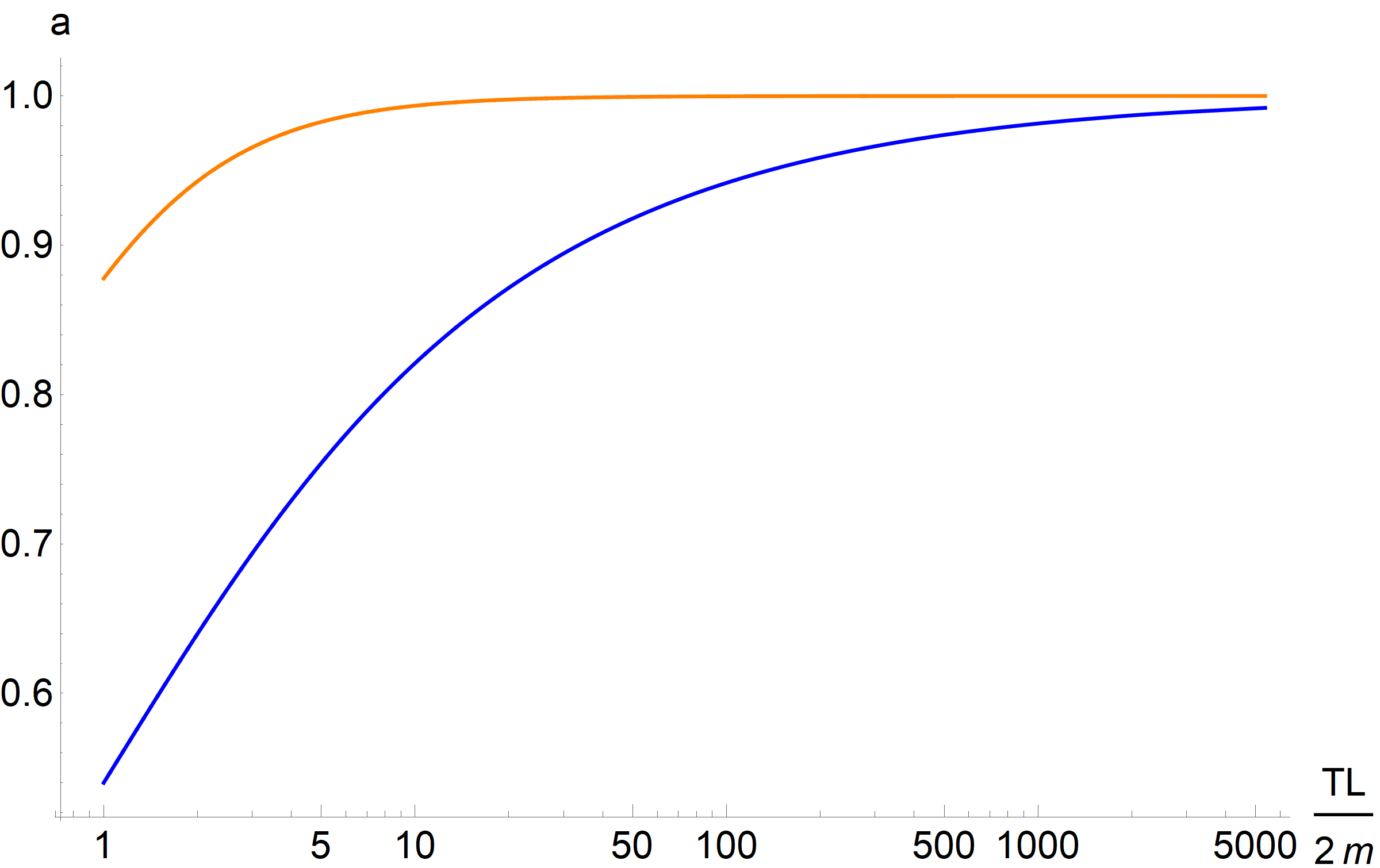}
\includegraphics[width=0.48\textwidth]{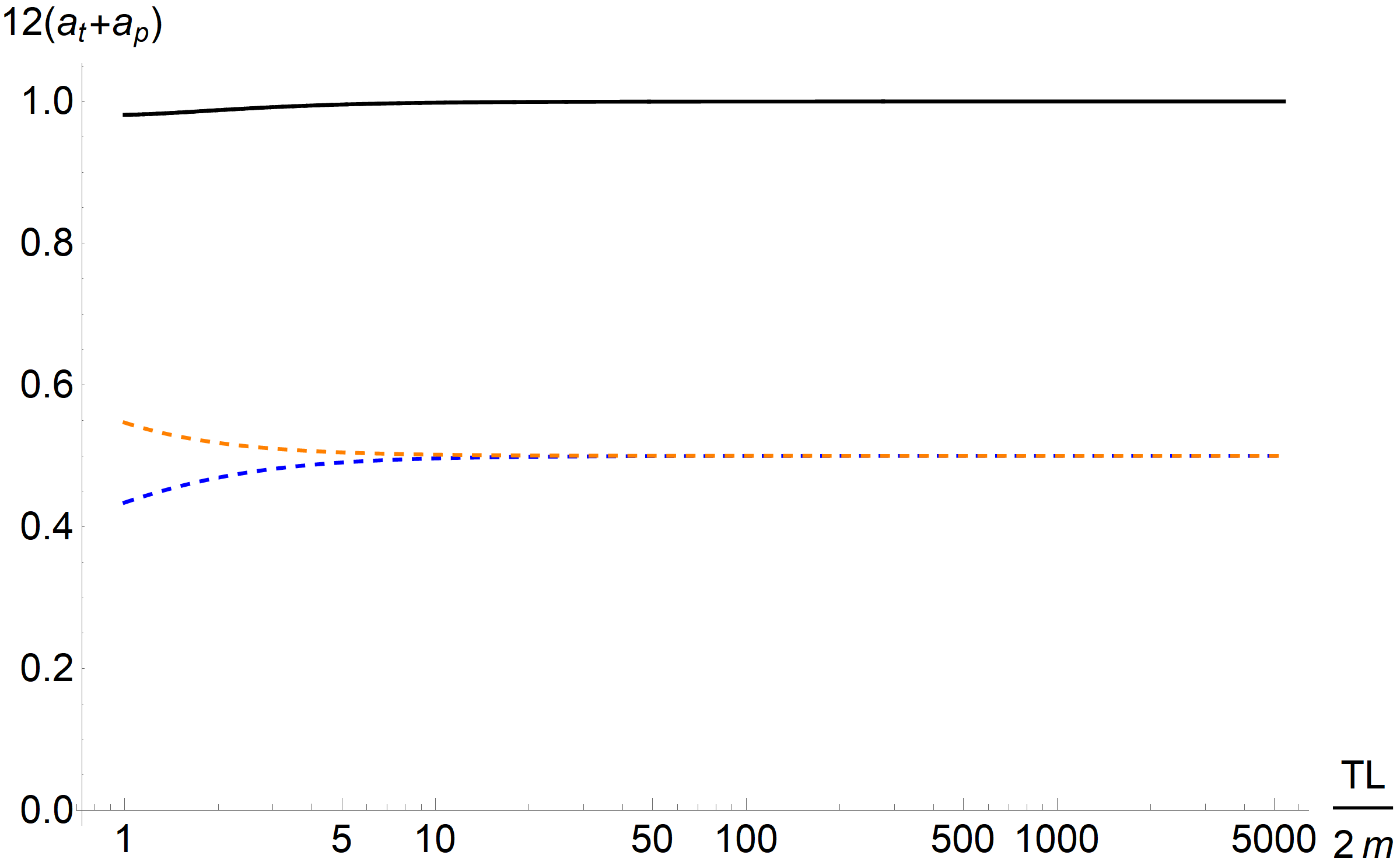}
\caption{\label{fig:intercept_full} \textbf{Left:} The intercept in \(D=4\) (lower, blue line) and in \(D=26\) (orange). For \(D=4\) the PS term is dominant. \textbf{Right:} Here we plot the contributions from the fluctuations separately, where the dashed lines are \(a_t\) (blue) and \(a_p\) (orange), and the solid line \(a_p+a_t\). We normalize this by \(\frac{1}{12}\), which is the value of \(a_t+a_p\) at \(TL/m\to\infty\).}
\end{figure}

If we take the result for  \(a_{PS}\) from above, our calculation of the full intercept in \(D\) dimensions is now complete. It is given in total by
\be a = (D-3)a_t + a_p + a_{PS} \ee
where \(a_t\) is the intercept of the transverse fluctuations computed in section \ref{sec:transverse}, \(a_p\) the planar intercept of section \ref{sec:planar}, and \(a_{PS}\), the result in the present section is proportional to \(26-D\). We plot the full intercept for \(D = 4\) in figure \ref{fig:intercept_full}, alongside the comparison to \(D=26\) where the PS term does not play a role. In contrast, for \(D=4\) the PS intercept is dominant, and results in larger corrections to \(a=1\) than given by the fluctuation modes.

\section{The quantum Regge trajectory} \label{sec:quantumRegge}
The full Hamiltonian for all the modes about the rotating solution, can be written as
\be H= \frac{2}{L}\sum_{n=1}^\infty \left(\alpha_{-n}^i\alpha_n^i + \alpha_{-n}^p\alpha_n^p\right)-k\left((D-3) a_t + a_p + a_{PS}\right) \ee
where \(i = 3,\ldots,D-1\) and \(p\) denotes the operators for the single planar mode. The operators \(\alpha_n\) are normal ordered and the intercept that we calculated in the previous sections is added. We also count the effect of the Polchinski-Strominger term by adding only its expectation value \(\langle H_{PS}\rangle = -k a_{PS}\). We assume that any extra excitations that might arise from it are not present at the leading order.

We can construct the spectrum of states the usual way by acting on the vacuum with \(\alpha_{-n}\) for \(n>0\). The vacuum will be a state representing the rotating solution with no fluctuations and annihilated by all \(\alpha_n\) with \(n>0\).

The worldsheet Hamiltonian was related to the target space energy and angular momentum through \(H = \delta E - k\delta J\). We want to look at the effect of the new modes on the Regge trajectories. For the massless case, we can write the quantum trajectory as 
\be J + N = \alp E^2  + a \label{eq:radial_massless}\ee
where \(N\) is the quantum radial excitation number. In the massive case we can define the number operators
\be N_t = \sum_{n=1}^\infty \alpha_{-n}^i\alpha_n^i \qquad N_p = \sum_{n=1}^\infty \alpha_{-n}^p\alpha_n^p \ee
and using them, write 
\be J - J_{cl}(E) = -\frac{1}{k}H = (D-3)a_t + a_p -\frac{2}{kL}(N_t + N_p) \ee
Thus the generalization of eq. \ref{eq:radial_massless} to the case of the rotating string with endpoint masses turns out to be
\be J + \frac{1}{\beta}(N_t+N_p) = J_{cl}(E) + a\label{eq:radial_massive}\ee
We now use \(a = (D-3)a_t + a_p + a_{PS}\). When acting with \(N_t\) or \(N_p\) on a generic state built from acting with creation operators on the vacuum, their value is
\be N_t = \sum_{n=1}^\infty \omega_n^{(t)}\ell N_n^{(t)} \qquad N_p = \sum_{n=1}^\infty \omega_n^{(p)}\ell N_n^{(p)} \ee
where \(N_n\) counts the number of the corresponding \(\alpha_{-n}\) used to construct the state. The eigenfrequencies (times \(\ell\)) are the solutions of the equations we wrote down in previous sections (eqs. \ref{eq:w_t_lin} and \ref{eq:w_p_lin}).

\begin{table}[t!] \centering
\begin{tabular}{|c|c|c|c|} \hline
\(N\) & State & \(N_t + N_p\) & No. of states \\ \hline

1 & \(\alpha_{-1}^i\vac\) & \(\omega_1^{(t)}\ell\)  & \(D-3\) \\ \cline{2-4}

 & \(\alpha_{-1}^p\vac\) & \(\omega_1^{(p)}\ell\) & \(1\) \\ \hline

 & \(\alpha_{-2}^i\vac\) & \(\omega_2^{(t)}\ell\) & \(D-3\) \\ \cline{2-4}

 & \(\alpha_{-2}^p\vac\) & \(\omega_2^{(p)}\ell\) & \(1\) \\ \cline{2-4}

2 & \(\alpha^i_{-1}\alpha^j_{-1}\vac\) & 2\(\omega_1^{(t)}\ell\) & \((D-3)^2\) \\ \cline{2-4}

 & \(\alpha_{-1}^i\alpha_{-1}^p\vac\) & \(\omega_1^{(t)}\ell + \omega_1^{(p)}\ell\) & \(D-3\) \\ \cline{2-4}

 & \(\alpha_{-1}^p\alpha_{-1}^p\vac\) & \(2\omega_1^{(p)}\ell\) & \(1\) \\ \hline

\end{tabular}
\caption{\label{tab:spectrum} The first two excited states and their masses. In total there are \(D-2\) states with \(N=1\) and \((D-2)^2+1\) with \(N=2\), as expected. The level \(N=1\) is split in two, and the level \(N=2\) is split in five. }
\end{table}

The last few equations tell us that, for a given \(J\) we can build the radial trajectories defined by eq. \ref{eq:radial_massive}. We can also think of eq. \ref{eq:radial_massive} as defining multiple parallel orbital trajectories for \(J\) as a function of \(E\), noting that the \(\omega_n\) and the intercept are themselves implicitly \(J\)-dependent through their dependence on \(\beta\). This also means that, unlike in the massive case, the spacing between parallel orbital trajectories is not constant when the string has massive endpoints.

We can define the \(N\)-th energy level as the collection of the states with a given value of \(N = \sum_n n(N_n^{(t)}+N_n^{(p)})\). Unlike the massless case, where \(N\) is exactly the number operator of eq. \ref{eq:radial_massless}, there are two types of degeneracies that are removed in the massive case, and not all states with equal \(N\) will have the same mass. First, the distinction between the planar and transverse modes means that states constructed from \(\alpha^{p}_{-n}\) have different masses from those constructed from \(\alpha_{-n}^i\), so we have to count \(N_t\) and \(N_p\) separately. Second, for both the planar and transverse fluctuations the eigenfrequencies no longer obey \(\omega_{n+m}=\omega_n+\omega_m\), which obviously holds in the massless case where \(\omega_n\ell = n\).

We list the first two levels of states in table \ref{tab:spectrum}. To calculate the mass of each state, we have to insert \(N_t+N_p\) into the Regge trajectory \ref{eq:radial_massive} and then solve for \(E\).

\section{Generalization to asymmetric string} \label{sec:asym}
The previous sections have all assumed that the two endpoint masses are identical. Here we let go of this assumption and generalize the results of the above for the case of two different masses \(m_1\neq m_2\) on the endpoints. We take a mass \(m_1\) at \(\sigma = \ell_1\), and \(m_2\) at \(\sigma = -\ell_2\).

The most extreme case of asymmetry is that of a string with one massless endpoint and an infinite mass on the other, corresponding to Neumann-Dirichlet boundary conditions for the string. In that case, one can write the usual mode expansion and find that \(\omega_n = n-\frac12\) (\(n = 1,2,\ldots\)), and the contribution to the intercept from a single mode with ND boundary conditions is\footnote{The sum can be renormalized using Zeta function regularization. One simple way but non-rigorous way to get it is to say that \(\frac12+\frac32+\frac52\ldots = \frac12(1 + 2 + 3 + \ldots) - (1 + 2 + 3 + \ldots) = -\frac12 \zeta(1) = \frac{1}{24}\).}
\be a_{ND} = -\frac{1}{2}\sum_{n=1}^\infty(n-\frac12) = -\frac{1}{48} \ee
as opposed to \(+\frac{1}{24}\) from modes with NN or DD boundary conditions. This example illustrates that the intercept can change drastically for asymmetric boundary conditions, and that, for the string with massive endpoints, we can expect to see the intercept flip sign when one endpoint mass is much larger that the other.

In the following we use the parametrization \(R(\sigma)=\sigma\) to compute the correction from the fluctuations.

The two classical boundary conditions are
\be \frac{T\ell_i}{m_i} = \beta_i^2\gamma_i^2 \label{eq:802}\ee
The two arms of the string rotate with the same angular velocity \(k\), so
\be \beta_i = k\ell_i \qquad \Rightarrow \qquad \frac{\beta_1}{\ell_1} = \frac{\beta_2}{\ell_2} \ee
This is used together with the boundary conditions to write \(\beta_2\) as a function of \(\beta_1\) and the ratio between the masses:
\be \beta_2 = \frac{\beta_1^2-1+\sqrt{\beta_1^4+\beta_1^2 (4r^2-2)+1}}{2 \beta_1 r} \qquad r \equiv \frac{m_1}{m_2} \label{eq:beta12}\ee
In the following we write our expressions in terms of \(\beta_1\) and \(\beta_2\), where the two parameters are always related by the last equation.

\paragraph{Transverse modes:} To generalize the the calculation of section \ref{sec:transverse} to the asymmetric case, we generalize the boundary conditions of eq. \ref{eq:boundary_t_lin} to
\be \frac{T k}{\gamma_1^2} f^\prime_n - (m_1\omega_n^2+\frac12 Tk^2\ell_1) = 0 \ee
\be \frac{T k}{\gamma_2^2} f^\prime_n + (m_2\omega_n^2+\frac12 Tk^2\ell_2) = 0 \ee
The generalization of the eigenfrequency equation is 
\begin{align} 
&\left(\frac{\beta_1}{\gamma_2}+\frac{\beta_2}{\gamma_1}\right) y\cos\left[y \left(\arccos(\beta_1)-\arccos(-\beta_2)\right)\right] +
  \nonumber \\ 
&\qquad \left(\frac{y^2}{\gamma_1\gamma_2}-\beta_1\beta_2\right)\sin\left[y \left(\arccos(\beta_1)-\arccos(-\beta_2)\right)\right] = 0
\end{align} 
where \(y \equiv \frac{\omega_n}{k}\). If we take the symmetric case, \(\beta_1=\beta_2=\beta\), the equation reduces to eq. \ref{eq:w_t_lin} (\(y\) in the above is \(x/\beta\) in the previous definition).

The intercept is again given by
\be a_t = -\frac12\sum_{n=1}^\infty \frac{\omega_n}{k} \ee
and can be computed by performing the integral (which generalizes eq. \ref{eq:a_t_finite})
\be a_t = -\frac{1}{2\pi}\int_0^\infty dy \log\left(1-e^{-2(\arcsin\beta_1+\arcsin\beta_2)y}\frac{(y-\beta_1^2\gamma_1)(y-\beta_2^2\gamma_2)}{(y+\beta^2_1\gamma_1)(y+\beta^2_2\gamma_2)}\right) \label{eq:a_t_gen}\ee

\paragraph{Planar mode:} In a similar generalization of eq. \ref{eq:a_p_finite}, the planar mode intercept is given by
\begin{align} a_p &= -\frac{1}{2\pi}\int_0^\infty dy \nonumber \\ &\log\left[1-e^{-2(\arcsin\beta_1+\arcsin\beta_2)y}\left(\frac{y^2-2y\gamma_1\beta_1+\gamma_1^2(1+\beta_1^2)}{y^2+2y\gamma_1\beta_1+\gamma_1^2(1+\beta_1^2)}\right)\left(\frac{y^2-2y\gamma_2\beta_2+\gamma_2^2(1+\beta_2^2)}{y^2+2y\gamma_2\beta_2+\gamma_2^2(1+\beta_2^2)}\right)\right] \label{eq:a_p_gen}\end{align}

\paragraph{PS term:} The result is simply generalized by inserting the different boundary values \(\delta_1\) and \(\delta_2\) into eq. \ref{PSm}
\be a_{PS} = \frac{26-D}{24\pi}(\delta_1+\delta_2) = \frac{26-D}{24\pi}(\arcsin\beta_1+\arcsin\beta_2) \label{eq:a_PS_gen}\ee

\begin{figure} \centering
(1) \includegraphics[width=0.45\textwidth]{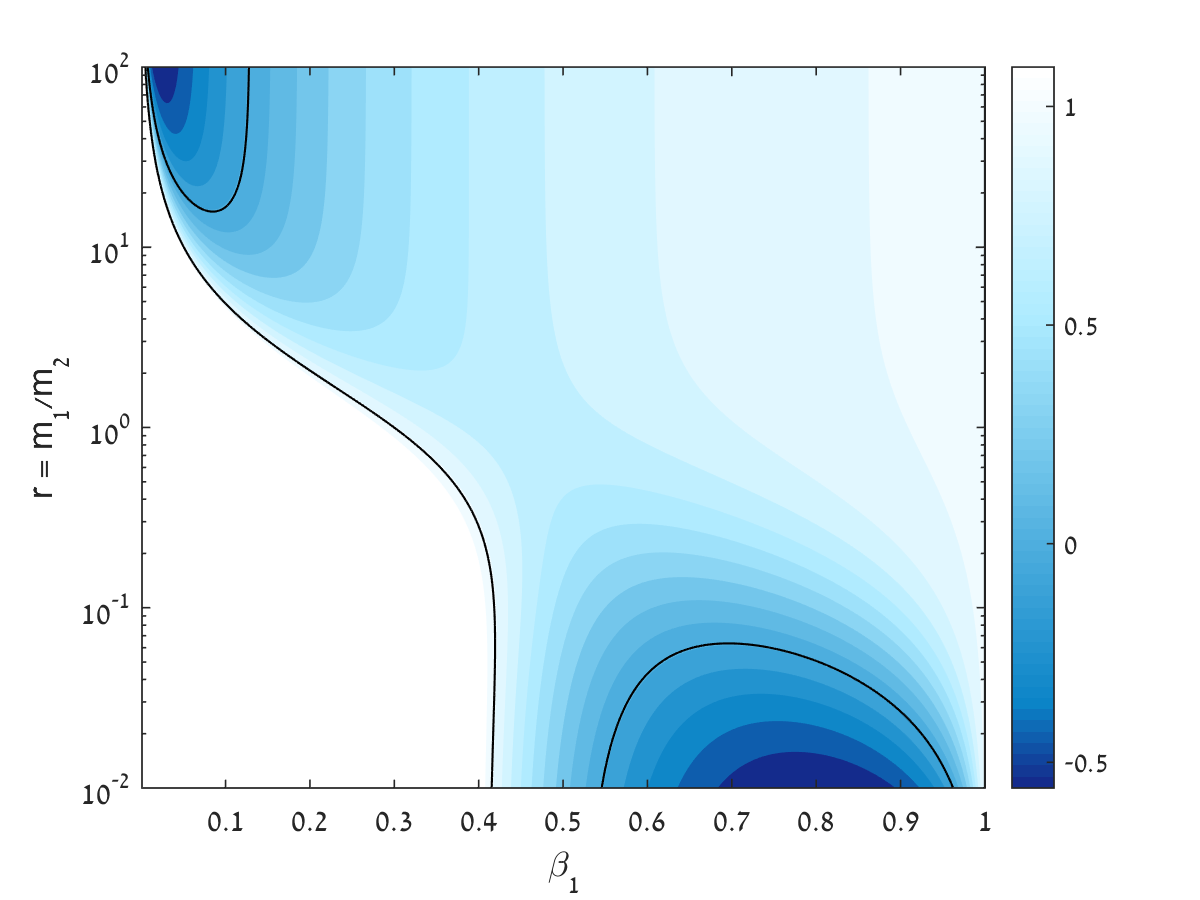}
(2) \includegraphics[width=0.45\textwidth]{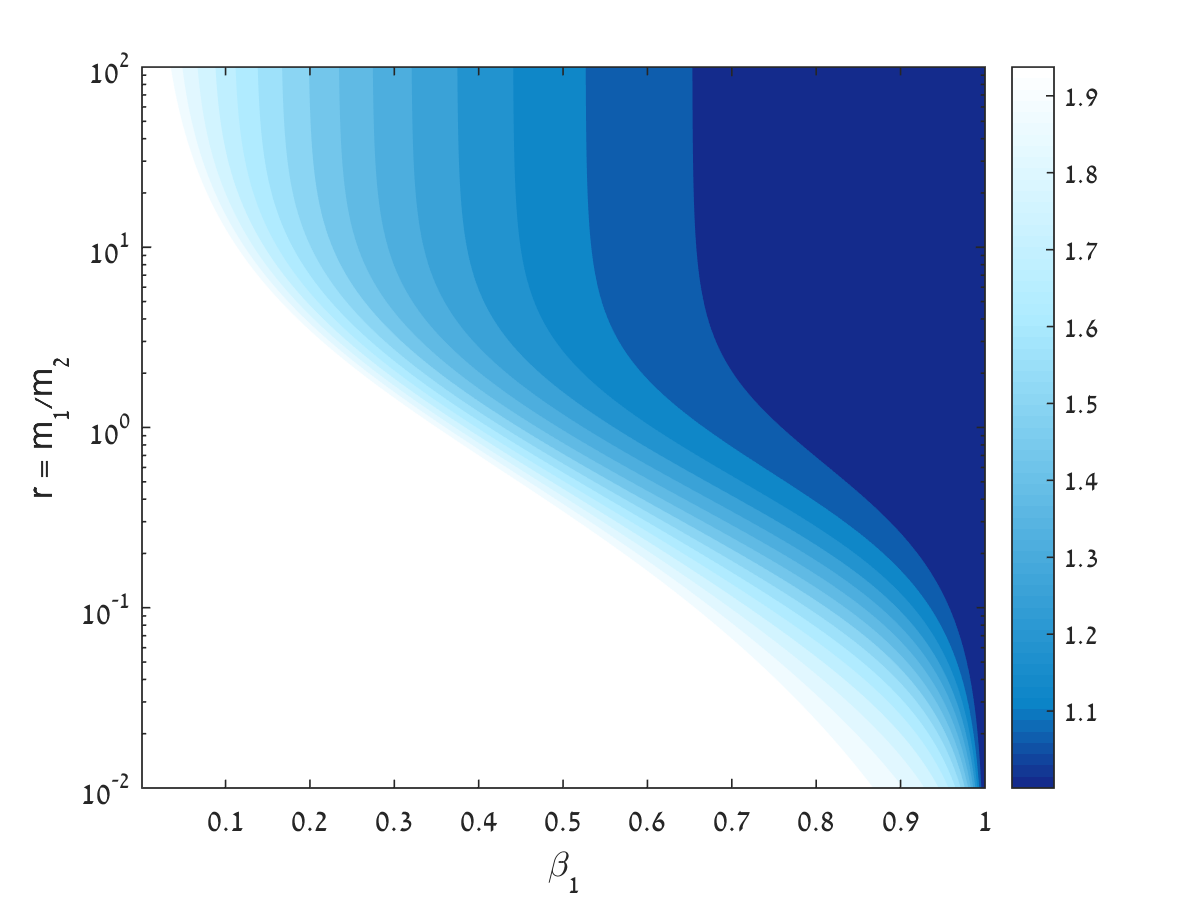} \\ 
(3) \includegraphics[width=0.45\textwidth]{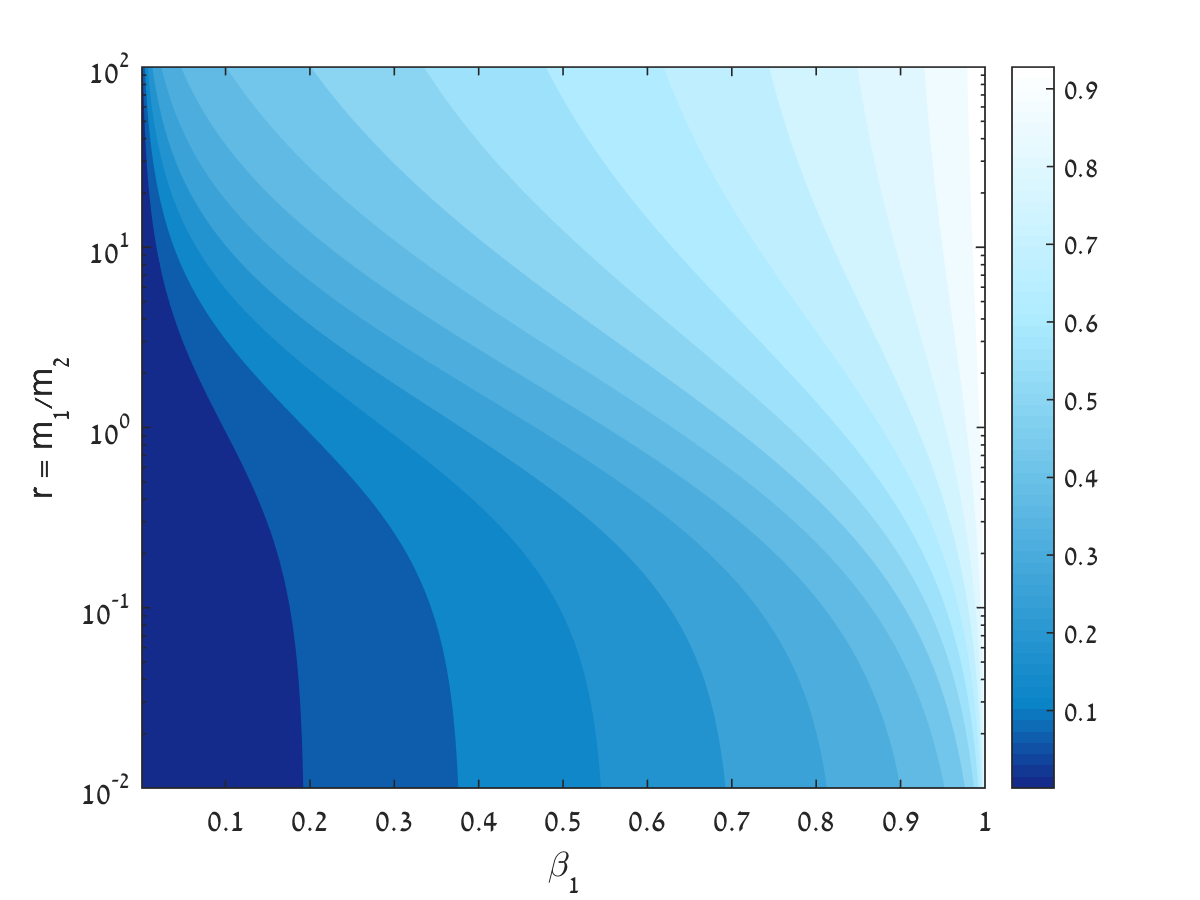}
(4) \includegraphics[width=0.45\textwidth]{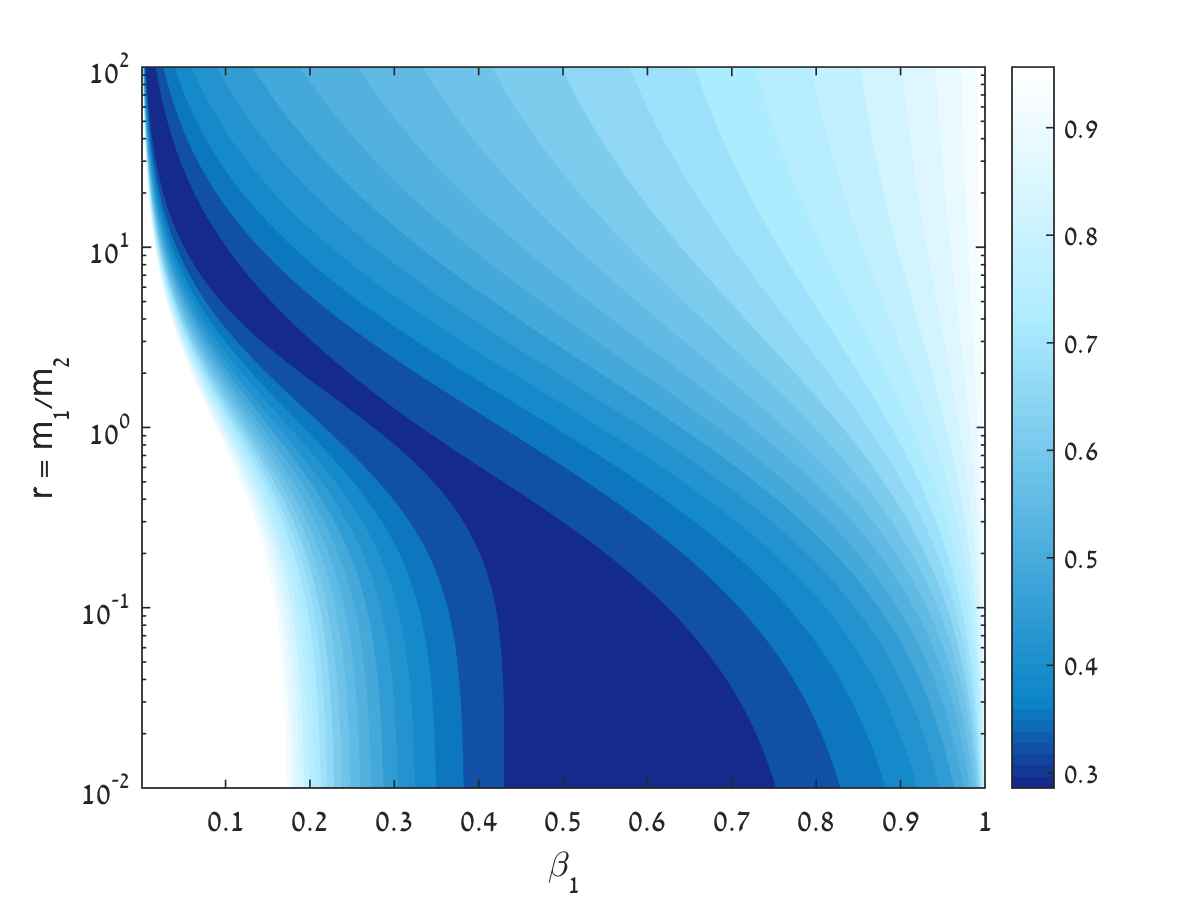} \\
(5) \includegraphics[width=0.45\textwidth]{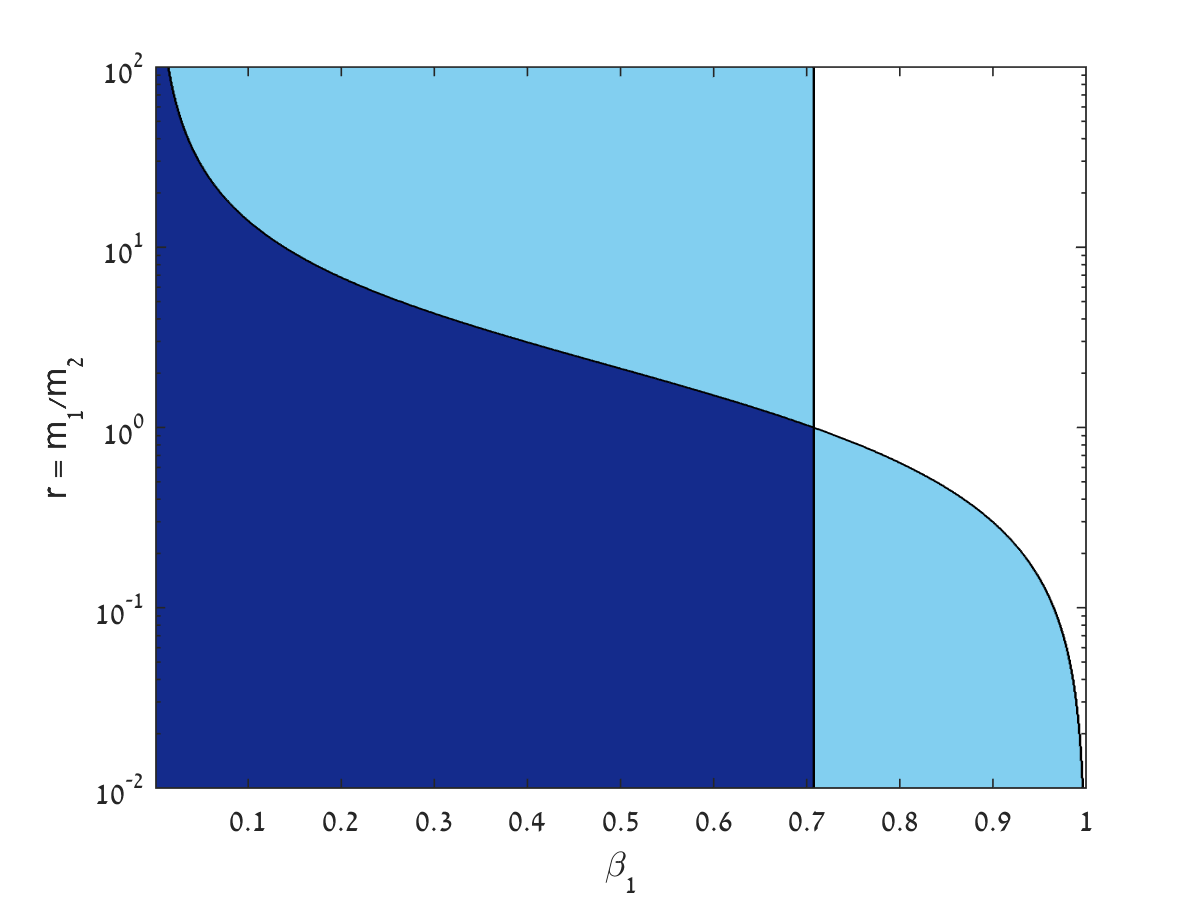} \\
\caption{\label{fig:intercept_asym} The intercept as a function of \(\beta_1\) and \(r = m_1/m_2\). (1) The transverse intercept \eqref{eq:a_t_gen}. In this plot the intercept changes sign and along the black line \(a_t = 0\). (2) The planar intercept \eqref{eq:a_p_gen}. (3) The PS intercept \eqref{eq:a_PS_gen}. (4) The full intercept with all three contributions in \(D=4\). (5) The domain of validity defined by \(\beta_1,\beta_2 > \frac{1}{\sqrt{2}}\) (see text). The white area is where both \(\ell_1\) and \(\ell_2\) are both (beginning to be) long and the approximation can be trusted. In the light blue area only one of the arms of the string is long, and dark blue means both are short. Note that in (1) and (2) the values in the plot are normalized by dividing by the \(\beta\to1\) result, so namely, a value of 1 in the plot means the contribution of the mode to the intercept is \(\frac{1}{24}\). In (3) the values should similarly be multiplied by \(\frac{D-26}{24}\). The values in (4) are the values of the intercept with no further normalization factors.}
\end{figure}

The last three equations summarize all that is needed to compute the intercept in the general case. They are written in terms of \(\beta_1\) and \(\beta_2\), which are related as in eq. \ref{eq:beta12} above. We plot them as a function of \(\beta_1\) and the ratio \(r = m_1/m_2\) in figure \ref{fig:intercept_asym}. The quadratic approximation is not expected to be valid for all values of the parameters. Therefore, we also draw the expected range of validity in the same figure. We require that the two arms of the string both satisfy \(T\ell_i / m_i > 1\), which, given eq. \ref{eq:802}, amounts to requiring \(\beta_i > \frac{1}{\sqrt2}\). The plots (1)-(4) in figure \ref{fig:intercept_asym} are best trusted then in the area specified by plot (5).

\clearpage

\section{Assessment of next to leading order corrections} \label{sec:nextorder}
In all previous sections, we have analyzed the system when the fluctuations are taken to be ``small'', by expanding the action, energy, and angular momentum to quadratic order in the fluctuations. Here we would like to show that we are in fact working on a long string expansion, and show for which values of the physical parameters it is valid.

We would like to show that, in expanding the action to quadratic order in the fluctuations, we are effectively using a long string expansion, with higher order terms suppressed by powers of \(L^{-1}\). Therefore, we are also consistent in using the effective string theory action of section \ref{sec:noncritical} which is itself subject to corrections in higher powers of \(L^{-1}\). More precisely, the first requirement that follows from writing the effective string action as we have, is that the string length should be much longer than the intrinsic length scale \(T^{-1/2}\), or \(T L^2 \gg 1\).

To verify that the corrections coming from the fluctuations are small for long strings, and to quantify what it means for the string to be long in this case, we look at the simplest next order terms, which are those of the contribution to the intercept from the transverse modes. We do it first in the gauge with \(R(\sigma)=\frac1k\sin(k\sigma)\).

The next terms are of order \(\lambda^4\). We write the string and point particle contributions to the transverse intercept, which have similar forms. The bulk terms are
\be a_{st,\delta z} = -\frac{T\lambda^2}{2k}\int_{-\ell}^\ell d\sigma \left(\delta\dot z^2+\delta z^{\prime2}+\frac{\lambda^2}{4\cos^2(k\sigma)}(3\delta\dot z^4-\delta z^{\prime4}-2\delta \dot z^2 \delta z^{\prime2})\right)\ee
and from the boundary we have
\be a_{pp,\delta z} = -\frac{m\lambda^2}{2k\cos\delta}\left(\delta \dot z^2 + \frac{3\lambda^2}{4\cos^2\delta}\delta\dot z^4\right)\ee
The condition that the next order be small for all terms is
\be \frac{\lambda^2}{\cos^2\delta}\delta \dot z^2 \ll 1 \label{eq:903}\ee
For the boundary part this is exactly the condition, while to reach this condition for the bulk part we simply note that \(\cos(k\sigma) \geq \cos\delta\) and make the assumption that \(\delta z^\prime \sim \delta \dot z\).

We can estimate the LHS of the inequality as follows. First, \(\cos^{-2}\delta = \gamma^2 = 1 + \frac{TL}{2m}\) by the boundary conditions. Next we evaluate the order of magnitude of the fluctuations, \(\lambda \delta z\), to be of the order of the characteristic length of the string defined by \(\ell_s \approx T^{-1/2}\) so \(\lambda^2\delta z^2\sim T^{-1}\), while each derivative adds factors of order \(k \sim 1/\ell\) or \(\omega_n \sim n/\ell\). The contribution from eigenfrequencies \(\omega_n\) with large \(n\) is regularized when summing over them, so we do not need to keep track of factors of \(n\). This results in the condition
\be \frac{1}{m L} \ll 1 \ee
Note that the assessment above is consistent with the leading order contribution being of order one, as was found in previous sections. The string part is of the order
\be \frac{T}{k}\int_{-\ell}^\ell d\sigma \lambda^2(\delta \dot z^2 +\delta z^{\prime2}) \sim \frac{T}{k}\times \ell \times \frac{k^2}{T} \sim k\ell = \mathcal O(1) \ee
with corrections from the endpoint terms that go like
\be \frac{m\lambda^2}{k\cos\delta}\delta \dot z^2 \sim \frac{mk}{T\cos\delta} \sim \frac{1}{\beta\gamma}\ee
Where in the last part we used the boundary condition, \(\frac{T}{mk} = \gamma^2\beta\).

An alternative assessment would have had the amplitude of the fluctuations to be proportional to the second length scale in the problem, \(\lambda \delta z \sim m^{-1}\). Then the condition on the length is \(mL \gg \frac{T}{m^2}\).

A second condition on the validity of the expansion is that our definition of the intercept as
\be a \equiv \langle\delta\left(J-J_{cl}(E)\right)\rangle \approx \langle\delta J - \frac{1}{k}\delta E\rangle \ee
should make sense. We want to check that the next term in the correction to \(J_{cl}(E)\) is small, namely that
\be \frac{\pa J}{\pa E}\delta E \gg \frac12 \frac{\pa^2 J}{\pa E^2}(\delta E)^2 \ee
The first derivative is \(\frac{1}{k} = \frac{L}{2\beta}\), and computing the second derivative leads to the condition
\be \frac{L}{\beta}\delta E \gg \frac{1}{T} \frac{1+\beta^2}{4\beta\sqrt{1-\beta^2}+2(1+\beta^2)\arcsin\beta}(\delta E)^2  \ee
\be T L \gg \frac{1+\beta^2}{4\sqrt{1-\beta^2}+2(1+\beta^2)\frac{\arcsin\beta}{\beta}}\delta E  \ee
The function on the RHS is always in the range \(\frac{1}{6}\) to \(\frac{1}{\pi}\), so does not add any constraints on \(\beta\), only that
\be T L \gg \delta E \ee
Now if \(\delta E \sim a/L\) where \(a\) is order one, then the constraint reads simply
\be TL^2 \gg 1 \ee
With the previous constraint of \(m L\gg 1\) we find that for the approximation to hold, our string needs to be long compared with both length scales in the problem, \(m^{-1}\) and \(T^{-1/2}\).


We can also see the nature of the approximation a little more explicitly by looking at the contribution of a single mode. We substitute
\be \delta z = \sqrt{\mathcal N}\frac{1}{\omega_n}\cos(\omega_n\tau) f_n(\sigma) \ee
into condition \ref{eq:903} and evaluate the RHS explicitly for the single mode. In the simple case of a symmetric string \(f_n\) is either \(\cos(\omega_n\sigma)\) or \(\sin(\omega_n\sigma)\). The normalization constant \(\mathcal N\) is \(\mathcal N = \frac{1}{2T\ell^2\lambda^2}\), and the eigenmodes normalized as in section \ref{sec:canonical_quantization}, with
\be \frac{1}{\ell}\int_{-\ell}^\ell d\sigma f_n^2(\sigma) + \frac{\gamma m}{T\ell}((f_n^+)^2 + (f_n^-)^2) = 1 \ee
We pick an even mode with \(f_n = c\cos(\omega_n\sigma)\), with \(c\) determined from the above equation. The result is (for \(x=\omega_n\ell\))
\be \frac{\lambda^2}{\cos^2\delta}\delta \dot z^2 = \frac{(\frac{\ell T}{2m}+1) \sin ^2\left(\frac{\tau  x}{\ell}\right)\cos ^2(x)}{\ell \left(2 m  \cos ^2(x) \sqrt{\frac{2  T\ell}{m}+4}+ T\ell (2 +\frac{\sin (2 x)}{x})\right)} \sim \frac{(\frac{T\ell}{2m}+1)}{m\ell \left(\cos^2x \sqrt{\frac{2 T\ell}{m}+4}+\frac{T\ell}{m}\right)} \label{eq:915}\ee
If we are in the regime \(T\ell/m \gg 1\), then
\be \frac{\lambda^2}{\cos^2\delta}\delta \dot z^2 \sim \frac{1}{m \ell} \ee
So the correction is indeed small when \(T L / m \gg 1\) and \(m L \gg 1\).\footnote{The passage from \(\ell\) to \(L\) in this gauge is \(L = 2\ell\frac{\sin\delta}{\delta}\) (see eq. \ref{eq:conditions_sin}), so \(\ell\sim L\).} When the two hold, then also \(TL^2\gg 1\).

To check that the above discussion was not gauge dependent, we see that we find the same condition of \(mL\gg 1\) using the gauge with \(R(\sigma)=\sigma\). We can look at the boundary part of the intercept and see that it is actually identical to that in eq. \ref{eq:903},
\be a_{pp,\delta z} = -\frac{\gamma m\lambda^2}{2k}\left(\delta \dot z^2 + \frac{3\gamma^2\lambda^2}{4}\delta\dot z^4\right)\ee
and so we find the same condition \(mL\gg1\) by demanding that \(\gamma^2\lambda^2\delta \dot z^2 \ll 1\) and estimating the size of the fluctuations as before.

\section{Summary and future prospects} \label{sec:summary}
In this paper we have computed the leading order quantum correction to the classical Regge trajectory of the string with massive endpoints. We have seen that quantum fluctuations around a rotating string solution for the string with massive endpoints with angular velocity \(k\), obey the relation
\be a \equiv \langle\delta\left(J-J_{cl}(E)\right)\rangle= \langle\delta J - \frac{1}{k}\delta E\rangle = -\frac{1}{k} \langle H_{ws}(\delta X^i)\rangle \,, \ee
That is, the quantum intercept is equal to the expectation value of the world sheet Hamiltonian, computed for the fluctuations. The Hamiltonian was seen to be diagonal for both the transverse and planar modes, with the intercept then being proportional to the sum of the eigenfrequencies of the fluctuations. The eigenfrequencies are dependent on the masses at the endpoints of the string through the boundary conditions of each mode, and ultimately we can write \(a\) compactly as a function of the endpoint velocities.

In the course of computing the intercept, we also computed the spectrum of fluctuations around the rotating string, which we used to define the radial trajectories of the string with massive endpoints, as
\be J + \frac{1}{\beta}(N_t+N_p) = J_{cl}(E) + a\label{eq:radial_massive2}\ee
In addition to the intercept, the spectrum of fluctuations is such that some of the degeneracies of the regular bosonic string with massless endpoints are removed, resulting in the splitting of levels.

The intercept of the rotating string is comprised of three parts. One is the contribution of the \(D-3\) fluctuation modes orthogonal to the plane of rotation, one from the mode in the plane, and another from the Polchinski-Strominger term in the effective action.

Each of the contributions is divergent, or has a term that diverges when the endpoint masses are taken to zero. The contributions to the intercept from the fluctuations around the rotating solution are given by the infinite sum over the mode's eigenfrequencies which is divergent. We have shown how to calculate it using a contour integral method, and offered a prescription for renormalizing the divergences by looking at the corresponding Casimir force and subtracting the constant terms in the \(L\to\infty\) limit. The contribution due to the PS term in the non-critical effective string action contained a term that was divergent in the massless limit. In that case, the finite endpoint masses can be thought of as a regulator to that divergence, and we used the same prescription to subtract it. We noted that all divergences are of a form that can be subtracted by adding appropriate counterterms and renormalizing the string tension and endpoint masses.

We showed that the approximation we used, in which the action is truncated to include only terms quadratic in the fluctuations is a long string approximation, so the results we have obtained are only relevant at high energies. In that sense, what we computed is the quantum correction to the asymptotic Regge trajectory at high energies. In the regime where our computation is to be trusted, the intercept is close to 1, which is the massless result of \cite{Hellerman:2013kba}, and receives some small corrections due to the masses.

The finite result for the intercept for the general case of two different endpoint masses was summarized in eqs. \ref{eq:a_t_gen}, \ref{eq:a_p_gen}, and \ref{eq:a_PS_gen}. In the symmetric case, we can write the full intercept as an expansion in \(\frac{2m}{TL}\) at high energies, combining eqs. \ref{eq:a_t_approx}, \ref{eq:a_p_approx}, and \ref{eq:a_PS_approx},
\begin{align} a &= (D-3)a_t + a_p + a_{PS} \approx 1 - \frac{26-D}{12\pi}(\frac{2m}{TL})^{1/2} + \frac{199-14D}{240\pi}(\frac{2m}{TL})^{3/2} \end{align}
The leading order term is from the PS term alone. Furthermore, for \(D=4\) the overall coefficient of \(26-D\) from the PS term is also larger than that of the fluctuations, so the PS term has the dominant contribution to the intercept overall. Therefore, one of the conclusions of this paper in that in the presence of endpoint masses the chief correction to the intercept comes from the modification of the expectation value of the PS term, rather than the contribution from the fluctuations of the string. In the critical dimension \(D=26\), there is a weaker dependence on the masses, as \(a_{D=26} \approx 1 - \frac{11}{16\pi}(2m/TL)^{3/2}\).

Following what was accomplished in this paper, there are several open questions for future inquiry. These include: 
\begin{itemize}

\item A more precise and in depth formulation of the effective string theory with boundary terms. In particular we have assumed that the PS term does not get any correction from the boundary term. The PS term including its coefficient was determined \cite{Polchinski:1991ax} so that one retrieves  the right OPE for the  worldsheet energy-momentum tensor. A similar treatment has to be performed in the presence of the mass term on the boundary. More generally, the question is whether also for the present case, where on the boundary scale invariance is broken explicitly, one has to follow the same procedure for curing the worldsheet scale anomaly in non-critical dimensions as for the massless case. In particular, one can write down additional possible equivalent classical actions for the system. For instance, one could write the Polyakov action for the string instead of Nambu-Goto. A natural question to ask is whether the quantization of these other actions will lead to the same quantum picture derived in this paper.  


\item  As was explained in the introduction the spectra of hadrons are characterized by several different intercepts depending on the masses of the endpoint particles as well as on the electrical charges and  spin and isospin of the corresponding trajectories. In this paper we have considered only the former dependence. We intend to study also the latter. It will be very interesting to compare the predictions of this theoretical analysis with the observational values of the intercept. The question is whether one can isolate the dependence on the endpoint masses and to what extent it is similar to the quantum Regge trajectories calculated here for the string with massive endpoints. A key question for the future work is how to account for the fact that the phenomenological intercept (which we define in the plane of $M^2$ and the orbital angular momentum) is always negative.

\item In \cite{Sonnenschein:2017ylo} we have considered the strong decay width of stringy hadrons. The intercept entered in various points of the analysis in particular in relating the result derived in \cite{Dai:1989cp} for the critical dimension and that for non-critical dimensions in \cite{Mitchell:1989uc}. Now that we have determined the quantization of the string with massive endpoints and the corresponding intercept, it is possible to go back and use the result of this computation to improve our understanding on the strong decay processes and their width.

\item The intercept plays an important role not only in the spectra but also in the scattering of hadrons. This is manifested for instance in the Veneziano amplitude. A natural question to ask is how is this formula modified for strings with massive endpoints. This is an open question that we currently investigate. It is very plausible that the differences between the intercepts associated with this model and that of the massless string case play an important role in the modifications of the scattering amplitudes. Moreover, in the analysis of scattering the asymptotic high energy (or large \(L\)) intercept computed here is more likely to be relevant than in other aspects of phenomenology.

\item Related to the last point, in a recent paper \cite{Sever:2017ylk} a universal correction to the Veneziano amplitude in a quite general setup of weakly interacting high spin particles was found using a bootstrap approach. The result was found in the asymptotic imaginary angle limit of large \(s\) and large \(t\). Moreover, from a string perspective the universal correction could be associated with endpoint masses and their modification of the classical Regge trajectory. Namely, the leading correction to the amplitude is associated with the leading order term \(\sim \alp m^{3/2} E^{1/2}\) in the expansion of the massive modified Regge trajectory \(J_{cl}(E)\). The quantum intercept we have computed now includes the next terms in the asymptotic expansion. The full Regge trajectory defined by \(J_{cl}(E) + a(m/TL)\) now contains more terms in the asymptotic expansion whose effects on the amplitude could be explored, the next to leading term being the zeroth order term in the intercept, which is just 1.

\item In this model we considered the impact of the massive endpoint only on the bosonic string. A related question is what is the impact of the massive endpoints  on a  fermionic string and in the context of the various superstring theories. In addition we intend to explore models where the endpoint particles are fermionic and carry non-trivial spin. These models presumably have more relevance for describing hadrons in terms of strings. 


\end{itemize}

\section*{Acknowledgments}
We would like to thank Ofer Aharony, Simeon Hellerman and Shimon Yankielowicz for taking part in the early stages of this project. We benefited from discussions with Michael Green, Vadim Kaplunovsky, Arkady Tseytlin, Gabriele Veneziano, and especially Ofer Aharony and  Shimon Yankielowicz, with whom we had  many  useful discussions.    We would also like to thank Ofer Aharony and Shimon Yankielowicz for their remarks on the manuscript. J.S. would like to thank the theory group of Imperial College London and the Leverhulme trust for supporting his stay at Imperial College where part of this work has been carried out. This work was supported in part by a center of excellence supported by the Israel Science Foundation (grant number 1989/14), and by the US-Israel bi-national fund (BSF) grant number 2012383 and the Germany Israel bi-national fund GIF grant number I-244-303.7-2013.

\bibliographystyle{JHEP}
\bibliography{intercept}

\clearpage

%
%
%
%

\end{document}